\documentclass[fleqn,usenatbib]{mnras}

\usepackage{newtxtext,newtxmath}
\usepackage[T1]{fontenc}
\usepackage{graphicx}   % Including figure files
\usepackage{amsmath}    % Advanced maths commands
\usepackage{comment}
\usepackage{epsfig}
\usepackage{xcolor}
\usepackage{tabularx}
\usepackage{threeparttable}
\usepackage{hyperref}
\hypersetup{colorlinks=true, citecolor=blue, linkcolor=blue}
\usepackage{xspace}
\let\oldAA\AA
\renewcommand{\AA}{\text{\oldAA}\xspace}

\newcommand{\redtxt}[1]{\textcolor{black}{#1}}
\newcommand{\realredtxt}[1]{\textcolor{black}{#1}}

\newcommand{\hei}{He\,{\sc i}}
\newcommand{\heii}{He\,{\sc ii}}

\newcommand{\oiii}{[O\,{\sc iii}]}

\newcommand{\civ}{C\,{\sc iv}}
\newcommand{\feii}{Fe\,{\sc ii}}

\newcommand{\ha}{H$\alpha$}
\newcommand{\hb}{H$\beta$}

\newcommand{\kms}{$\rm km~s^{-1}$}

\newcommand{\jwst}{\textit{JWST}}
\newcommand{\cloudy}{\textsc{Cloudy}}

\newcommand{\blackthunder}{BlackTHUNDER}

\newcommand{\target}{Abell 2744-QSO1\xspace}

\title[BlackTHUNDER - A non-stellar Blamer break in a LRD]{\centering BlackTHUNDER -- A non-stellar Balmer break in a black hole-dominated little red dot at $z=7.04$}
\author[Ji et al.]{Xihan Ji,$^{1,2}$\thanks{E-mail: \href{mailto:xj274@cam.ac.uk}{xj274@cam.ac.uk}}
Roberto Maiolino,$^{1,2,3}$
Hannah \"{U}bler,$^{4}$
Jan Scholtz,$^{1,2}$
Francesco D'Eugenio,$^{1,2}$
Fengwu Sun,$^{5}$
\newauthor
Michele Perna,$^{6}$
Hannah Turner,$^{1,2}$
\redtxt{Stefano Carniani,$^{7}$}
Santiago Arribas,$^{6}$
Jake S. Bennett,$^{5}$
Andrew Bunker,\redtxt{$^{8}$}
\newauthor
Stéphane Charlot,$^{9}$
Giovanni Cresci,$^{10}$
Mirko Curti,$^{11}$
Eiichi Egami,$^{12}$
Andy Fabian,$^{13}$
Kohei Inayoshi,$^{14}$
\newauthor
Yuki Isobe,$^{1,2,15}$
Gareth Jones,$^{1,2}$
Ignas Juodžbalis,$^{1,2}$
Nimisha Kumari,$^{16}$
Jianwei Lyu,$^{12}$
\newauthor
Giovanni Mazzolari,$^{17,18}$
Eleonora Parlanti,$^{8,4}$
Brant Robertson,$^{19}$
Bruno~Rodr\'iguez~Del~Pino,$^{6}$
\newauthor
Raffaella Schneider,$^{20,21,22,23}$
Debora Sijacki,$^{13,1}$
Sandro Tacchella,$^{1,2}$
Alessandro Trinca,$^{24,21,22}$
\newauthor
Rosa Valiante,$^{21,22}$
Giacomo Venturi,$^{8}$
Marta Volonteri,$^{9}$
Chris Willott,$^{25}$
Callum Witten,$^{26}$
Joris Witstok$^{27,28}$
\\
Affiliations are listed after the references.
}

\begin{document} 
\label{firstpage}
\pagerange{\pageref{firstpage}--\pageref{lastpage}}
\maketitle
 
\begin{abstract}
Recent observations from \jwst\ have revealed an abundant population of active galactic nuclei (AGN) and so-called ``Little Red Dots'' (LRDs) at $2\lesssim z \lesssim 11$, many of which are characterized by V-shaped UV-to-optical continua with turnovers around the Balmer limit.
The physical nature of these LRDs is unclear, and it remains debated whether the peculiar spectral shape originates from AGN, compact galaxies, or both.
We present the analysis of new NIRSpec-IFU data from the BlackTHUNDER \jwst\ Large Programme and
archival NIRSpec-MSA data of a lensed LRD at $z=7.04$. 
The spectra confirm the presence of a smooth Balmer break and a broad H$\beta$ tracing the Broad Line Region (BLR) of an AGN.
The small velocity dispersion of the H$\beta$ narrow component indicates a small dynamical mass of the host galaxy of $M_{\rm dyn}<4 \times 10^8~M_{\odot}$, which implies that the stellar population cannot contribute more than 10\% to the optical continuum.
We show that the Balmer break can be well described by an AGN continuum absorbed by very dense ($n_{\rm H}\sim 10^{10}~{\rm cm^{-3}}$) and nearly dust-free gas along our line-of-sight (possibly gas in the BLR or its surrounding). The same gas is expected to produce H$\beta$ absorption, at a level consistent with a tentative detection ($3\sigma$) in the high-resolution spectrum.
Such a non-stellar origin of the Balmer break may apply to other LRDs, and would alleviate the issue of extremely high stellar mass surface densities inferred in the case of a stellar interpretation of the Balmer break.
We note that this is a rare case of a black hole that is overmassive relative to both the host galaxy stellar {\it and} dynamical masses. We finally report indications of variability and the first attempt of AGN reverberation mapping at such an early epoch.
\end{abstract}

%
%-------------------------------------------------------------------

\section{Introduction}

The first two years of observations with the James Webb Space Telescope (\jwst) have revealed a large number of accreting black holes (Active Galactic Nuclei, AGN) at $2\lesssim z \lesssim 11$ \citep[e.g.,][]{kocevski2023,ubler2023a,Uebler24a,harikane2023,Perna23,Perna23b,larson_ceersagn_2023,maiolino2023,maiolino2023b,matthee2024,greene2024,juodzbalis_dormantagn_2024, scholtz2023}. A significant fraction of these AGN were classified as Type 1, through the detection of a broad component in permitted emission lines (primarily H$\alpha$ and/or H$\beta$) with a full-width-at-half-maximum (FWHM) $\gtrsim 1000$ \kms, and without a counterpart in forbidden emission lines (in particular \oiii$\lambda 5008$), indicating that the broad component emerges from the Broad Line Region of AGN \citep[BLR,][]{kocevski2023,ubler2023a,harikane2023,maiolino2023,taylor24,tripodi24}. An even more abundant population of narrow-line (Type 2) AGN has also been identified by the detection of high-ionization emission lines or by using narrow-line diagnostic diagrams (e.g., \citealp{scholtz2023,mazzolari2024,mazzolari_ceers_2024,chisholm_nlagn_2024}). All these studies have opened the exploration of intermediate-/low-luminosity AGN with typically estimated $L_{\rm bol}\lesssim 10^{45}~{\rm erg~s^{-1}}$ and $M_{\rm BH}\lesssim 10^{7-8}~M_{\odot}$ at $z\gtrsim4$, populating a parameter space not sampled by observations in the pre-\jwst\ era.

However, the new population of AGN revealed by \jwst\ has properties that are quite different compared to the general population of AGN in the local Universe as well as luminous and distant quasars (QSOs). In fact, most of the \jwst\ selected AGN are characterized by significant X-ray weakness, many of them showing no detection of hard X-ray emission at $2-10$ keV in the rest frame \citep{yue_agn_2024,ananna_agnxray_2024,lyu_miriagn_2024,Maiolino2024_Xrays}, as well as radio weakness, at $5$ GHz in the rest frame (\citealp{mazzolari_radioagn_2024,juodzbalis_agnabs_2024}).
In terms of continuum emission, many \jwst-selected AGN with multi-epoch photometry seem to lack variability in the rest-frame optical and UV \citep{Maiolino2024_Xrays,kokubo_harikane2024,zhang_lrdvar_2024}.
In terms of emission lines, the Type 1 AGN selected by \jwst\ show a significantly weaker \feii\ emission bump in the optical compared to low-redshift AGN with similar luminosities \citep{trefoloni_feii_2024}.
In addition, the loci of \jwst-selected AGN in widely used optical narrow-line diagnostic diagrams such as BPT/VO diagrams \citep{bpt,vo87} are clearly offset from those of low-redshift AGN \citep{harikane2023,ubler2023a,maiolino2023b}.
%lack of prominent iron emission bump \citep{trefoloni_feii_2024}, lack of variability \citep{Maiolino2024_Xrays,kokubo_harikane2024,zhang_lrdvar_2024}, and optical narrow line diagnostics that are offset from the locus of local AGN \citep{harikane2023,maiolino2023b,ubler2023a}. 
These peculiarities have raised concerns about the true identities of 
these \jwst-selected sources and
prompted the question of non-AGN scenarios for explaining this population of objects \citep{kokubo_harikane2024,Maiolino2024_Xrays,yue_agn_2024}. However, it has been shown, on various physical and observational grounds, that currently proposed non-AGN scenarios are not tenable for most of these sources and produce more difficulties (\citealp{killi_lrd_2023,juodzbalis_agnabs_2024,Maiolino2024_Xrays}; Ji et al. in prep).
Alternatively, the above peculiarities of \jwst-selected AGN might be explainable in the context of high (near-to-super Eddington) gas accretion around the central black hole and/or dense gas obscuration due to changes in the BLR structure \citep[e.g.,][]{schneider_2023,king_2024,im24,inayoshi_spedd_2024,madau_spedd_2024,Maiolino2024_Xrays,pacucci_spedd_2024,trefoloni_feii_2024,trinca2024, Madau2025}.

A fraction of this new population of AGN uncovered by \jwst\ exhibits further peculiar features by having
red optical colors and compact morphologies in their NIRCam images, and ``V-shaped'' spectral energy distributions (SEDs) with turnovers at roughly 4000 \AA\ in the rest frame \citep[e.g.,][]{greene2024, wangbingjie_lrd_2024, Setton2024}. % \redtxt{[refs]}.
Based on the above observational facts, these sources are usually referred to as ``Little Red Dots'' (LRDs).
%spectral slopes typical of the so-called `Little Red Dots' (LRDs). 
%These are high redshift systems characterized by compact sizes, red optical rest-frame slopes and blue UV rest frame slopes, yielding a typical V-shaped spectrum (refs).
Notably, not all \jwst-selected AGN are found to be LRDs, and not all LRDs are identified as AGN.
While the fraction of AGN among LRDs is debated (ranging between 20\% and 80\%, e.g., \citealp{greene2024,kocevski2024,perez-gonzalez_lrd_2024}), which likely depends on the specific criteria for selecting LRDs, the opposite has been established quite carefully: only about 10\%-30\% of the new population of Type 1 AGN discovered by \jwst\
have colors and/or slopes typical of LRDs \citep{hainline2024}, although the fraction probably depends on luminosity and redshift.

An intriguing feature of spectroscopically confirmed LRDs is the presence of a prominent break at around 4000 \AA\ in the rest frame \citep[e.g.,][]{kocevski2024,wangbingjie_lrd_2024,wang_break_2024,furtak2024,greene2024}. If this feature is interpreted as the Balmer break associated with stellar populations, one usually obtains very high stellar masses and, combined with the compact sizes of LRDs ($R_{\rm e}\lesssim100$ pc), extremely high stellar densities \citep[$\sim10^4-10^7~{M_\odot~{\rm pc^{-2}}}$ from a maximum AGN contribution to no AGN contribution, see e.g.,][]{baggen_2024} not observed in galaxies at lower redshift  \citep[e.g.,][]{akins_lrd_2024,baggen_2024,guia_lrd_2024,mayilun_lrd_2024,wangbingjie_lrd_2024,wang_break_2024}.
In fact, such a high density is only comparable locally to the nuclear star clusters (NSCs) at the center of galaxies \citep[e.g.,][]{hopkins_2010,Pfister_nsc_2020}.
For example, the NSC in the Milky Way (MW) has a stellar density of $\sim 10^{6.4}~{M_\odot~{\rm pc^{-2}}}$ but is only within the central 0.5 pc \citep{neumayer_2020}.
%No other system so massive within a few 100 pc and with such high stellar density has been ever observed at lower redshift and in the local Universe, 
In addition, the implied stellar mass densities from LRDs within the cosmic volume might be in tension with the standard $\rm \Lambda CDM$ cosmology \citep{akins_lrd_2024,inayoshi_ombh_2024,wang_break_2024}.
The above stellar Balmer break interpretation hence raises the questions of what physical mechanisms make these systems so massive and what could dissolve these systems into the galaxies at low redshift.

Another interesting feature found in $10-20$\% of the Type 1 AGN newly discovered by \jwst\ (including LRDs) is the presence of H$\alpha$ and/or H$\beta$ absorption along with the broad emission components \citep[e.g.,][]{matthee2024,wangbingjie_lrd_2024,wang_break_2024,juodzbalis_agnabs_2024,kocevski2024,lin_aspire_2024,labbe_monster_2024,taylor24}. These absorption features are relatively narrow (with typical $\rm FWHM \sim 100-200$ \kms) yet deep (with typical equivalent widths of $\rm EW\gtrsim 4$ \AA).
%distinguishes them from photospheric absorption in stars
%that are much broader due to pressure broadening \redtxt{(value)}.
The centroids of the absorption are generally slightly blueshifted (typically by $\sim$100-200 \kms), but there are cases in which the absorption is nearly at the rest frame of the narrow-line emission or even redshifted \citep[e.g.,][]{matthee2024,
labbe_monster_2024}. 
%These findings are remarkable. Indeed, in contrast to other absorption transitions, 
If the Balmer absorption observed in LRDs comes from photospheric absorption in stars, based on the extremely high stellar masses inferred from the Balmer break ($\gtrsim 10^{10}~M_\odot$) and the small sizes ($<100$ pc) of LRDs, the expected FWHM should be $\gtrsim 1000$ \kms\ \citep{baggen_2024}, which is significantly broader than the observed widths. Additionally, the large EW of the H$\alpha$ absorption (relative to the continuum) is generally far larger than observed in any stellar population. 
%In many of these cases, the H$\alpha$ absorption is so deep to become negative when the broad H$\alpha$ emission is subtracted. 
Furthermore, many of these absorption features show velocity offsets relative to both the narrow and broad \ha. Therefore, the Balmer absorption in these systems cannot have a stellar origin.

The findings of the Balmer absorption lines are surprising because, unlike lines that are often seen in absorption such as \civ$\lambda \lambda 1548,1551$,
\ha\ and \hb\ are not resonant lines (i.e., the lower level of the emission, $n=2$ of hydrogen, is not even metastable and is extremely short-lived).
Hence, to see Balmer lines in absorption, it requires the $n=2$ level of hydrogen to be populated and to keep it populated.
By studying an AGN with deep Balmer absorption selected by \jwst\ at $z=2.26$,
\citet{juodzbalis_agnabs_2024} showed that, in order to produce the observed absorption, one needs (in addition to temperatures typical of the warm ionized interstellar medium (ISM), which is $\sim 10^4$ K) gas densities higher than $10^8~{\rm cm^{-3}}$. 
The required densities are much higher than the densities found in the ISM of any galaxies at $0\lesssim z\lesssim 10$ \citep[][]{ji_agnbound_2020,DaviesRL2021,isobe_ne_2023,lisijia_ne_2024}.
In fact, such high densities are more typical of the clouds in the BLRs of AGN \citep{netzer1990}.
%inconsistent with the ISM in any kind of galaxy, but are typical of the densities in the BLR clouds or its immediate surroundings. 
Given the identifications of AGN in these sources, a natural explanation is that we are seeing absorption from clouds within or very close to the BLRs.
%in these objects we are likely seeing the nuclear region absorbed by clouds of the BLR or in its vicinity. 
%Being these clouds within the dust sublimation radius they are mostly dust free (although some dust may still be present, as discussed in ...), hence explaining the detection of broad-line and modest dust reddening in these systems. 
Since the BLR clouds are supposed to lie within the dust sublimation radius and are thus likely dust free \citep[although some dust might still be present in the outer BLRs, e.g.,][]{shields2010}, the broad-line emission is not extinguished by dust despite the potentially large column densities of these clouds \citep[$\sim 10^{23}~{\rm cm^{-2}}$,][]{netzer1990}.
The small velocities and velocity dispersions of the absorption lines imply that such nuclear gas is not rapidly outflowing. 

%It is also interesting to note that at least 20\% of AGN discovered by \jwst\ presents H$\alpha$ or H$\beta$ in absorption. 
It is worth noting that the current fraction of the Balmer absorption among \jwst-discovered AGN and LRDs is about 10-20\%, but this is likely a lower limit.
This is because detecting \ha\ (not to mention \hb) in absorption requires high signal-to-noise (S/N) and high spectral resolutions, as the absorption can be hidden by the narrow component in the emission of the same line.
For instance, \citet{deugenio_restoutflow_2025} report a case where H$\alpha$ absorption is seen when using R2700 spectroscopy, while the same absorption is undetected in the R1000 spectrum. Since most of the \jwst\ spectra for Type 1 AGN are obtained with low-resolution NIRSpec/PRISM ($R\sim 100$) and medium-resolution NIRSpec/grating ($R\sim 1000$) or NIRCam/grism ($R\sim 1600$) spectroscopy, the Balmer absorption lines in these AGN might well go undetected.

Within the same context, \citet{im24} have pointed out that the presence of H$\alpha$ and/or H$\beta$ in absorption must also entail the absorption in higher-order Balmer lines and the presence of a Balmer break.
In addition, \citet{im24} point out that at least part of the Balmer break seen in LRDs (with broad line detections) discovered by \jwst\ could have a non-stellar origin, and the non-stellar part must be associated with the same dense gas responsible for H$\alpha$ and/or H$\beta$ absorption.
As speculated by \citet{im24}, the Balmer breaks observed in LRDs can potentially be entirely associated with non-stellar gas absorption, although thus far, no full-spectral fitting attempts have been made for individual spectra of \jwst\ selected LRDs confirmed as broad-line sources.
%They showed that the Balmer breaks observed in LRD can potentially be entirely associated with gas absorption, although they do not attempt to fit the Balmer break in the spectra of individual objects.
The non-stellar explanation for the Balmer breaks in LRDs has the advantages of alleviating the issue associated with extreme stellar densities as well as the need for special dust attenuation laws \citep[see e.g.,][]{li_lrdatt_2024,mayilun_lrd_2024}.
Still, these aspects remain to be tested with actual observations.
%If the Balmer breaks are entirely or even partially associated with gas absorption, then the inferred stellar masses would be  reduced and this would greatly alleviate the issue of the extreme stellar densities found in some LRDs.

In the context of broad-line LRDs in observations, \target\ is one of the most intriguing and enigmatic objects. 
It was first identified by \citet{furtak_abell2744_2023} as a triply imaged red point-like source lensed by the cluster Abell 2744, as part of observations from the UNCOVER survey \citep[Ultra-deep NIRSpec and NIRCam ObserVations
before the Epoch of Reionization; PIs: I.~Labbé, R.~Bezanson;][]{Bezanson2024}. 
Subsequent low-resolution (PRISM) NIRSpec-MSA spectroscopy confirmed a spectroscopic redshift at $z=7.04$ and identified the presence of broad H$\beta$, which indicates the presence of an AGN \citep{furtak2024}. This source also has a V-shaped continuum, with a red optical slope and a blue UV slope in the rest frame \citep{furtak2024}. 
The spectrum shows, as in other LRDs, the presence of a turnover around 4000 \AA, resembling the shape of a Balmer break. 
However, as pointed out recently by \citet{mayilun_lrd_2024}, the spectral shape of \target\ cannot be well reproduced with a stellar continuum and an associated Balmer break without invoking a large stellar mass of $M_*\approx4\times10^9~M_\odot$, an unusually steep dust attenuation law, and a dust attenuated AGN continuum in the rest-frame optical.
The derived stellar mass, combined with the small size of the system ($R_e<30$ pc), places \target\ among the galaxies with the highest stellar mass surface density observed to date \citep[$>10^6~{M_\odot~{\rm pc^{-2}}}$,][]{mayilun_lrd_2024}.
Even with the best-fit model that includes the above assumptions, the reduced $\chi^2$ of the fit \textit{excluding} emission lines and the continuum region close to Ly$\alpha$ emission is $2.85$ \citep[Table 2 of][the full range is $\chi^2_{\nu}=2.74-4.32$ based on different model assumptions]{mayilun_lrd_2024}.
On the other hand, the same authors point out that an AGN continuum absorbed by dust cannot reproduce the shape of the turnover either, unless a very peculiar extinction curve is assumed.

In this work, we combine previous NIRSpec-MSA spectroscopic observations of \target\ with newly obtained high-resolution (grating; $R\sim 2700$) and low-resolution (PRISM; $R\sim 100$) NIRSpec-integral field unit (IFU) spectroscopic observations from the \blackthunder\ Large Programme (Black holes in THe early Universe aNd their DensE surRoundings; PID 5015; PIs: H.~\"Ubler, R.~Maiolino).
Among other findings, we illustrate that the shape of the continuum can be fit by an AGN continuum absorbed by dense gas along the line-of-sight, which reproduces the smooth Balmer break seen in the low-resolution spectrum and the tentative H$\beta$ absorption seen in the high-resolution spectrum.

The layout of the manuscript is as follows.
In Sections~\ref{sec:data} and \ref{sec:photometry}, we describe the observational data and the data reduction.
In Section~\ref{sec:R2700_measure}, we describe our spectral measurements.
%for the new \blackthunder\ spectrum.
%We comment on the lens modelling in Section~\ref{sec:lens_model}. 
%In Sections~\ref{sec:mass_acc}, \ref{sec:mdyn}, and \ref{sec:scaling}, 
In Section~\ref{sec:bh_properties}, we present our derivation of the black hole parameters and discuss them in the context of scaling relations.
In Section~\ref{sec:balmer_break}, we describe our new method to fit the UV-to-optical \jwst/NIRSpec spectra of \target\ and present our fitting results.
In Section~\ref{sec:variability}, we present our analysis on the variability of \target\ based on multiple observations.
We discuss the implications of our results in Section~\ref{sec:discussion} and draw our conclusions in Section~\ref{sec:conclude}.
Throughout this work, we assume a flat $\rm \Lambda CDM$ cosmology with $h=0.674$ and $\Omega _{\rm m} = 0.315$ \citep{planck2020}.

% \begin{itemize}
%     \item AGN selected by \jwst\ are puzzling.
%     \item LRD with a ``V'' spectral shape is of great interest. Fitting it as a stellar Balmer break predicts very density stellar disks.
%     \item Abell 2744-QSO1 is a particularly interesting case where the Balmer break appear smooth. despite many fitting attempts, the spectral decomposition of Abell 2744-QSO1 into stellar and AGN components remain uncertain
%     \item alternatively, IM24 recently suggested that the dense gas in the BLR can produce a Balmer break without involving massive stellar populations. this also explains the Balmer absorption observed in some \jwst\ AGN
%     \item in this work we aim to explore further the idea of IM24 by fitting the Balmer break in Abell 2744-QSO1 with a gas obscured AGN continuum. our works combine the most recent high-resolution observations from Black Thunder with the previous PRISM observations of Abell 2744-QSO1 and fit both the continuum and emission lines
% \end{itemize}

\section{Observational data}
\label{sec:data}

\begin{figure*}
    \centering
    \includegraphics[width=0.8\textwidth]{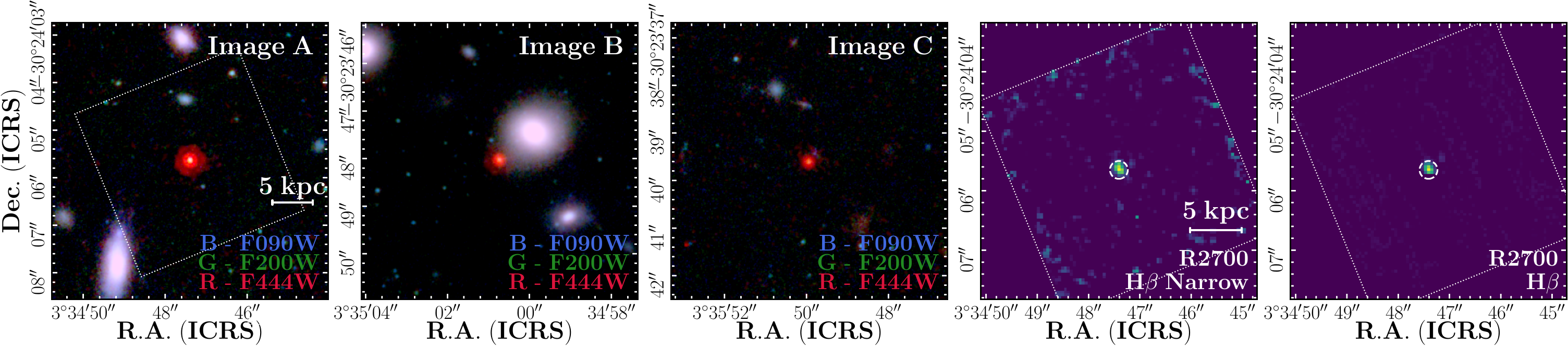}
    \includegraphics[width=0.8\textwidth]{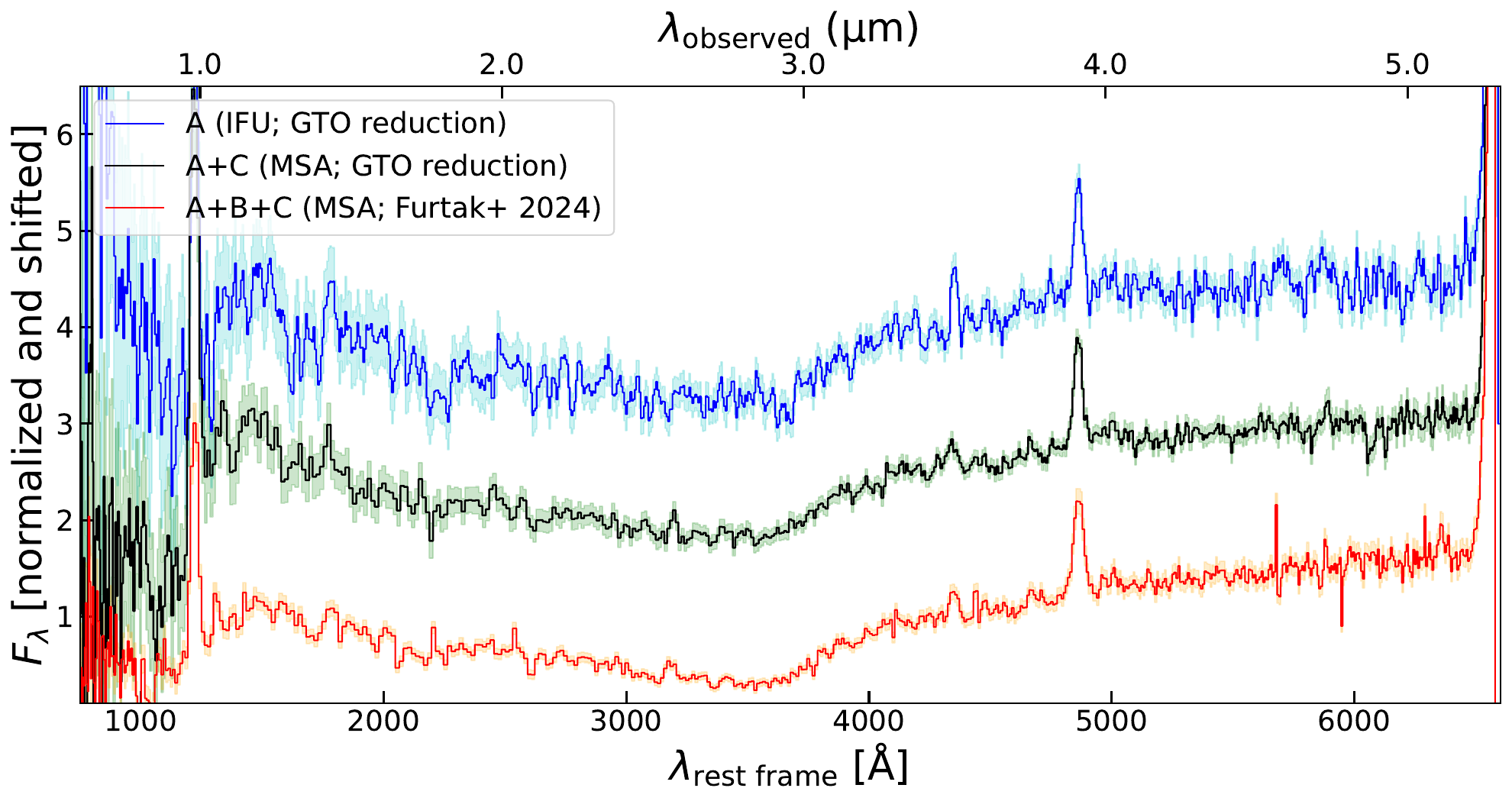}
    \includegraphics[width=0.8\textwidth]{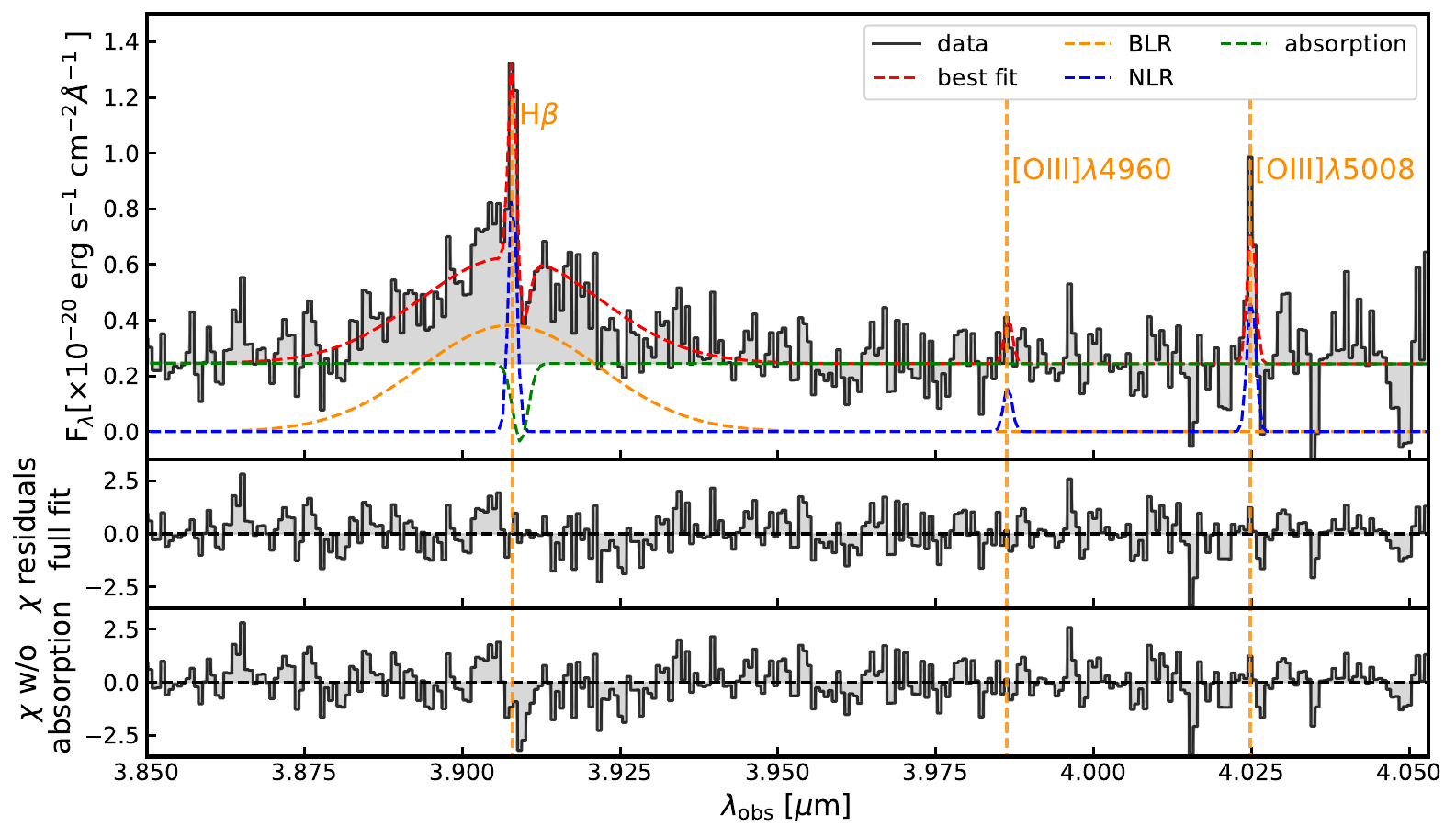}
    \caption{Observations of the triply imaged \target.
    \textit{Top:} Color-composite NIRCam images as well as NIRSpec/IFU maps.
    The NIRCam images of three lensed images from the UNCOVER DR4 \citep{uncover_dr4} are shown, where the field of view of the IFU observation for image A is indicated by the dotted white rectangular aperture.
    For IFU maps, fluxes of the narrow and the whole \hb\ are shown and the extraction aperture for the PRISM and grating spectra is indicated by the dashed white circle.
    \textit{Middle:} IFU and MSA PRISM spectra combining different lensed images normalized to the flux density at $\lambda = 4260$ \AA in the rest frame.
    The spectra are manually shifted in the $y$ axis for presentation purposes.
    From top to bottom, the spectra correspond to the central region of image A extracted from a {circular} aperture with a diameter of $0.\!\!^{\prime\prime}25$ using the \blackthunder\ IFU observations with the GTO reduction, the A and C images combined from the MSA observations with the GTO reduction (image B is removed due to the background subtraction issue described in Section~\ref{sec:data}), and the A, B, and C images combined from the MSA observations with \citet{furtak2024}'s reduction, respectively.
    The shaded regions correspond to the $1\sigma$ pipeline uncertainties.
    %The solid black spectrum combines images A and C reduced by the GTO pipeline, with the shaded green region representing the $1\sigma$ pipeline uncertainty.
    %The light red spectrum is taken from \citet{furtak2023} with their reduction that combines images A, B, and C.
    \textit{Bottom:} The high-resolution (R2700) \blackthunder\ IFU spectrum of image A extracted from a circular aperture of $0\farcs25$ zoomed around the location of \hb\ and \oiii$\lambda \lambda 4960,5008$.
    The dashed lines are the best-fit emission line and continuum models, which show a tentative absorption in \hb\ at a significance of $3\sigma$. 
    The two bottom panels show the residuals normalized by uncertainties ($\chi={\rm residual}/\sigma$) of fits with and without the \hb\ absorption, respectively.
    The fit with the \hb\ absorption produces improved $\chi$ residuals near the centroid of the broad \hb\ ($\sim 3.91~{\rm \mu m}$).
    %\redtxt{[showcase the difference between images and motivate the usage of A+C [this can go in the appendix]}
    }
    \label{fig:spec_qso1}
\end{figure*}

\begin{table*}
        \centering
        \caption{Spectral measurements for \target\ based on the new \blackthunder\ IFU R2700 spectrum extracted from the central $0\farcs25$ of image A (shown in the bottom panel of Figure~\ref{fig:spec_qso1}) as well as the \redtxt{GTO reduced R100 spectrum}.
        The velocities are measured with respect to the narrow component. The fluxes are corrected for aperture losses and are scaled to the flux of the IFU PRISM spectrum. The fluxes are not corrected for lensing magnification.
        }
        \label{tab:measurements}
        \begin{tabular}{l c c c c}
                \hline
                Line & $v$ [\kms] & FWHM [\kms] & Flux [$10^{-20}~{\rm erg~s^{-1}~cm^{-2}}$] & Equivalent width [\AA] \\
                \hline
                \multicolumn{5}{|c|}{\redtxt{R2700 spectrum}} \\
                \hline
                \hb\ (narrow emission) & - & $<75$ ($<112$ conserv.) & 
                %$14.6_{-2.5}^{+2.96}$ 
                {$12.8_{-2.2}^{+2.6}$}
                & $7.5_{-2.1}^{+2.5}$ \\
                \hb\ (broad emission) & $17_{-43}^{+41}$ & $2658_{-292}^{+351}$ & 
                %$135_{-9}^{+12}$
                {$118_{-8}^{+10}$}
                & $72.6_{-7.7}^{+10.1}$ \\
                \hb\ (absorption) & $101_{-77}^{+65}$ & $185_{-72}^{+69}$ & %$8.4_{-2.6}^{+2.3}$ & 
                %$-8.4_{-2.3}^{+2.6}$ 
                {$-7.3_{-2.0}^{+2.3}$}
                & $5.5_{-1.7}^{+2.2}$ \\
                \oiii$\lambda5008$ & - &  $<75$ ($<112$ conserv.)& 
                %$8.7_{-2.1}^{+2.1}$ 
                {$7.6_{-1.8}^{+1.8}$}
                & $4.52_{-1.2}^{+1.4}$ \\
                \hline
                \multicolumn{5}{|c|}{\redtxt{R100 spectrum}} \\
                \hline
                \redtxt{Ly$\alpha$} & \redtxt{-} & \redtxt{unresolved} & \redtxt{$491\pm 16$} & \redtxt{$155\pm 11$} \\
                \hline
        \end{tabular}
\end{table*}

\subsection{BlackTHUNDER NIRSpec-IFU spectra}
\label{sec:BT_data}

We used integral field spectra from the BlackTHUNDER Large Programme (PID 5015; PIs: H.~\"Ubler, R.~Maiolino). The programme consists of observations with the NIRSpec-IFU mode \citep{jakobsen2022,boker2022,boker2023}, both at high (G395H, hereafter R2700) and low (PRISM) spectral resolution, of a sample of 20 Broad-Line AGN previously identified by \jwst\ MSA observations at $z>5$ \footnote{The sample was selected from published sources at the time of the proposal writing and excluded those already having IFU observations.}.

Image A of Abell-2744-QSO1 was observed for 7.4 hours with the high-resolution grating G395H and 2 hours with the PRISM on December 10, 2024. Image A is less magnified than image B. However, image B is contaminated by a bright foreground galaxy, closer than $1''$  in projection.
14 dithers with the ``medium'' pattern were adopted for both observations. For the grating observation, 26 groups and the NRSIRS2 readout mode were adopted, while for the PRISM observation, we used 34 groups and the NRSIRS2RAPID pattern \citep{rauscher_2017}.

We downloaded raw data files from the Barbara A.~Mikulski Archive for Space Telescopes (MAST) and subsequently processed them with the {\it JWST} Science Calibration pipeline\footnote{\url{https://jwst-pipeline.readthedocs.io/en/stable/jwst/introduction.html}} version 1.15.0 under the Calibration Reference Data System (CRDS) context jwst\_1281.pmap. To increase data quality, we made the following modifications to the default reduction steps \citep[see][for details]{Perna23}:
count-rate frames were corrected for $1/f$ noise through a polynomial fit; 
and during calibration in Stage 2, we removed regions affected by failed open MSA shutters and strong cosmic ray residuals.
Remaining outliers were flagged in individual exposures using an algorithm similar to {\sc lacosmic} \citep{vDokkum01} and we rejected the 95\textsuperscript{th} (99.5\textsuperscript{th}) percentile of the resulting distribution for the grating (PRISM) observations. 
%We need to mask any major pixel outliers that were not flagged by the data reduction pipeline. Although these pixels do not cause significant problems during the emission line fitting of galaxy-integrated spectra, these outliers can become a problem during the fitting of faint emission lines in the data. 
The cubes were combined using the `drizzle' method and reconstructed using a $0.\!\!^{\prime\prime}05$ spaxel scale. 
To identify residual outliers not flagged by the procedures summarized previously, we used the \texttt{ERROR} extension of the data cubes. We flagged any pixels whose error is 10$\times$ above the median error value of the cubes. We verified that this choice of threshold does not have any impact on the emission line maps by using the 5 and 20 thresholds without any changes to our results. 

To perform background subtraction, we mask the location of the source based on its H$\beta$ emission ($2\sigma$ S/N contours). Then, we estimate the background using \texttt{astropy.photutils.background.Background2D} (2D background estimator) task with 5$\times$5 spaxels box window, for each individual channel in the data cubes. We visually inspected the resulting background spectra and found no evidence of narrow features (e.g.,~emission or absorption lines). Therefore, we smoothed the background in spectral space using a median filter with a width of 25 channels to reduce any noise effects. The final estimated background is subtracted from the flux data cubes. %\redtxt{[JS: I can see that the BKG was also estimated differently on the top. I think I got cube with BKG and then subtracted it myself using method above. Let me know if I should change something.]}

%old text% Background was extracted in regions far from the source and taking the median of the sky and then subtracting it from the extracted spectrum, after renormalizing to the adopted aperture.

As a basis for our scientific analysis of the NIRSpec-IFU data, we extracted the integrated PRISM spectrum with an extraction aperture of diameter $0.\!\!^{\prime\prime}25$. We also estimated the wavelength-dependent aperture losses by extracting the spectrum with a much larger diameter of $0.\!\!^{\prime\prime}5$. The latter spectrum is much more noisy, but can be used to estimate the total flux of the source at each wavelength, which was then used to correct the flux (as a function of wavelength) of the smaller (and higher S/N) aperture spectrum.
The average correction factor around \hb\ is 1.8.
{The integrated grating spectrum is extracted from the same circular aperture of $0.\!\!^{\prime\prime}25$ and a similar process for aperture-loss correction is applied. 
However, the flux extracted from the grating spectrum is roughly a factor of 1.9 higher than that of the PRISM spectrum near the location of \hb, possibly due to a flux calibration issue.
We also compare the flux of the IFU PRISM spectrum with that of the MSA PRISM spectrum (image A with the GTO reduction; see also Sections~\ref{sec:archival_spec_data} and \ref{sec:variability}) and find good consistency.
Therefore, we scale the flux of the grating spectrum to match that of the PRISM spectrum.
}

Previous works have noted that the error extension in the combined NIRSpec-IFU cube can be underestimated \citep[e.g.,][]{ubler2023a, RodriguezDelPino2024}. To obtain realistic uncertainties on our aperture-integrated spectra, we rescale the noise spectrum with a measurement of the noise obtained from the sigma-clipped rms of the integrated spectra in regions free of line emission.

{The top panel of Figure~\ref{fig:spec_qso1} shows the field of view of the IFU observation and the extraction aperture on top of the NIRCam images (to be introduced in the next subsection) and IFU images.
The middle panel shows the extracted 1D PRISM spectrum for image A from \blackthunder\ together with MSA spectra from images A+C and images A+B+C, which we introduce in Section~\ref{sec:archival_spec_data}.
}
The bottom panel of Figure~\ref{fig:spec_qso1} shows a portion of the extracted spectrum around H$\beta$ and \oiii$\lambda \lambda 4960,5008$, illustrating the well resolved broad component of H$\beta$, a clear narrow component of H$\beta$ and a narrow \oiii$\lambda 5008$.
H$\beta$ also presents a tentative detection of an absorption feature, which will be discussed in Section~\ref{sec:R2700_measure}.

%\begin{itemize}
%    \item Black Thunder observations, reduction (errors), analyses
%    \item Furtak's source: UNCOVER observations and GTO JADES reductions (errors); other observations in the Appendix
%\end{itemize}

\subsection{Archival data}
\label{sec:archival_data}

\subsubsection{Imaging}
\label{sec:archival_imaging_data}

\target\ was initially discovered through NIRCam imaging observations of UNCOVER in November 2022 \citep{furtak_abell2744_2023,Bezanson2024}. 
Cycle-1 DDT-2756 (PI: Chen) also obtained NIRCam imaging observations covering the three images of \target\ on October 20 and Dec 6, 2022. The repeated filters are F115W, F150W, F200W, F277W, F356W and F444W. 
In November--December 2023, Cycle-2 program GO-3516 (All the Little Things, ALT; \citealt{naidu24}) obtained repeated NIRCam imaging observations in the F356W band at two PAs. 
We have processed these imaging data through the same \jwst\ pipeline version 1.11.2 and CRDS context \verb|jwst_1174.pmap|.
In addition to the standard \jwst\ pipeline, we also include many frequently adopted customized steps, such as $1/f$ noise subtraction, additional bad pixel masking, wisp removal, and global background subtraction.
For the UNCOVER observations, two epochs of DDT-2756 and ALT images are mosaicked separately with a common world coordinate system (WCS) at a pixel size of $0\farcs03$, and the astrometry is tied to Gaia DR3 \citep{gaiadr3}.

\subsubsection{Spectroscopy}
\label{sec:archival_spec_data}

NIRSpec-MSA PRISM spectroscopy of the three lensed images of \target\ were obtained between 31 July and 2 August 2023 \citep{furtak2024,Bezanson2024}. The observations are discussed in detail in \citet{furtak2024}. Here we only briefly summarize that the observations used a 2-points dithers pattern with a 3-shutters nodding strategy for background subtraction. The total exposure times for the three images ranged from 9.4h to 16.4h.

We also use the spectrum of the three images extracted and combined by \citet{furtak2024} and present the analysis in Appendix~\ref{appendix:other_reductions}. However, we also re-process the MSA PRISM spectra of the three images with the NIRSpec GTO pipeline, following the same procedure described in \cite{deugenio_dr3_2024}.
We note that, especially for exploring variability, (in contrast with some previous studies) the spectra must be extracted from the full shutter, as it is the only extraction that guarantees accurate path losses and diffraction losses corrections. Additionally, the spectra should be extracted from the individual 2D dithered spectra and then the extracted 1D spectra should be combined. Extracting 1D spectra from the 2D-combined spectra is deprecated as the combination of the 2D spectra does not preserve the path losses and diffraction losses of each individual spectrum.

% The spectra obtained from the GTO pipeline are roughly consistent with those obtained in \citet{furtak2023}, although they also present some differences, as discussed in the following.
When the spectra of the three images are compared, image B shows a bluer spectrum at wavelengths shorter than $\sim 3~\mu {\rm m}$, when compared to the spectra of the A and C images. {This can be clearly seen in Appendix~\ref{appendix:other_reductions},} where the three spectra are normalized to the flux densities at $\lambda = 3.6~\mu {\rm m}$. The reason for this difference in shape is unlikely due to variability but instead to improper background subtraction of the spectrum of image B because of the extended emission of a foreground, large and bluer galaxy, located about $0.\!\!^{\prime\prime}9$ in projection from image B. We therefore do not use image B in our analysis, and combine the spectra of images A and C, normalized and inverse variance weighted.

The GTO-processed and combined spectra of images A and C are shown in Figure~\ref{fig:spec_qso1}, where they are compared with the combined spectrum in \citet{furtak2024} as well as the PRISM spectrum of image A from the \blackthunder\ programme  as described in Section~\ref{sec:BT_data}.
The spectra are generally consistent with each other, but they also present some deviations. One difference is that the noise estimated for the GTO-processed spectra is higher than the one inferred by \citet{furtak2024}. The lack of one of the spectra (image B) cannot explain the difference in the noise. The reason, as when comparing many other spectra between the GTO and other pipelines, is that the noise (ERROR) extension of the GTO pipeline incorporates a factor to account for the effect of correlated noise \citep{dorner+2016}, which is instead typically not accounted for by the other pipelines. The finer grid of the GTO pipeline also partly contributes to the higher noise.
The additional difference is that at short wavelengths ($\lambda <1.3~\mu \rm {m}$) our A + C combined and GTO-pipeline-processed spectrum is bluer than the \citet{furtak2024} spectrum; the difference likely originates from the problematic background subtraction in image B used in the combined \citet{furtak2024} spectrum.
However, this spectral region is not critical for the modelling of the source, except when trying to explore the contribution of the nebular continuum, as discussed later. Indeed, in Appendix~\ref{appendix:other_reductions} we show that fitting the combined spectra by \citet{furtak2024} provides an equally good fit -- the reduced $\chi ^2$ is higher, but primarily as a consequence of the lower estimation of the errors in that spectrum.

\section{Photometry from imaging}
\label{sec:photometry}

According to \cite{furtak_abell2744_2023} and \cite{mayilun_lrd_2024}, {the three images of \target\ are unresolved}. As already mentioned, image B is located in the halo of a bright foreground galaxy, located about $0.\!\!^{\prime\prime}9$ in projection. This may cause some problems in the photometry and, as already discussed, in the background subtraction of the MSA spectrum. The foreground galaxy and its halo are much bluer than \target, hence the contamination effect is likely most significant in the blue bands.
As we are predominantly interested in the spectroscopic aspects of \target, we mostly rely on the UNCOVER photometry provided by \cite{furtak_abell2744_2023}.

However, for exploring variability, we also extracted photometry from the UNCOVER, DDT-2756, and ALT images in the F356W and F444W filters, covering the wavelength redder than the Balmer break. 
F356W is the only band used in these three programmes.
Circular aperture photometry is conducted with radius $r=0\farcs15$.
Local background is measured using a circular annulus and subtracted.
We correct for aperture loss assuming the \texttt{webbpsf} model and compute photometric uncertainty from random-aperture experiments in source-free regions. 
The F356W and F444W flux densities of all three images are similar to those reported by \citet{furtak2024}, and we do not find photometric variability at a significance level higher than $3\sigma$ for any of the images, similar to previous studies \citep{zhang_lrdvar_2024}.
Yet, variations of up to 25\% can be potentially consistent with photometric errors and calibration/flat-field uncertainties for image C in DDT-2756 epochs.
% at one year apart in the observed frame.

% The resulting photometry, along with the synthetic photometry from the spectra, are reported in the Appendix.

%\section{R2700 IFU spectral measurements}
\section{Spectral measurements}
\label{sec:R2700_measure}

% As mentioned, the low-resolution spectra extracted from the IFU cube and the MSA spectra, from the GTO reduction (excluding image B, which has a problematic background subtraction) and the one presented in \cite{furtak2023}, are generally consistent with each other within the uncertainties (Figure~\ref{fig:spec_qso1}).
% As already discussed, the noise extension of the \cite{furtak2023} combined spectrum is much lower than the GTO-reduced spectrum (and the difference cannot be reconciled by the spectrum of image B which is not included in the combined GTO spectrum). This difference in the \texttt{ERROR} extensions is common to many other MSA spectra and due to the fact that the GTO pipeline is more conservative, by incorporating a correction factor to the individual pixels flux errors to take into account the effect of correlated noise \citep{dorner+2016}.

% \subsection{IFU R2700 analysis}

The spectral fitting of the low-resolution PRISM spectra will be discussed extensively in Section~\ref{sec:balmer_break}. 
\redtxt{For Ly$\alpha$, which is not modelled in Section~\ref{sec:balmer_break}, we measured its basic properties.
To be consistent with the grating-based measurements, we measured Ly$\alpha$ observed in the image A.
We remain agnostic about the physical nature of the continuum during the fit by using a power-law to describe the UV continuum, where the region near the Ly$\alpha$ damping wing is described by a step function convolved with the instrumental line spread function (LSF) of the NIRSpec PRISM spectrum.
%for MSA PRISM spectra from \citet{degraaff2024}.
%\footnote{Available at \href{https://jwst-docs.stsci.edu/jwst-near-infrared-spectrograph/nirspec-instrumentation/nirspec-dispersers-and-filters}{jwst-docs website}.}. 
For Ly$\alpha$, we fitted it as a Gaussian function, which is LSF-dominated due to the low resolution of PRISM ($R\lesssim 50$ near 1 $\mu$m). Therefore, the kinematics of Ly$\alpha$ remain largely unconstrained.
}

%Here 
\redtxt{Next}, we focus on the analysis and fitting of the various components of the H$\beta$ line (as well as the tentative detection of \oiii) in the high resolution BlackTHUNDER spectrum.
We fitted the integrated aperture R2700 spectra of this source (see bottom panel of Figure \ref{fig:spec_qso1}) as a series of Gaussian profiles for emission and absorption lines, and a power-law to describe the continuum, using the \texttt{Fitting} routines in \texttt{QubeSpec} code\footnote{\url{https://github.com/honzascholtz/Qubespec}}. We fit the H$\beta$ and \oiii$\lambda \lambda 4960,5008$. To each of the emission lines we fit a single Gaussian profile to describe the narrow component of the emission lines, tying their redshift (centroid) and intrinsic FWHM to a common value to reduce the number of free parameters, leaving the flux of each Gaussian profile free. We fixed the \oiii$\lambda 5008$/\oiii$\lambda 4960$ flux ratio to be 2.99, given by the Einstein coefficients \citep{Dimitrijevic07}. The broad line region and the absorption associated with it is modelled as a Gaussian profile with its own redshift and FWHM. The FWHM of each emission line is convolved with the line spread function of NIRSpec from the JDOCS\footnote{Available at \href{https://jwst-docs.stsci.edu/jwst-near-infrared-spectrograph/nirspec-instrumentation/nirspec-dispersers-and-filters}{jwst-docs website}.}.

\redtxt{We note that recently, it was found that the broad \ha\ observed in LRDs is better described by two Gaussians \citep[e.g.,][]{deugenio_qso1_2025,Brazzini_2025,Linxiaojing_2025,ji_lord_2025} or an exponential \citep{Rusakov_2025,ji_lord_2025}, when enough S/N is achieved.
The double-Gaussian broad lines were also observed in some AGN at $z\sim 2$ \citep{Santos_2025}.
For \target, specifically, the broad-line profile is discussed extensively by \citet{deugenio_qso1_2025}, \citet{Maiolino_qso1_2025}, and \citet{juodzbalis_qso1_2025}, who show that, by extending the nominal wavelength coverage of the NIRSpec spectrum with a customized reduction to cover \ha, the broad \ha\ is better described by two Gaussians. \citet{Maiolino_qso1_2025} also suggest that the broad H$\beta$ might have an intermediate component, which, however, seems to be spatially resolved and thus likely associated with a (low-metallicity) outflow. \citet{juodzbalis_qso1_2025} reach the same conclusion for the intermediate Gaussian component of H$\alpha$, which they found to be spatially resolved.
Given these results, we also attempted fitting the broad \hb\ with two Gaussians, and we used the Bayesian Inference Criterion \citep[BIC,][]{Liddle_2007} to check whether the double-Gaussian model is preferred.
We found $\Delta {\rm BIC}=-46<0$, which indicates given the current S/N of the data, the double-Gaussian model is not preferred statistically for the broad \hb.
%In addition, we note that \citet{juodzbalis_qso1_2025} performed a direct dynamical measurement of the black hole mass in \target, and they found the dynamical black hole mass is most consistent with the single-epoch mass estimate based on \hb.
With the above results and given the current data quality, we used the single-Gaussian fitting result for the broad \hb.
Later in Section~\ref{sec:bh_properties}, we show that the \hb-based single-epoch black hole mass is consistent with the recent dynamical measurement of the black hole mass performed by \citet{juodzbalis_qso1_2025}.
}

We note that, while for NIRSpec-MSA spectroscopy, compact sources result in a spectral resolution significantly higher than the nominal spectral resolution (because the source is smaller than the shutter width), this is not the case for NIRSpec-IFU spectroscopy. This is due to the anamorphic magnification in the IFU optical path, which makes each slice of the IFU image slicer have a width of $0\farcs1$ projected on sky, while maintaining the 2-spectral pixel sampling. For a compact, unresolved source the effective resolution might still be slightly higher than the nominal spectral resolution (which is for a uniform illumination of each slice), but the effect is not as large as in the case of the MSA.

The best-fit parameters are estimated with a Bayesian approach, where the posterior probability distribution is estimated using the Markov-Chain Monte-Carlo (MCMC) ensemble sampler - \textsc{emcee} \citep{emcee} with 10,000 steps. For each of the variables, we need to define set priors for the MCMC integration. The prior on the redshift of each component is set as a truncated Gaussian distribution, centered on the systemic redshift of the galaxy with a sigma of 300 km s$^{-1}$ and boundaries of $\pm 1000$ km s$^{-1}$. The prior on the intrinsic FWHM of the narrow-line component is set as a uniform distribution between 10-500 km s$^{-1}$, while the prior on the amplitude of the line is set as a uniform distribution in log-space between 0.5$\times$rms of the spectrum and the maximum of the flux density in the spectrum. 

The final best-fit parameters and their uncertainties are calculated as the median value and the 68 \% confidence interval of the posterior distribution. We note that all the quantities derived from R2700 spectral fitting are calculated from the posterior distribution to account for any correlated uncertainties in the spectrum between each component.

Despite the longer exposure, the high-resolution spectrum appears more noisy as a consequence of the continuum and broad H$\beta$ being spread over many more resolution elements. {The broad H$\beta$ is well detected
%and its width can be measured much better than in the PRISM spectrum.
% SUGGESTED CHANGE
and its FWHM agrees with the value reported in \citet{furtak2024}.}
However, 
the grating spectrum clearly reveals a narrow component of H$\beta$, which is barely resolved and it cannot be deblended using the PRISM alone.
Formally, the fit provides a $3\sigma$ upper limit on the width of $\rm FWHM < 75$ \kms\ (after deconvolving for the instrumental LSF). However, in the rest of this paper, to be even more conservative, we take as an upper limit on the line width the nominal spectral resolution of the IFU at this wavelength (i.e., $< 112$ \kms, although the actual resolution is probably slightly better). 
There is also a detection of \oiii$\lambda 5008$ at a significance of $4\sigma$, which matches the velocity of the narrow component of the H$\beta$ emission.
 
The profile of broad H$\beta$ also shows the tentative detection of H$\beta$ absorption with a FWHM of $185^{+69}_{-72}$ \kms\ and a velocity shift of $101^{+65}_{-77}$ \kms\ with respect to the narrow component, indicative of an inflow. As discussed in Introduction, the presence of Balmer absorption is not uncommon among \jwst-discovered AGN, and at least 20\% of the systems are found with this feature in H$\alpha$ or H$\beta$ when they are observed at medium to high resolution \citep[e.g.,][]{matthee2024,wangbingjie_lrd_2024,wang_break_2024,juodzbalis_agnabs_2024,kocevski2024,lin_aspire_2024,labbe_monster_2024,taylor24}.

The coloured dashed lines in the bottom panel of Figure~\ref{fig:spec_qso1} show the best fit of that portion of the spectrum with the following components: a broad component of H$\beta$ (orange), a narrow component of H$\beta$ and \oiii\ (blue), and a power-law continuum and absorption component of H$\beta$ (green). The total fit is shown with a red dashed line.
The central row of the same panel shows the $\chi$ residuals of the full fit, while
the bottom row shows the residuals when not accounting for H$\beta$ absorption.
%, which better highlights the significance of the absorption.
We summarize the spectral measurements for the 
\redtxt{image A of \target}
%new \blackthunder\ R2700 spectrum 
in Table~\ref{tab:measurements}.

\section{Black hole properties and scaling relations}
\label{sec:bh_properties}

In this Section, we describe the derivations of the black hole properties in \target\ as well as the related scaling relations. We also discuss implications for the stellar mass of \target.

\subsection{Black Hole mass and accretion rate}
\label{sec:mass_acc}

The black hole mass can be estimated assuming the local virial relations between BH mass, widths of the lines from the BLR and the continuum or broad line luminosity \citep[e.g.,][]{Greene2005, Vestergaard2006,Reines2015}. It is not obvious that the same relations apply at high redshift. In particular, it has been speculated that in the super-Eddington regime (which may apply to some high-$z$ AGN), the black hole masses may be overestimated \citep[e.g.,][]{lambrides2024,Lupi2024}. However, recently the GRAVITY+ collaboration has obtained a direct (interferometric) measurement of the BLR size and of the BH mass in a quasar at \redtxt{$z=2.33$} \citep{Abuter2024}, which is a cosmic epoch much closer to \target than to the present epoch. According to their direct measurements, this black hole is accreting at a highly super-Eddington rate ($L/L_{\rm Edd}\sim 7-20$). Despite the high accretion rate, the BH mass estimated from the virial relation using broad \ha\ is consistent with the direct interferometric measurement within a factor of 2.5, well within the scatter of the virial relation \citep[and this small offset can be mitigated even further by applying the correction factor proposed by][]{Abuter2024}. When using broad H$\beta$ the agreement is worse (a factor of 5), but still not in dramatic contrast with the virial relations, taking into account their scatter and also the fact that this test case is actually a highly super-Eddington accreting black hole.

We take the FWHM of the broad component of H$\beta$ measured in the grating spectrum. Then we estimate the luminosity of H$\alpha$ by assuming a typical \ha/\hb\ ratio of 3 \citep[different from the Case B value of 2.86 due to the collisional enhancement in AGN-ionized gas, see e.g.,][]{dong_balr_2008} and correcting for a visual dust extinction of $A_{\rm V}=2.1$ estimated from our spectral fitting result in Section~\ref{sec:balmer_break} (see Table~\ref{tab:best_fit}) assuming a Small Magellanic Cloud (SMC) extinction curve with $R_{\rm V}=2.505$ \citep{gordon2003}, which has been adopted for describing dust attenuation in high-$z$ galaxies \citep[e.g.,][]{reddy_2015}.
This leads to $\rm H\alpha/H\beta=7.6$.
As a consistency check, we measure the Balmer decrement from the MSA PRISM spectrum of image A with \textsc{pPXF} \citep{cappellari2004,cappellari2017}.
Although the profile of \ha\ is only partially covered as can be seen in Figure~\ref{fig:spec_qso1}, we assumed it is Gaussian and fixed its kinematics to those of \hb.
The resulting ratio from the fit is $\rm H\alpha/H\beta=8.0\pm 0.8$, consistent with our estimation above.
%We also made a consistency check using the Balmer decrement measured from the PRISM spectrum, although the profile of \ha\ is only partially covered as can be seen in Figure~\ref{fig:spec_qso1}.
We note that \citet{furtak2024} also reported the PRISM-based Balmer decrement for images A+B+C of \target\ assuming Gaussian line profiles, which is $\rm H\alpha/H\beta=7.5\pm 0.4$ (based on their Table 1).
%\redtxt{[provide our own fitting consistent with Section~\ref{sec:measure}?]}
%Assuming an intrinsic ratio of 3 and the same SMC extinction curve, we obtained $A_{\rm V}=2.05\pm 0.26$ mag, consistent with our spectral fitting result.
% Adopting a \citet{calzetti2000} extinction curve would lead to a larger attenuation of $A_{\rm V}= 2.41\pm 0.25$ mag.
%\RMcomm{I think we can provide an additional estimation, or consistency check, by the ratio of the portion of visible H$\alpha$ and H$\beta$}. 
We also correct the luminosity for a lensing magnification of {$\mu _A=5.8$} for image A.
We describe the measurements of the lensing magnification in Appendix~\ref{sec:lens_model}, which are broadly consistent with measurements of \citet{furtak_abell2744_2023} and \citet{furtak2024}.
%as estimated in Section~\ref{sec:lens_model}.
We then use the virial relation proposed by \cite{Reines2015}, to estimate a black hole mass of {$\log{(M_{\rm BH}/M_{\odot})}=7.59$}. The formal uncertainties would be about 0.2~dex. However, we prefer to be conservative by giving an uncertainty of 0.3 dex to take into account the dispersion and uncertainties in the virial relations \citep[e.g.,][]{Ho14}.

\redtxt{One caveat of the above analysis is the assumed intrinsic ratio of the Balmer decrement, which can be significantly larger than 3 in LRDs \citep[e.g.,][]{lambrides2024,Chang_matthee_2025}.
If we simply used the \hb\ luminosity to derive the BH mass following \citet{Greene2005}, and applied the dust attenuation correction based on the narrow-line attenuation of $A_{\rm V}=0.66\pm 0.40$ mag derived by \citet{Maiolino_qso1_2025}, we obtained a black hole mass of $\log{(M_{\rm BH}/M_{\odot})}=7.2\pm 0.3$. 
This estimation is consistent with the value estimated above within the combined $1\sigma$ uncertainty, especially considering that the BLR is likely more dust attenuated compared to the NLR.
Interestingly, \citet{juodzbalis_qso1_2025} recently made a dynamical measurement of the BH mass for \target\ and obtained $\log{(M_{\rm BH}/M_{\odot})}=7.7\pm 0.3$, which is more consistent with our first single-epoch BH mass estimation.
We refer the readers to \citet{juodzbalis_qso1_2025} for further discussions on BH mass measurements in \target.
}

We also estimate the bolometric luminosity of the AGN by using the scaling relations with broad emission line fluxes presented by \cite{Stern12}; using the H$\beta$ line (corrected as discussed above) gives $L_{\rm AGN} \approx 2.6\times 10^{44}~{\rm erg~s^{-1}}$, hence implying $L/L_{\rm Edd} \approx 0.05$. 
As we discuss in Sections~\ref{sec:mdyn}, \ref{sec:balmer_break} and \ref{sec:variability}, the rest-frame optical light is likely dominated by the AGN continuum. 
Therefore, we can also infer the bolometric luminosity from the optical continuum ($L_{5100}$), corrected for extinction and lensing magnification, and again adopting the scaling relation in \cite{Stern12}, resulting in $L_{\rm AGN} \approx 1.2\times 10^{45}~{\rm erg~s^{-1}}$, hence $L/L_{\rm Edd} \approx 0.24$. This is close to the estimation by \citet{furtak2024} of $L/L_{\rm Edd} \approx 0.3$ based on MSA PRISM spectra of \target, where they used a bolometric conversion to get $L_{\rm bol}$ from line luminosities.
\redtxt{The two estimations of the bolometric luminosity differ by 0.66 dex, larger than the combined scatter of 0.43 dex between the bolometric conversions of \citet{Stern12}.
The difference might arise due to the intrinsic difference between the SED of \target and the average SED of \citet{Stern12}'s sample. 
Qualitatively, if the accretion disk in \target has a softer ionizing spectrum, the bolometric luminosity based on hydrogen Balmer lines that scale with the ionizing luminosity would give a lower bolometric luminosity compared to the 5100 \AA continuum-based value.
In addition, if the observed Balmer lines have non-negligible optical depths, the actual bolometric conversion factor would be higher than assumed.
Finally, as discussed in \citet{Maiolino2024_Xrays}, if the covering fractions of BLRs in \jwst-selected AGN are systematically different from the low-reshift AGN, the Balmer line-based bolometric conversions would also be different.}
\redtxt{Regardless}, since in either case this black hole is accreting at substantially sub-Eddington, it is unlikely that it is in the regime where the virial scaling relations are drastically affected.

%\RMcomm{Note, I still have to correct the grating spectrum for aperture losses; the correction is small, but it may slightly increase the BH mass}

%\RMcomm{maybe we should also attempt to derive the BH mass by fitting the available half of H$\alpha$ from the PRISM spectrum}

\begin{figure}
    \centering
    \includegraphics[width=\columnwidth]{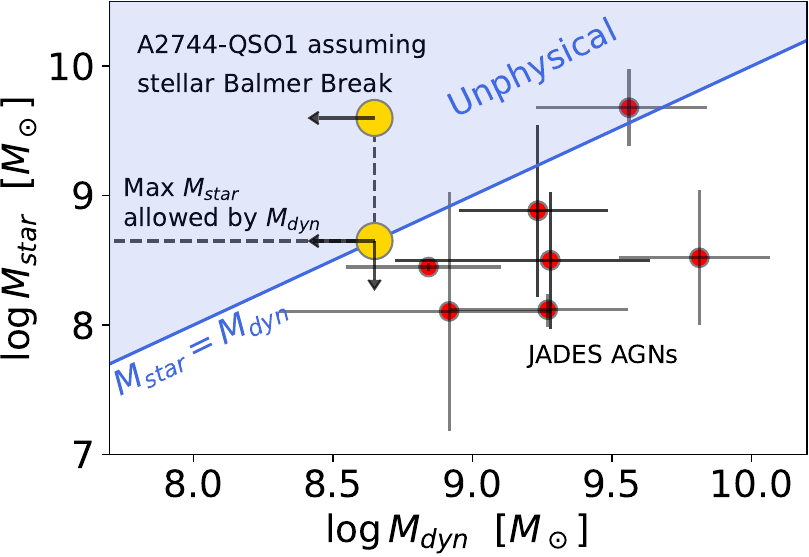}

    \caption{
    Comparison of stellar and dynamical masses for AGN at high-$z$ discovered by \jwst\ and observed with high spectral resolution. The shaded region is the ``unphysical'' area where the stellar mass is larger than the dynamical mass. Red small symbols are objects in the JADES sample presented in \citet{maiolino2023b} (updated with new estimated stellar masses in \citealp{juodzbalis_jadesbh_2025}). The upper golden symbol shows the dynamical mass estimated by the BlackTHUNDER high resolution spectrum of Abell2744-QSO1, where the stellar mass is the one inferred by \citet{mayilun_lrd_2024} assuming that the optical continuum and Balmer break are entirely dominated by stellar light; clearly this scenario leads to an unphysically high stellar mass, about an order of magnitude larger than the dynamical mass. The maximum stellar mass allowed by the dynamical mass is shown with the lower golden symbol.
    }
    \label{fig:Mdyn_Mstar}
\end{figure}

\subsection{A small dynamical mass of the host galaxy}
\label{sec:mdyn}

As discussed above,
the very small velocity dispersion inferred for the narrow component of H$\beta$ (and for \oiii$\lambda \lambda 4960,5008$; see Table~\ref{tab:measurements}) indicates that the dynamical mass of the host galaxy must be quite low. In this section, we quantify this aspect starting with the conservative upper limit on the line width of FWHM$<$112~\kms.

%As discussed in Sect.\ref{sec:R2700_measure}, formally the fit provides a $3\sigma$ upper limit on the width of 75~km/s (i.e. velocity dispersion $<$32~km/s). However, we prefer to be even more conservative by taking as upper limit the instrumental resolution, i.e. FWHM$<$112~km/s (i.e. velocity dispersion $<$47~km/s).
% Formally, the narrow component of H$\beta$ is measured to have a FWHM of $55^{+44}_{-22}$ \kms. This is below the nominal spectral resolution of NIRSpec with the IFU and the G395H disperser at the redshifted wavelength of H$\beta$ ($R=2670$). While it is possible to obtain intrinsic widths below the resolution limit if the S/N is high enough, as indeed obtained by the formal fitting, we prefer to be conservative and use an upper limit on the width given by the spectral resolution. This translates into an FWHM of the narrow component of H$\beta$ of 112 \kms.

In order to constrain the dynamical mass, we use the same approach as \cite{ubler2023a} and \cite{maiolino2023b} by estimating the dynamical mass through the equation
\begin{equation}\label{eq:mdyn}
    M_{\rm dyn} = K(n)K(q)\frac{\sigma^2 R_e}{G},
\end{equation}
where $K(n)=8.87-0.831n+0.0241n^2$ with S\'ersic index $n$, following \cite{Cappellari06}, $K(q)=[0.87+0.38{\rm e}^{-3.71(1-q)}]^2$, with axis ratio $q$ following \cite{vdWel22}, and $R_e$ is the effective radius.
Given that we do not have information on $q$, we take it equal to one, which gives an even more conservative estimate of the upper limit on the dynamical mass; taking $q=0$ would reduce the value by a factor of about two. As for the size, we use the measurements from \citet{furtak_abell2744_2023}, who report the source to be unresolved both in the rest-frame UV and optical.
%, and in terms of nebular emission lines, based on the \blackthunder\ cube. 
Therefore, we take the upper limit of 30 pc on the effective radius of the source inferred by \citet{furtak_abell2744_2023}.
In this equation, $\sigma$ is the stellar velocity dispersion. As we do not have this information, we adopt the same approach as in some previous papers \citep{ubler2023a, maiolino2023b} of using the nebular gas velocity dispersion and applying the empirical scaling between integrated stellar and gaseous velocity dispersions derived by \cite{Bezanson18b} at $z\sim1$. There are only a few systems in which this relation can be tested at higher redshift \citetext{\citealp{Carnall+2023,DEugenio23,Pascalau_2025}}, and it seems to hold, although in systems much more massive than \target\ ($M_{\star}\gtrsim 10^{10.5}~M_\odot$). The correction from the \cite{Bezanson18b} relation entails an increase of 0.18~dex in the velocity dispersion. As for the S\'ersic index, we assume $n=1$, as is the case for the majority of star-forming galaxies at such high redshift \citep[e.g.,][]{Ormerod24}, including AGN host galaxies \citep{maiolino2023b}. 
Since we assume an upper limit on both the velocity dispersion and the size, we obtain a very conservative upper limit on the dynamical mass, which we estimate to be $M_{\rm dyn} < 4.4\times 10^8~M_{\odot}$.
Alternatively, if we assume the system is dispersion-dominated and apply Eq. A2 of \citet{diaz-santos_mdyn_2021}, which is $M_{\rm dyn}^{\rm disp}=3.4\sigma ^2 R_e/G$, we obtain $M_{\rm dyn}^{\rm disp} < 1.2\times 10^8~M_{\odot}$. Interestingly, the latter upper limit would imply that the dynamical mass of the system is dominated by the black hole\footnote{We note that rotation-dominated broadening would disappear at very low inclination. However, since the inclination should be random in observations but many LRDs likely have $M_{\star}/M_{\rm dyn}>1$, the low-inclination scenario seems implausible.}.
If we used the formal $3\sigma$ upper limit from the fit on the FWHM of the line ($<$75~\kms), then the upper limits on the dynamical masses would be even tighter, specifically $M_{\rm dyn} < 2.0\times 10^8~M_{\odot}$ if using Equation~\ref{eq:mdyn}, and $M_{\rm dyn}^{\rm disp} < 5.5\times 10^7~M_{\odot}$ if using the dispersion-dominated equation.

\redtxt{One caveat of the above calculations is the origin of the narrow lines, that is, whether they are dynamically influenced by the whole galaxy or only by the black hole if they are sufficiently concentrated spatially.
Interestingly, as recently pointed out by \citet{juodzbalis_qso1_2025} and \citet{Maiolino_qso1_2025}, the narrow lines in \target are actually spatially resolved in NIRSpec IFU observations, whose dynamics were subsequently modelled by \citet{juodzbalis_qso1_2025} using the code \textsc{moka3D} \citep{moka3d}.
\citet{juodzbalis_qso1_2025} found that the kinematics of narrow \ha\ are best modelled by a point source-like mass of $5\times 10^{7}~M_\odot$, likely dominated by the central black hole.
While an extended exponential disk component is not preferred by the data, enforcing such a component would give an upper limit on the extended dynamical mass of $2\times10^{7}~M_\odot$, much smaller than our dynamical mass estimate.
Therefore, no matter whether we used the rough dynamical mass estimate by assuming an unresolved stellar component and stellar velocity dispersion scaled from the gas velocity disperison, or took the more sophisticated dynamical modelling result of \citet{juodzbalis_qso1_2025}, we obtained relatively small dynamical mass upper limits in a range of $10^{7.3-8.6}~M_\odot$.
}

We note that the derived dynamical mass, even taking the most conservative upper limit, is an order of magnitude lower than the stellar mass inferred by \citet{mayilun_lrd_2024} when assuming that the optical continuum and Balmer break are entirely dominated by stellar light, which is $4\times 10^9~M_\odot$\footnote{It is worth noting that the fiducial fit by \citet{mayilun_lrd_2024} also includes contributions from AGN to the optical continuum. Even in this case, the inferred stellar mass is $\sim4\times 10^9~M_\odot$.}.
This discrepancy is illustrated visually in Figure~\ref{fig:Mdyn_Mstar}. This result highlights that most of the optical light, as well as the Balmer break, cannot have a stellar origin.
Even our most conservative upper limit on the dynamical mass is only 10\% of the stellar mass inferred from stellar dominated Balmer break and optical continuum.
Our finding indicates that no more than 10\% of the optical light originates from stars.
%; this is a conservative upper limit, inferred by the fact that even our most conservative upper limit on the dynamical mass is 1/10 of the stellar mass inferred when assuming that the optical continuum is stellar. 
In Section~\ref{sec:balmer_break}, we further illustrate that an AGN continuum and dense gas along the line-of-sight (LOS) can explain well the shape of the optical continuum and, in particular, the Balmer break, hence greatly alleviating the issue of the high stellar mass and extremely high stellar densities inferred when assuming a stellar origin of the Balmer break.

\subsection{An over-massive Black Hole}
\label{sec:scaling}

The upper limits on the dynamical mass also translate into a tight upper limit on the stellar mass. These upper limits can be compared with the black hole mass inferred from the virial relations and with the local scaling relations between black hole mass and galaxy properties.

\begin{figure*}
    \centering
    \includegraphics[width=0.9\textwidth]{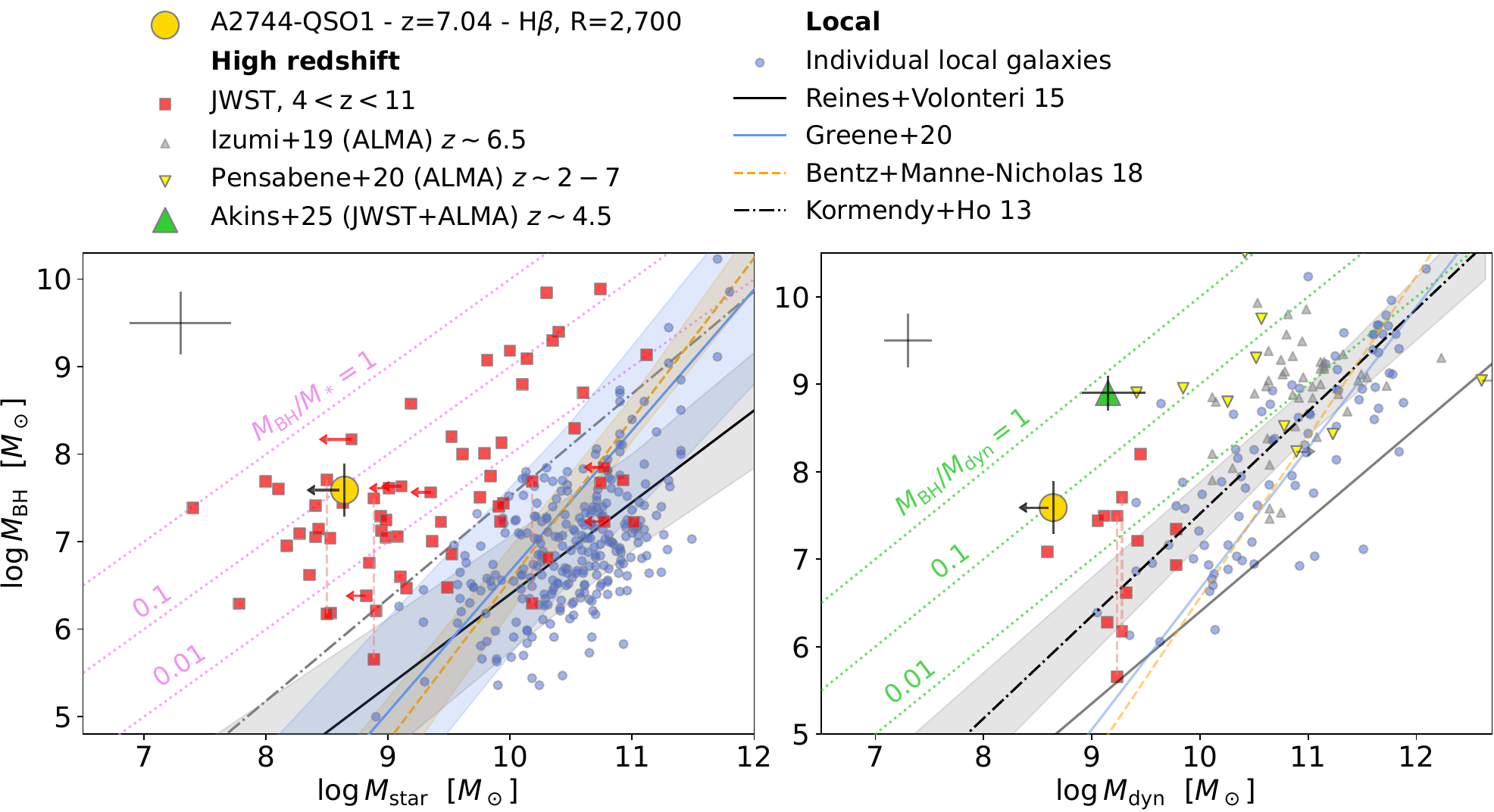}

    \caption{
    Scaling relations between black holes and their host galaxies, specifically BH mass versus galaxy stellar mass (left) and versus galaxy dynamical mass (right). The small blue symbols show local galaxies, and the straight lines and shaded regions illustrate the best fit local scaling relations (see text for details). The red points show AGN at 4$<$z$<$11 for which the black hole mass and host galaxy stellar/dynamical mass has been measured with \jwst\ data, as reported by \citet{Bogdan23,Carnall+2023, Goulding23, Ding23, harikane2023, kocevski2023,Kokorev2023, ubler2023a, maiolino2023b,Stone_2024,Yue_bhmass_2024,wang_break_2024,trefoloni_feii_2024,juodzbalis_jadesbh_2025}. 
    \redtxt{The mean uncertainties of masses are indicated by the black cross on the top left location of each panel. Additionally, we plot high-$z$ sources from \citet{Izumi_2019,Pensabene_2020,Akins_ci_2025}.}
    The golden symbols show Abell2744-QSO1 for which we have taken the upper limit on the host galaxy dynamical mass as conservative upper limit on the stellar mass. The black hole in Abell2744-QSO1 is clearly overmassive both in terms of stellar and dynamical mass, when compared with the local relations.
    }
    \label{fig:mbh_gal}
\end{figure*}

The left panel of Figure~\ref{fig:mbh_gal} shows the relationship between BH mass and host galaxy stellar mass. Local galaxies (blue points) are taken from \citet{Reines2015}; these are disc galaxies, hence with S\'ersic index $n$ comparable to the host galaxies of high-$z$ AGN; additionally, the black hole masses in this sample have been measured using the same virial relations as adopted in our work, hence are fully consistent. We also show the galaxies reported in \citet{Greene20}, which also have S\'ersic index $n\sim 1$, although in this case the BH masses are estimated with a different calibration.  The black solid line and blue solid line show the relations inferred from those two local studies, and the gray and blue shaded regions show their dispersions. We also show the relation obtained by \cite{Bentz2018_BHscalingrel} for local AGN with reverberation mapping (orange line and shaded region), mostly in disc galaxies.
For completeness we also show with a dot-dashed black line the \cite{Kormendy13} M$_{\rm BH}$--M$_{\rm spheroid}$ relation for early type galaxies, but we caution that this is inadequate as for the vast majority of high-$z$ galaxies, including AGN hosts, the stellar light has a disk-like profile. The red points are black hole and host galaxy stellar mass measurements from various \jwst\ studies at $z>4$ 
\redtxt{\citep{Bogdan23,Carnall+2023,Goulding23, Ding23, harikane2023, kocevski2023,Kokorev2023, ubler2023a, maiolino2023b,Stone_2024,Yue_bhmass_2024,wang_break_2024,trefoloni_feii_2024,juodzbalis_jadesbh_2025}.}  
%\RMcomm{I should probably update and/or differentiate symbols by source, but it would make the legend probably far too long}. 
Note that candidate dual/merging BHs are connected with red dashed lines \citep{maiolino2023b}. Most of the high-$z$ AGN discovered by \jwst\ are overmassive relative to the local $M_{\rm BH}$--$M_{\star}$ relations, by even orders of magnitude. This was already highlighted by previous studies \citep[e.g.,][]{harikane2023, ubler2023a, maiolino2023b, juodzbalis_dormantagn_2024, Marshall24}. The golden large symbol shows the location of \target. As no solid detection of the stellar light has been obtained yet, here we use the the dynamical mass (based on the conservative upper limit on the line width) as a conservative upper limit on the stellar mass. \target\ deviates from the local relations by orders of magnitude. This was already pointed out by \citet{furtak2024}, by setting an upper limit on the stellar mass by assuming the case that all observed optical light is associated with stellar emission. Here we confirm their finding but in this case using the dynamical mass as an independent constraint on the stellar mass.

The right panel of Figure~\ref{fig:mbh_gal} shows the M$_{\rm BH}$-M$_{\rm dyn}$ relations. In this case, for local galaxies we use the early-type sample of \cite{Kormendy13} (blue points and dot-dashed best-fitting relation), where we use their stellar masses as a proxy of the dynamical masses. The justification is that these early-type galaxies have very little gas content and the total dynamical mass is nearly coincident with the spheroidal stellar mass component\footnote{We note that below $\log(M_\mathrm{\star}/\mathrm{M}_\odot)=10.5$, the dark-matter fraction decreases with decreasing stellar mass \citep{cappellari+2013,toloba+2014,eftekhari+2022}. This would make the true $M_\mathrm{BH}$--$M_{\rm dyn}$ relation steeper, making \target even more overmassive.}. However, for completeness we also show the stellar relation for the disc samples (solid black and solid blue lines), although these are inadequate for assessing the dynamical masses, given the large amount of gas and also dark matter expected in these early galaxies. The red points shown here are AGN at $z>4$ discovered by \jwst, whose dynamical masses could be estimated based on high spectral resolution data \citep{ubler2023a, maiolino2023b}. In this case, the high-$z$ BHs are closer to the local relation. This is partly due to the fact that the local \cite{Kormendy13} relation is higher than those by \cite{Reines2015} and \cite{Greene20}, although, given that the \cite{Kormendy13} relation is based on direct dynamical measurements of BH masses, it is likely biased high (i.e., for black holes massive enough that their sphere of influence can be resolved).
Also, black holes in non-classical bulges are excluded from the fit of \citet{Kormendy13}.
However, regardless of the adopted local relation and its potential biases, \redtxt{most} high-$z$ AGN do not have black holes masses that are in excess of 0.1 times the dynamical mass, which instead is the case for the stellar masses. 
\redtxt{One exception is the LRD with $M_{\rm dyn}$ measured by \citet{Akins_ci_2025} with the tentative detection of [C\,{\sc i}](2-1) using ALMA, where $M_{\rm BH}/M_{\rm dyn}$ is close to 1, although [C\,{\sc i}](2-1) might be spatially offset from the center of the NIRCam image. 
}
Regarding Abell2744-QSO1 (large golden symbol), the upper limit on its dynamical mass is \redtxt{lower than most} \jwst\ AGN and it indicates that its black hole mass is heavily overmassive also relative the black hole-dynamical mass relation. As already pointed out in the previous section, using the formal $3\sigma$ upper limit on the width of the narrow H$\beta$, or adopting the dispersion-dominated scenario, would yield an even lower upper limit on the dynamical mass, approaching the mass of the black hole itself.
\redtxt{In fact, the recent work of \citet{juodzbalis_qso1_2025} has shown that the dynamical mass of QSO1 is indeed dominated by the central black hole.}

For other \jwst-discovered AGN the overmassive nature on the M$_{\rm BH}$--M$_{\star}$ relation and being more aligned with the M$_{\rm BH}$--M$_{\rm dyn}$ relation could have been interpreted in terms of the host galaxy having the adequate baryon mass, but where most of the gas had been inefficient in forming stars. Yet, in the case of Abell2744-QSO1 the black hole is overmassive in all regards, also relative to the dynamical mass of the host galaxy.
%Such overmassive nature requires scenarios in which black holes form as already very massive ``heavy seeds'' \citep[e.g.,][]{Inayoshi2022,natarajan_2024}, remaining outliers in the population through sustained growth across cosmic times \citep{Hu2022,scoggins2024}, and/or experience repeated bursts of super-Eddington accretion \citep{volonteri_2015,schneider_2023,bennett_2024,trinca2024}.

\section{A Balmer break originating from dense gas absorption
}
\label{sec:balmer_break}

As discussed in Section~\ref{sec:mdyn}, interpreting the optical continuum and the Balmer break with a stellar origin would result in a stellar mass that is an order of magnitude larger than the dynamical mass. Additionally, the implied stellar density would be far larger than any other system seen at lower redshift. However, the growing number of \jwst-discovered AGN showing either H$\alpha$ and H$\beta$ absorption \citep[visible only in medium- and high-resolution spectra,][]{matthee2024,juodzbalis_agnabs_2024,kocevski2024,wangbingjie_lrd_2024,lin_aspire_2024,labbe_monster_2024} indicates that many (probably most) of these systems require extremely dense gas ($n_{\rm H}>10^8~{\rm cm}^{-3}$) with high column densities ($N_{\rm H}>10^{21}~{\rm cm}^{-2}$) along the LOS, possibly associated with the BLR clouds, which is the only way to keep the short-lived $n=2$ level of hydrogen populated and subsequently result into the observed Balmer absorption \citep{juodzbalis_agnabs_2024}. As shown in Figure~\ref{fig:spec_qso1}, Abell2744-QSO1 has a tentative detection of H$\beta$ absorption as well. \cite{im24} pointed out that H$\alpha$ and/or H$\beta$ absorption must also necessarily come with absorption of higher Balmer orders and including a Balmer break.

In this section, we explore whether the Balmer break observed in \target, as well as its other spectral properties, can be explained in the scenario where the optical light is dominated by an AGN continuum with dense absorbing gas along the line of sight.
{As we mentioned in Section~\ref{sec:data}, the background subtraction of image B is likely problematic.
Thus, we used the spectrum from the combined images of A and C obtained through the GTO reduction as the fiducial spectrum.
Still, we checked the results based on the spectrum from the combined images of A, B, and C images reduced by \citet{furtak2024}, as well as the integrated IFU spectrum from image A observed in \blackthunder, as detailed later.
}

\subsection{Balmer break fitting: Method}
\label{sec:method_BB}

\begin{figure}
    \centering
    \includegraphics[width=\columnwidth]{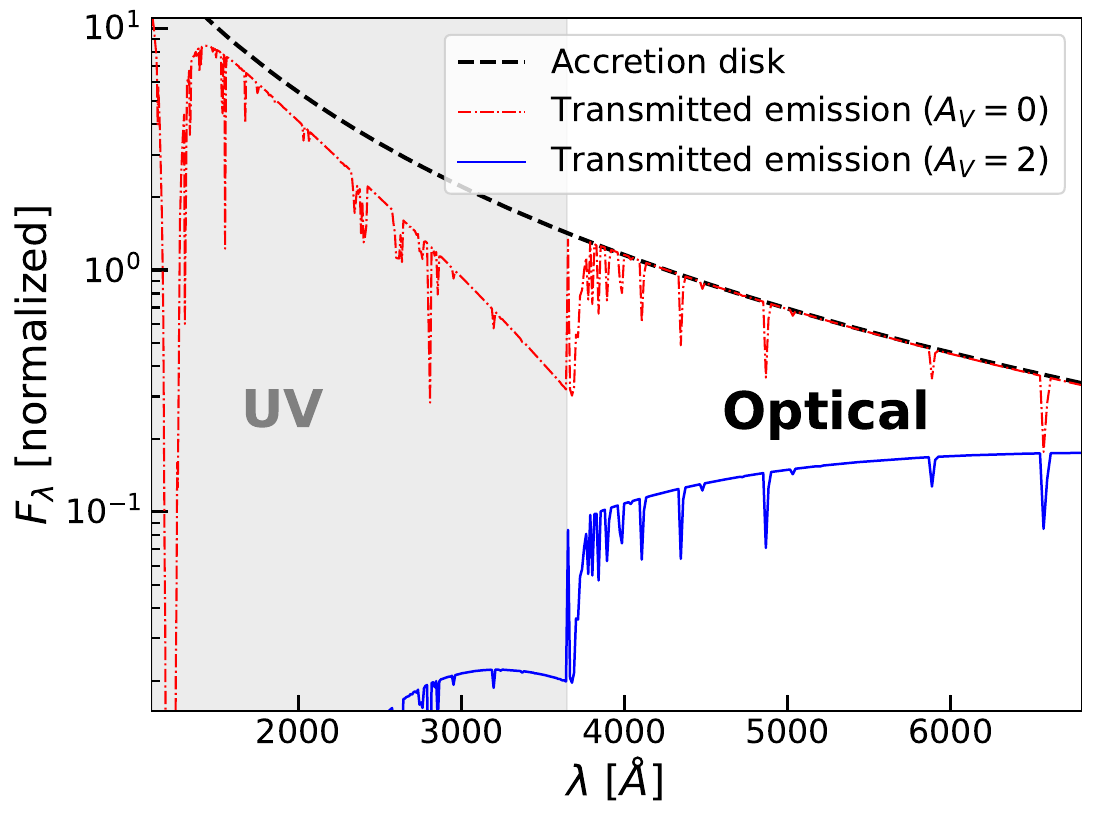}
    \caption{Example of an AGN accretion disk emission with $M_{\rm BH}=10^7~M_{\odot}$ and $\lambda _{\rm Edd}=0.1$ attenuated by a slab of gas with $n_{\rm H}=10^{10}~{\rm cm^{-3}}$, $N_{\rm H}=10^{23}~{\rm cm^{-2}}$, $U=10^{-1.5}$, and $v_{\rm turb}=120$ \kms\ (definitions of parameters given in Section~\ref{subsec:params_dependency}).
    The simulation is performed with \cloudy.
    The dashed line is the intrinsic continuum normalized at $\lambda =4260$ \AA.
    The dash-dotted red line is the dust-free attenuated continuum, which exhibits prominent absorption features including Balmer lines and the Balmer break.
    The solid blue line is the attenuated continuum further obscured by a dust screen with $A_{\rm V}=2$, which shows an uprising optical continuum in addition to the Balmer break resembling the rest-frame optical part of LRDs.
    The spike at the UV-optical interface is an artifact caused by the limited number of resolved energy levels of hydrogen.
    }
    \label{fig:model_demo}
\end{figure}

\begin{table*}
        \centering
        \caption{Input parameters for \textsc{Cloudy} simulations of BLR absorption and emission. The fiducial parameters are shown in the bold font.}
        \label{tab:models}
        \begin{tabular}{l c}
            \hline
                \hline
                Parameter & Values \\
                \hline
                SED & $\bf M_{\rm BH}=10^7~M_\odot$, $\lambda _{\rm Edd}={\bf 0.1}, 1, { 10}$ \citep{pezzulli2017}\\
                log($U$) & $-2.5$, $-2.25$, $-2.0$, $-1.75$, ${\bf -1.5}$, $-1.25$, $-1.0$\\
                $\log (Z/Z_{\odot})$ & $-1$\\
                Solar abundance set & \citet{grevesse2010} solar abundance set\\
                $\log (n{\rm_H} / {\rm cm}^{-3})$ & 8, 8.5, 9, 9.5, {\bf 10}, 10.5, 11, 11.5, 12\\
                $\log (N{\rm_H} / {\rm cm}^{-2})$ & $21, 21.5, 22, 22.5, 23, 23.5,\bf 24$\\
                $v_{\rm turb}$ [\kms] & 
                {20, 40, 60, 80, 100, {\bf 120}}\\
                Dust & No dust\\
                Resolved hydrogen energy levels & 50\\
                Atomic data set & CHIANTI \citep[v7,][]{chianti_old,chianti_v7}\\
                \hline
        \end{tabular}
\end{table*}
\begin{comment}
\begin{itemize}
    \item computation of the gas-absorbed AGN continuum; comparison with the IM24 models
    \item nebular continuum from the AGN; sub-Edd model and super-Edd model; way of co-adding; covering fraction as a free parameter
    \item Models for the UV part; power law; physically motivated choice: SF continuum, AGN continuum (unobscured)
    \item fitting method; MCMC
\end{itemize}
\end{comment}

% The purpose of this work is to expand the ideas of \citet{im24} and \citet{juodzbalis_agnabs_2024} and see if one can fit the spectrum of \target\ with an AGN dominated component in the optical including the Balmer break.
% To achieve this,

We formulate our spectral model as follows
\begin{equation}
\begin{aligned}
    F_{\lambda; \rm model}=b\lambda ^m
    + 10^{-0.4A_\lambda}(F_{\lambda; \rm Acc.~disk}^{\rm att.}+C_f F_{\lambda; \rm neb.~BLR})\\ 
    + F_{\lambda; \rm neb.~narrow},     
    \end{aligned}
    \label{eq:full_model}
\end{equation}
based on a series of physically motivated assumptions.
In the equation above, $F_{\lambda; \rm Acc.~disk}$ represents the flux of the accretion disk attenuated by a slab of dust-free gas in the BLR, $C_fF_{\lambda; \rm neb.~BLR}$ represents the nebular emission from the BLR observed along the LOS, $F_{\lambda; \rm neb.~narrow}$ represents the nebular emission from the narrow-line component, $10^{-0.4A_{\lambda}}$ is a dust attenuation factor for the BLR emission, and $b\lambda ^m$ is a power-law continuum. We explain in detail the physical meaning of each term below.

First, we assume the AGN emission that dominates the optical spectrum has two major components, which represent the attenuated continuum emission and the nebular emission, respectively.
The first component, $F_{\lambda; \rm Acc.~disk}$, originates in the AGN accretion disk and is attenuated by dense and dust-free gas along the LOS.
For the accretion disk emission, we used theoretical SEDs parametrized by the black hole mass, $M_{\rm BH}$, and the Eddington ratio, $\lambda _{\rm Edd}$, computed by \citet{pezzulli2017}.
According to the estimation of \citet{furtak2024}, \target\ has a black hole mass of $M_{\rm BH}=4^{+2}_{-1}\times10^7~{M_\odot}$ and an Eddington ratio of $\lambda _{\rm Edd}=L_{\rm bol}/L_{\rm Edd}\sim 0.3$.
In comparison, our estimations involving the new \blackthunder\ observations give $M_{\rm BH}=4^{+4}_{-2}\times10^7~{M_\odot}$ and $\lambda _{\rm Edd}=L_{\rm bol}/L_{\rm Edd}\sim 0.05-0.24$ (see Section~\ref{sec:mass_acc}).
We note that \target\ shows no detection of \heii$\lambda 4686$ in the optical \citep{furtak2024}, which might imply the lack of high energy photons with $h\nu > 54~\rm eV$ (although low \heii$\lambda 4686$/\hb\ down to 0.1 and lower has also been observed in local AGN, e.g., \citealp{vandenberk2001,tozzi_2023}). 
As recently suggested by \citet{lambrides2024}, the lack of high ionization lines in \jwst-selected Type 1 AGN might imply near-to-super Eddington accretion and the Eddington ratio might have been underestimated in these systems.
Therefore, we considered theoretical SEDs with $M_{\rm BH}=10^7~M_\odot$ and $\lambda _{\rm Edd}=0.1,1,10$ and compared their fitting results.
In the sub-Eddington regime ($10^{-2}\lesssim \lambda _{\rm Edd}\leq 1$), the SED is computed assuming the standard disk model with a geometrically thin and optically thick disk \citep{ShakuraSunyaev1973}.
%In the super-Eddington regime, the SED becomes a geometrically thick slim disk that efficiently traps hard ionizing photons, which makes the SED softer in the UV \citep{pezzulli2017}.
%However, as shown by \citet{inayoshi_spedd_2024}, the emission within the photon-trapping radius is not completely removed in super-Eddington accretion and the UV SED might actually become harder due to the contribution from the hotter part of the accretion disk within the photon-trapping radius.
%We discuss the effect of different SEDs later in this section.
In the super-Eddington regime, the accretion disk becomes geometrically thick due to inefficient radiative cooling, where instead energy advection via photon trapping dominates and suppresses the emergent radiation flux from the innermost disk region \citep{Abramowicz1988}.
\cite{pezzulli2017} accounted for the photon trapping effect on the SED by removing the emergent radiation flux within the characteristic radius, resulting in a softer UV SED. However, the emission within the photon-trapping radius is not entirely suppressed in super-Eddington accretion. In fact, the UV SED could become harder compared to the sub-Eddington case due to contributions from the hotter optically-thick regions of the accretion disk extended within the innermost stable circular orbit (e.g., \citealp{KubotaDone19,inayoshi_spedd_2024}; however, see \citealp{pacucci_spedd_2024} for a different model). We discuss the effect of different SEDs later in this section and in Appendix~\ref{appendix:other_params}.

Figure~\ref{fig:model_demo} shows the shape of the accretion disk continuum with $M_{\rm BH}=10^7~M_\odot$ and $\lambda _{\rm Edd}=0.1$ in the UV-to-optical regime. We adopt this value as the baseline, given that the estimated accretion rate is sub-Eddington and in the range $\lambda _{\rm Edd}\sim 0.05-0.24$.

To obtain the gas-absorbed SED, we passed the intrinsic disk emission through a slab of gas with high densities and column densities.
This calculation is done with the photoionization code \cloudy\ \citep[v17.03,][]{ferland2017}.
As already shown by \citet{im24}, at densities of $n_{\rm H}\sim 10^{9-11}~{\rm cm^{-3}}$, collisional excitation efficiently populates neutral hydrogen to the excited state of $n=2$, leading to a strong Balmer break.
At such high densities, the gas likely originates from the BLR or its proximity \citep{juodzbalis_agnabs_2024}.
Therefore, we did not consider dust attenuation within the dense gas, as the BLR clouds are usually within the dust sublimation radius (\citealp{netzer_1993,gaskell_2009}; for discussions of BLR models with dust, see e.g., \citealp{shields2010}).
In Figure~\ref{fig:model_demo}, we show a continuum attenuated by a slab of gas with $n_{\rm H} = 10^{10}~{\rm cm^{-3}}$ and $N_{\rm H} = 10^{23}~{\rm cm^{-2}}$.
One can see a clear Balmer break starting around the Balmer limit of 3646 \AA, as well as a series of absorption features underlying the locations of Balmer lines.
Another notable feature is a spike of emission near the Balmer limit, which is also seen in the models of \citet{im24}.
This spike is due to the fact that not all energy levels of hydrogen are resolved in the simulation.
To save computational time, we include the first 50 energy levels from the ground level in the calculation.
We note that the remaining spike does not have any significant impact on the fitting result.
We also considered dust attenuation external to the BLR, which is described by the attenuation factor $10^{-0.4A_{\lambda}}$
(note that potentially some of this dust obscuration can also come from some dust content in self-shielded clouds in the outer BLR, as suggested by \citealp{shields2010}; regardless, the dust location would not affect the fitting results).
%\citep[note that potentially some of this dust obscuration can also come from some dust content in self-shielded clouds in the outer BLR, as suggested by ][ regardless, the dust location would not affect the fitting results]{shields2010}.
The extinction curve is assumed to be an SMC bar extinction curve with $R_V=2.505$ \citep{gordon2003}.
Recently, \citet{mayilun_lrd_2024} suggested that an unusually steep dust attenuation law is required to fit the spectrum of \target. However, as we show later, the SMC extinction curve is sufficient to provide a sensible fit with the framework adopted here.
In Figure~\ref{fig:model_demo}, we show an example where the gas-obscured continuum is attenuated by a dust screen with $A_V = 2$, which leads to a rising continuum in the optical resembling the optical spectrum of LRDs \citep[see e.g.,][]{kocevski2024}.

In addition to the continuum emission, we considered the nebular emission from the BLR, $F_{\lambda;\rm neb.~BLR}$, including the nebular continuum and emission lines.
The nebular emission is also computed with \cloudy\ with the same physical condition as the continuum absorbing gas.
We considered a wide range of model parameters as summarized in Table~\ref{tab:models}.
The effects of various model parameters and how we narrowed down the ranges for these parameters are described in the next section.
The nebular emission is multiplied by a covering fraction, $C_f$, before being added to the continuum emission as the total emission from the BLR.
The covering fraction is defined as $C_f \equiv \Omega/4\pi$, where $\Omega$ is the solid angle of the BLR clouds with respect to the central accretion disk.
This parameter is included out of a geometrical consideration to describe how much of the nebular emission is produced and captured along the LOS (see \citealp{agn3_2006} and Section 16.44 of Hazy 1, C17\footnote{\url{https://gitlab.nublado.org/cloudy/cloudy/-/wikis/home}}).
Again, we considered the BLR nebular emission is further processed by external dust extinction with the same factor of $10^{-0.4A_\lambda}$ as the continuum emission.

Then, we assume the UV continuum is a featureless power law, representing either an AGN SED (e.g., scattered light by extended gas) or a stellar SED with young stellar populations.
For the moment, we do not consider the case where the physical process generating the UV continuum produces a Balmer break, for which the reason is justified as the following.
As shown by \citet{mayilun_lrd_2024}, assuming the UV continuum has a Balmer break and has a stellar origin, the AGN component needs to be sub-dominant over the full UV-to-optical range. This leads to three difficulties, which are the high intrinsic equivalent widths for broad emission lines \citep[see e.g.,][]{Maiolino2024_Xrays}, an extremely high stellar mass density (not observed in any lower redshift galaxy), and an unusual dust attenuation law \citep{mayilun_lrd_2024}.
Assuming the UV continuum has a Balmer break but has an AGN origin also faces difficulties, which are discussed later in Section~\ref{subsec:uv_fit}.
%\redtxt{[reasoning: AGN UV break would mean a strong DLA; SF UV break has been attempted previously by Ma+24 but not a good fit either and has several difficulties]}
In Section~\ref{subsec:uv_fit}, we also discuss in detail the possible physical origin of the UV continuum based on the observed Ly$\alpha$ damping wing. However, at the level of the fitting, as this is not the focus of the paper, we remain agnostic on the origin of the UV continuum and, as said, we simply empirically fit it with a power-law.

Finally, for the narrow-line emission, we only include \hb\ and H$\gamma$ as no other lines are significantly detected in the PRISM spectrum.
While \citet{furtak2024} claimed detection of other lines in the PRISM spectrum such as \oiii$\lambda \lambda 4960, 5008$ and S\,{\sc ii}$\lambda 6347$, as shown in Section~\ref{sec:data}, presence of these lines depends on data reduction and thus we do not consider them as confirmed detections.
The flux of the narrow \hb\ in the model is fixed to that of the broad \hb\ by the observed ratio of $\rm H\beta _{narrow}/H\beta _{broad}$ in Table~\ref{tab:measurements}.
%by multiplying the observed ratio of $\rm H\beta _{broad}/H\beta _{narrow}$ to the flux of the transmitted broad \hb\ in the model.
Then the flux of the narrow H$\gamma$ is obtained by multiplying the flux of the narrow \hb\ by a Case B factor of 0.469 \citep{draine2011}.
Since the narrow-line emission is significantly weaker than the broad-line emission, whether we include the dust attenuation for the narrow-line emission or not has little impact on the fitting results.

We used Equation~\ref{eq:full_model} to describe the spectral range of $\lambda = 1600-6400$ \AA\ to avoid the proximity of the Ly$\alpha$ damping wing and the partially covered \ha\ line.
Later in Section~\ref{subsec:uv_fit}, we extend the spectral range to $\lambda = 1250-6400$ \AA\ to fit part of the Ly$\alpha$ damping wing.
%\redtxt{[convolution]}
Before fitting, we convolve the BLR nebular emission with a FWHM measured from the broad \hb, which is $2658$ \kms (the intrinsic FWHM of the \cloudy\ models is negligible compared to the value).
We did not convolve the attenuated continuum emission or the narrow-line emission with the measured widths, which are much narrower compared to the LSFs of the PRISM spectrum ($\sim 1-3\times 10^3$ \kms).
We then convolved the full model with the LSF for point sources observed with the MSA given by \citet{degraaff2024}, and resampled the model to the wavelength grid of the PRISM spectrum using the \textsc{Python} function \textsc{spectres} \citep{spectres}.

There are a total of five free parameters to be fitted once the parameters for the \cloudy\ simulations are fixed.
These parameters are the normalization of the power law, $b$, the power-law slope, $m$, the covering fraction, $C_f$ [$\in (0,1)$], the magnitude of the V-band attenuation, $A_V$, and the overall normalization of the BLR component, $f_{\rm BLR}$ (the normalization of the narrow-line component has been tied to the BLR component based on the flux ratio in Table~\ref{tab:measurements}).
We used the Markov chain Monte Carlo (MCMC) method with \textsc{emcee} \citep{emcee} to estimate the best-fit values of the parameters. We set the likelihood function to
\begin{equation}
    \log L_{\rm likelihood} = \sum_{\lambda} -0.5(F_{\lambda;\rm obs.}-F_{\lambda;\rm model})^2/\sigma_{\lambda}^2,
\end{equation}
and assumed flat priors for all model parameters.
We ran 5000 steps of MCMC for each fit and took the medians of the posterior distributions as the best-fit parameters. The $1\sigma$ uncertainties are estimated using the 68 percent confidence interval of the posterior distributions.

\subsection{Balmer break fitting: Results}
\label{sec:results_BB}

In this section, we present our fitting results.
As we mentioned in the previous section, there are a total of 5 free parameters we need to constrain in addition to the \cloudy\ parameters.
While the \cloudy\ simulations cover a large parameter space, as shown by \citet{im24}, the Balmer break is enhanced in certain regimes of the parameter space.
Therefore, we divide the section into two major parts.
In the first part, we reexamine the dependencies of the Balmer break and Balmer absorption on various \cloudy\ parameters, similar to what \citet{im24} did.
In the second part, we perform a series of full spectral fits and check the effects of other model parameters in Equation~\ref{eq:full_model}.

\begin{figure*}
    \centering
    \includegraphics[width=\columnwidth]{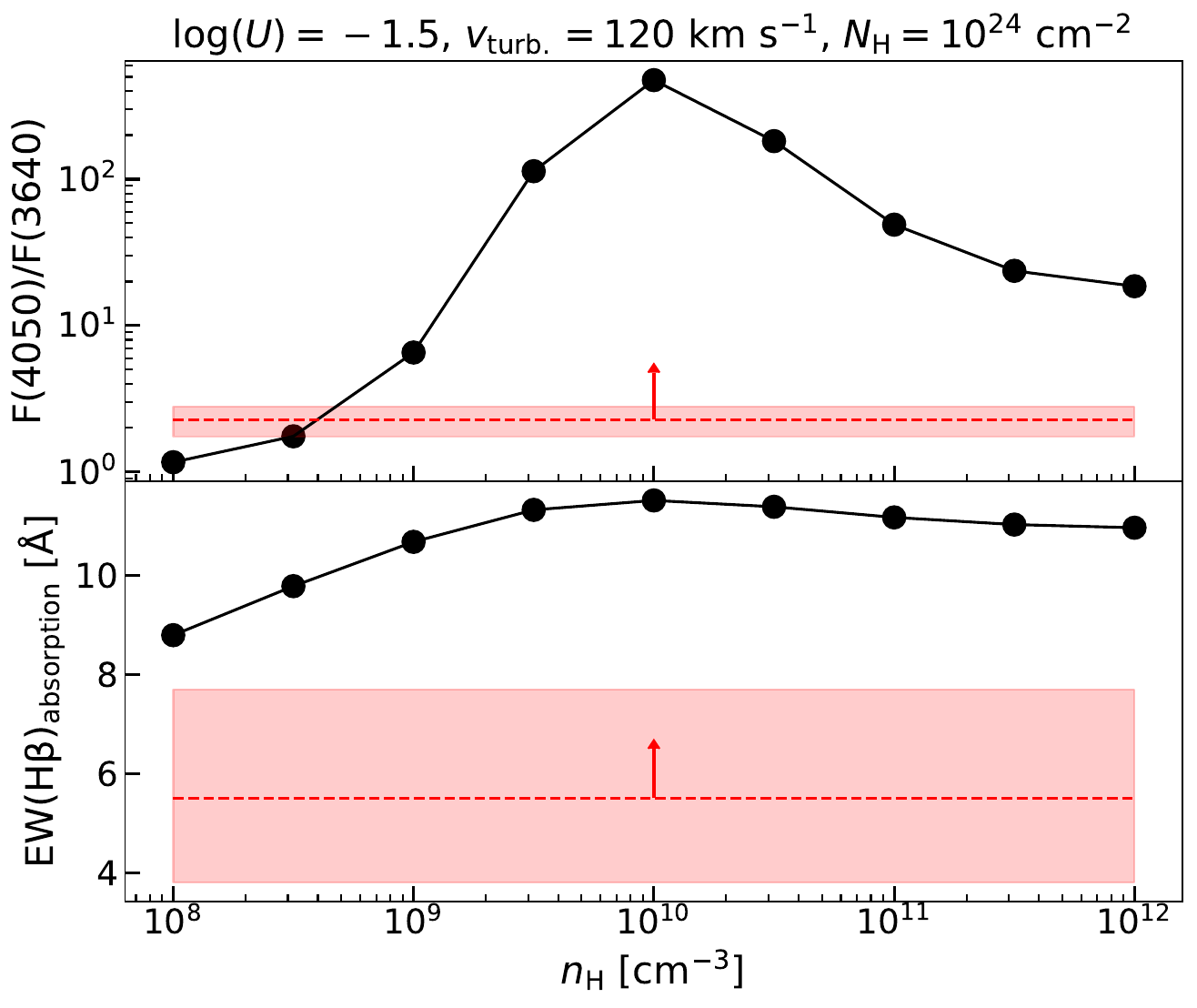}
    \includegraphics[width=\columnwidth]{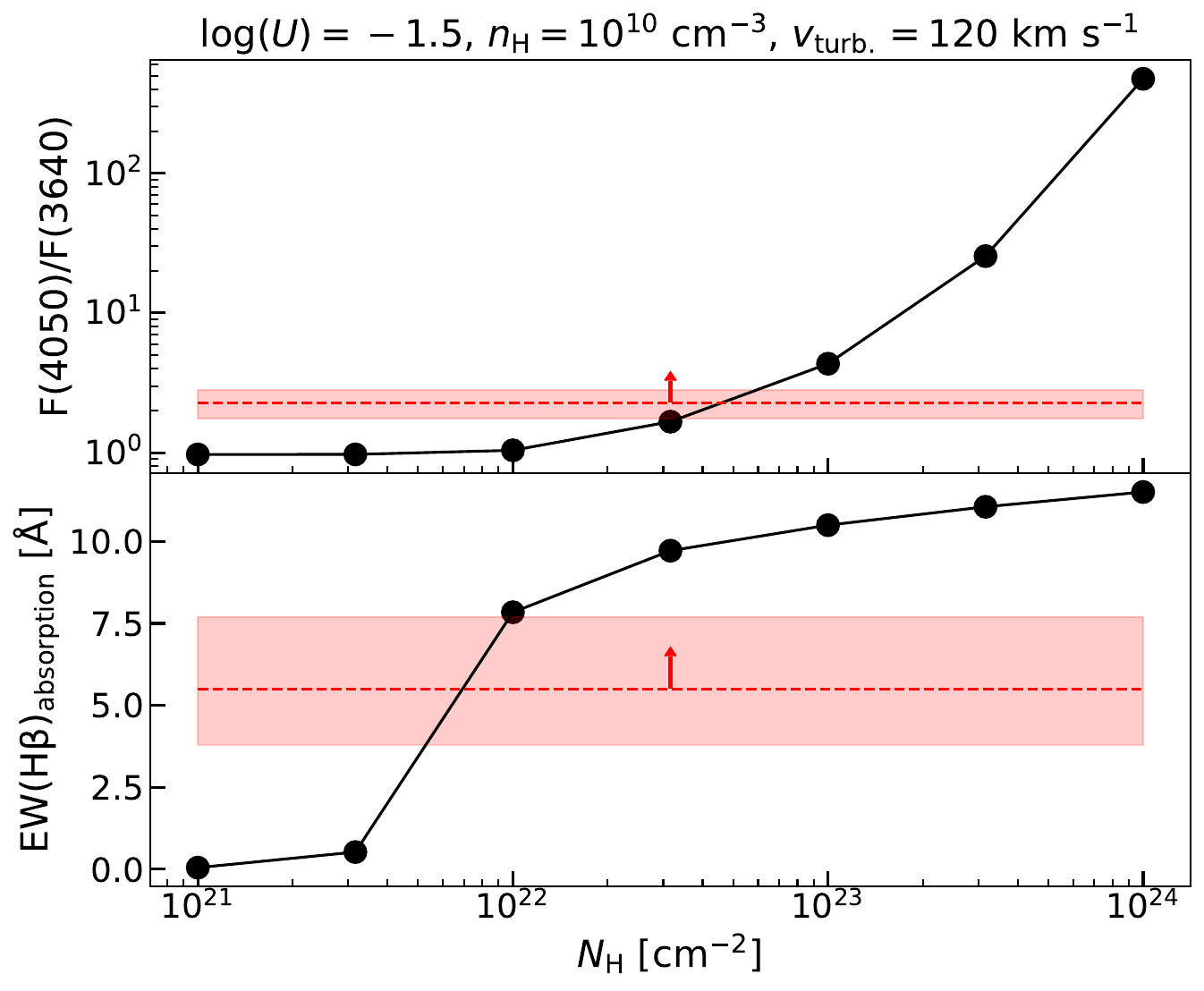}
    \includegraphics[width=\columnwidth]{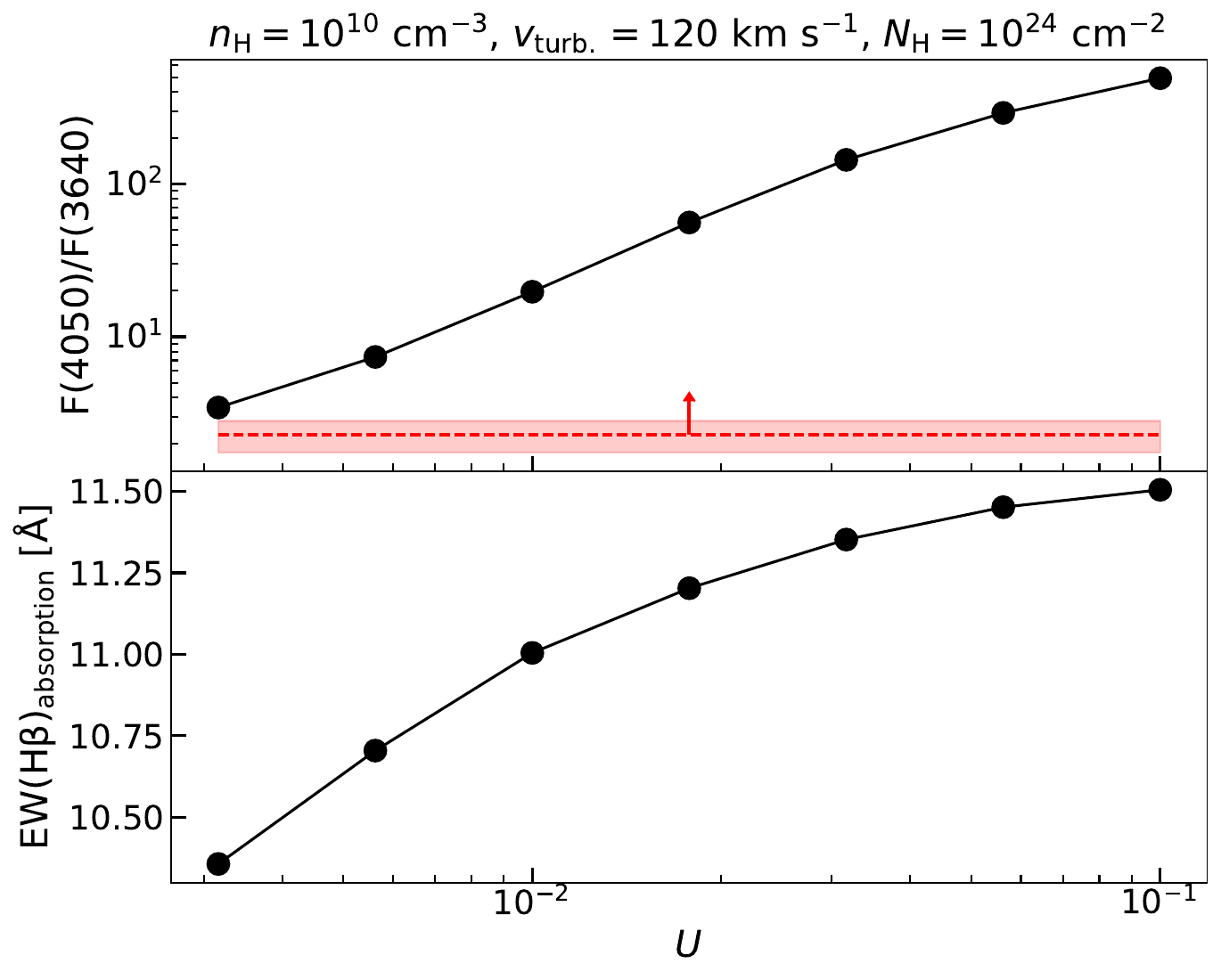}
    \includegraphics[width=\columnwidth]{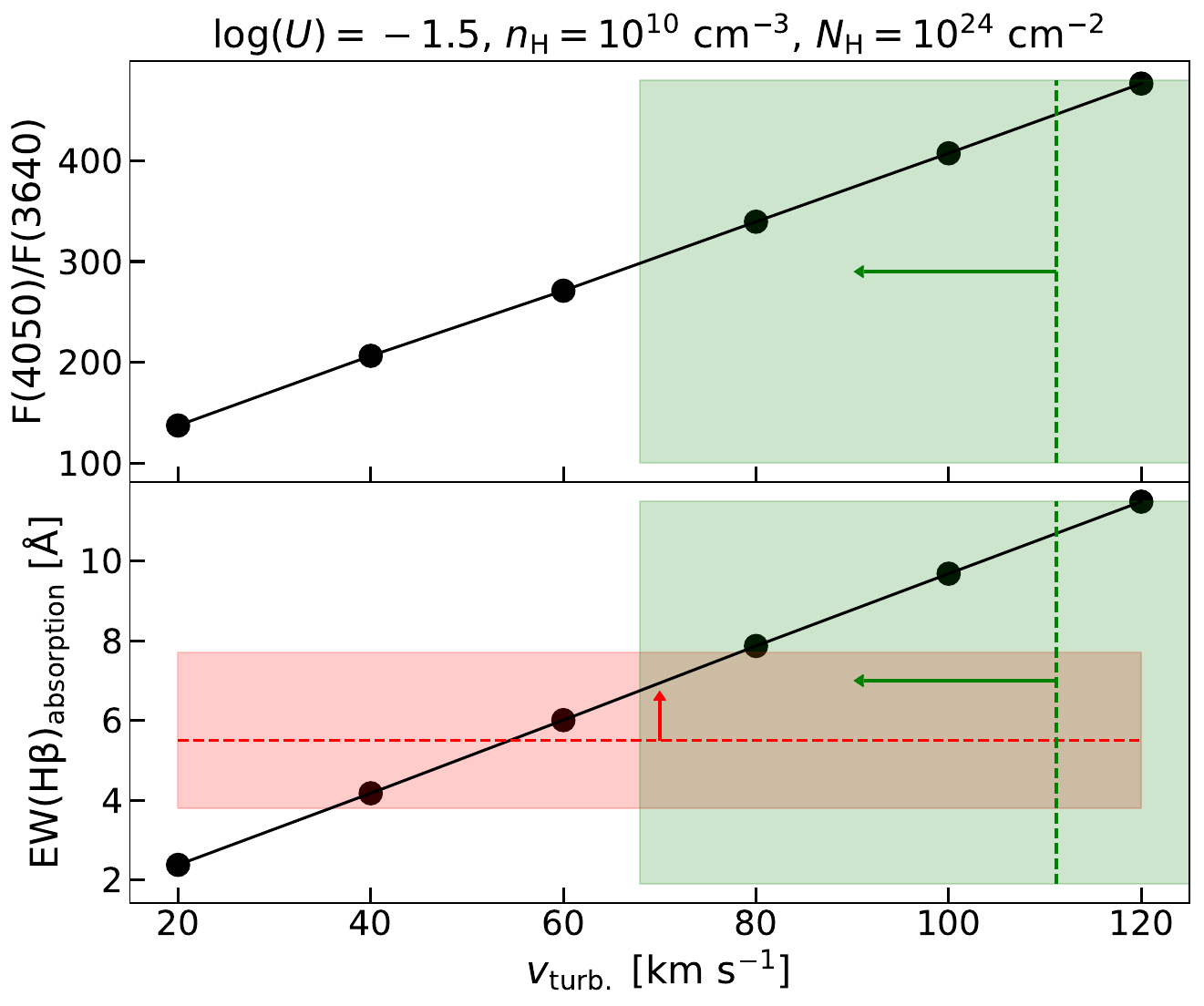}
    \caption{Dependencies of the strength of the Balmer break and the equivalent width of the \hb\ absorption on various model parameters for an AGN continuum attenuated by a slab of dust-free gas computed with \cloudy. From top left to bottom right, we vary the hydrogen density, $n_{\rm H}$, the hydrogen column density, $N_{\rm H}$, the ionization parameter, $U$, and the microturbulence velocity, $v_{\rm  turb}$, respectively.
    The horizontal red dashed lines and shaded regions represent lower limits and their 68\% confidence intervals constrained by the observed spectra of \target.
    The vertical green dashed lines and shaded regions represent the upper limit on $v_{\rm turb}$ and its 68\% confidence interval.
    For the strength of the Balmer break, the constraint is a lower limit since the intrinsic break is lifted by the extension of the UV continuum.
    Similarly, for EW(\hb), the constraint is a lower limit since the optical tail the UV spectrum is not subtracted.
    The upper limit on $v_{\rm turb}$ is obtained from the measured width of the \hb\ absorption.
    %We used these constraints to pre-select model parameters.
    The inclusion of the \hb\ absorption tightens the constraints on the microturbulence velocity, $v_{\rm turb}$, assuming it originates in the same absorber that produces the Balmer break.
    }
    \label{fig:dependency}
\end{figure*}

\begin{comment}
    
\begin{figure*}
    \centering
    \includegraphics[width=\textwidth]{Figs/fit_ledd_combined.pdf}
    
    \caption{Comparison between three fits to the PRISM spectrum of \target\ using different AGN SEDs with $\lambda _{\rm Edd}=0.1,1,\rm and~ 10$ zoomed in the rest-frame optical. The observed spectrum is from the GTO reduced A+C images and is plotted in solid black. The bottom panels show $\chi$ of the fits calculated for each pixel.
    In each panel, we also show the reduced $\chi ^2$ of the fit.
    The main differences between the fits are the nebular emission in H$\gamma$, \heii$\lambda 4686$, and \hei$\lambda 5876$.
    Overall the super-Eddington case with $\lambda _{\rm Edd}=10$ provides the best fit, but only marginally relative to the sub-Eddington fit with $\lambda _{\rm Edd}=0.1$.
    }
    \label{fig:fit_ledd}
\end{figure*}

\end{comment}

\begin{figure*}
    \centering
    \includegraphics[width=0.9\textwidth]{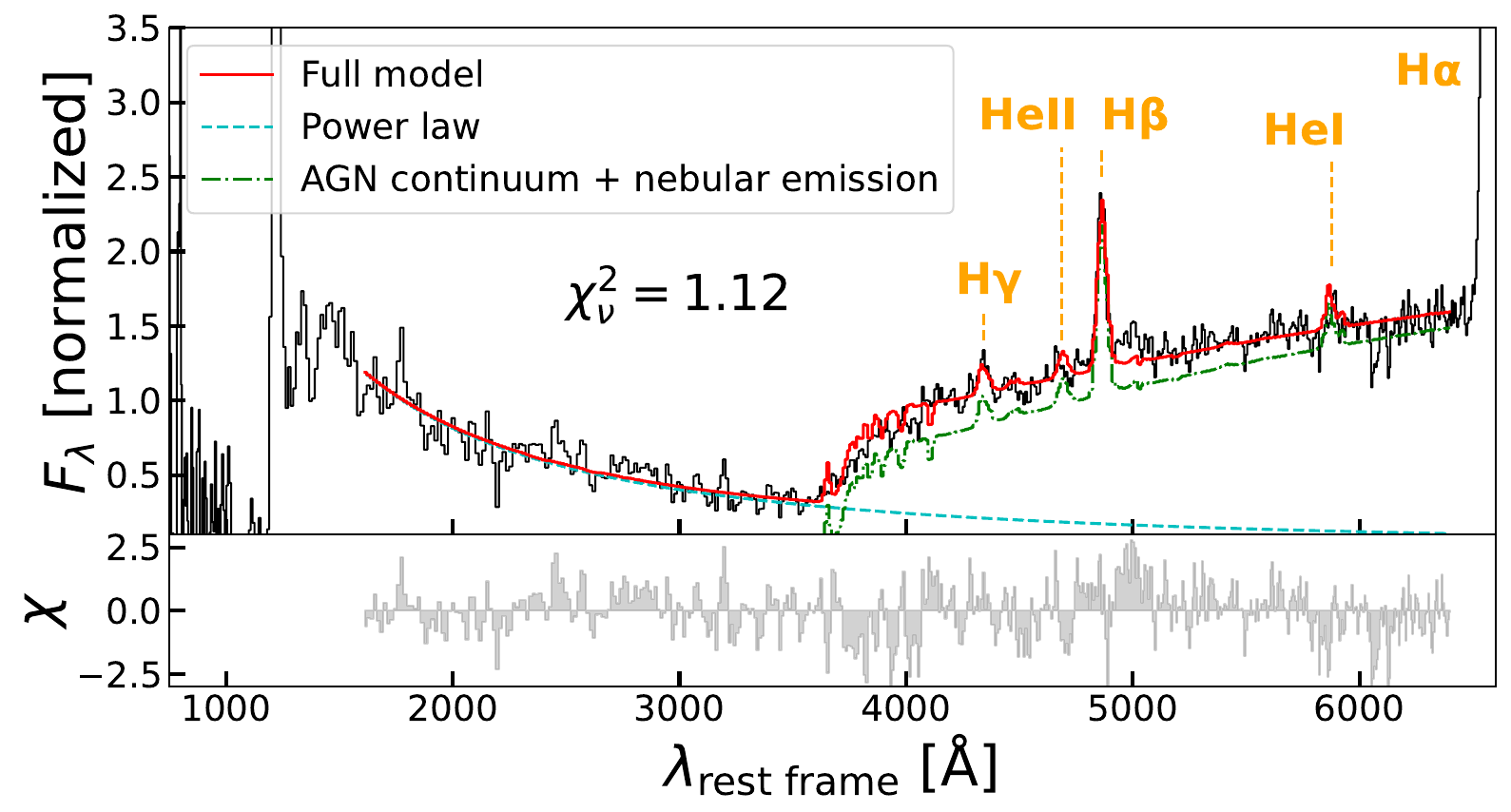}
    \includegraphics[width=1.2\columnwidth]{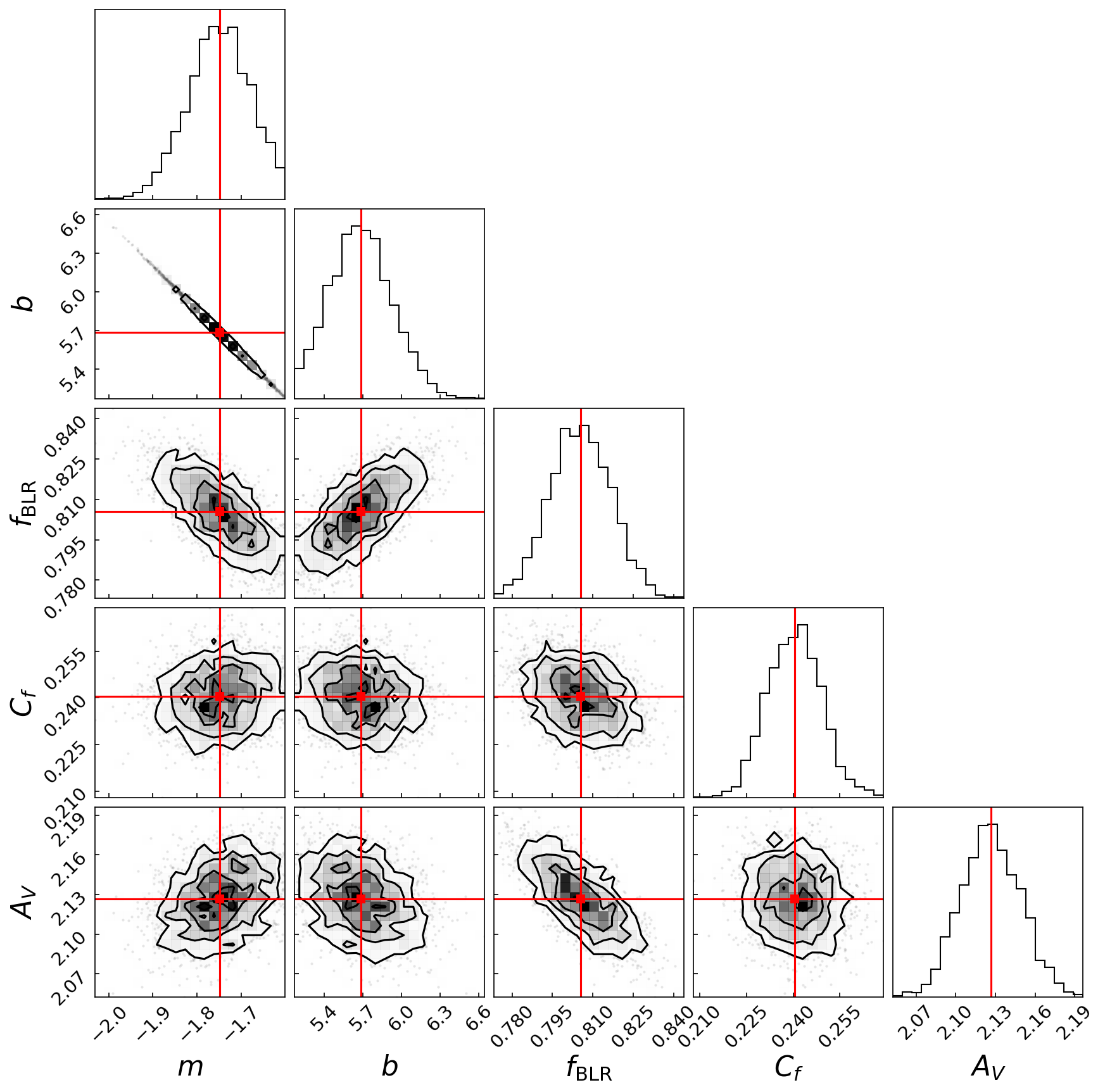}
    \caption{Fiducial fit for the PRISM spectrum of the A+C images reduced by the JADES GTO pipeline.
    \textit{Top:} Best fit model for \target, where the UV component and the optical component are plotted as the dashed line and the dash-dotted line, respectively.
    \textit{Bottom:} Posterior distributions of the model parameters including the UV power-law slope, $m$, the UV power-law normalization, $b$, the normalization of the optical continuum, $f_{\rm BLR}$, the covering fraction of BLR clouds, $C_f$, and the magnitudes of the V-band extinction for the AGN emission, $A_{\rm V}$.
    The medians of the distributions are used as the best estimations and are indicated by the solid red lines. 
    %\redtxt{[update corner plot]}
    }
    \label{fig:fit_demo}
\end{figure*}

\begin{figure*}
    \centering\includegraphics[width=\textwidth]{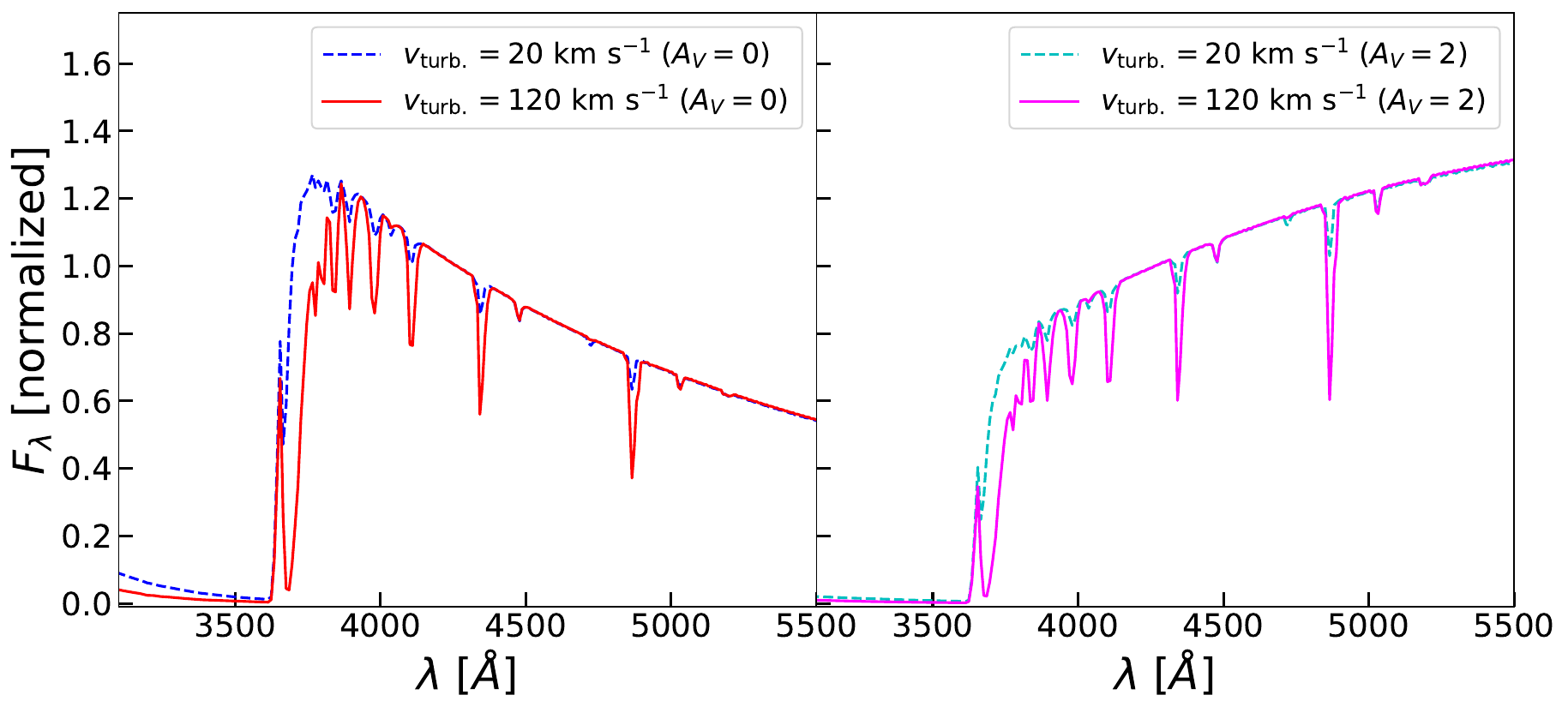}
    \includegraphics[width=\textwidth]{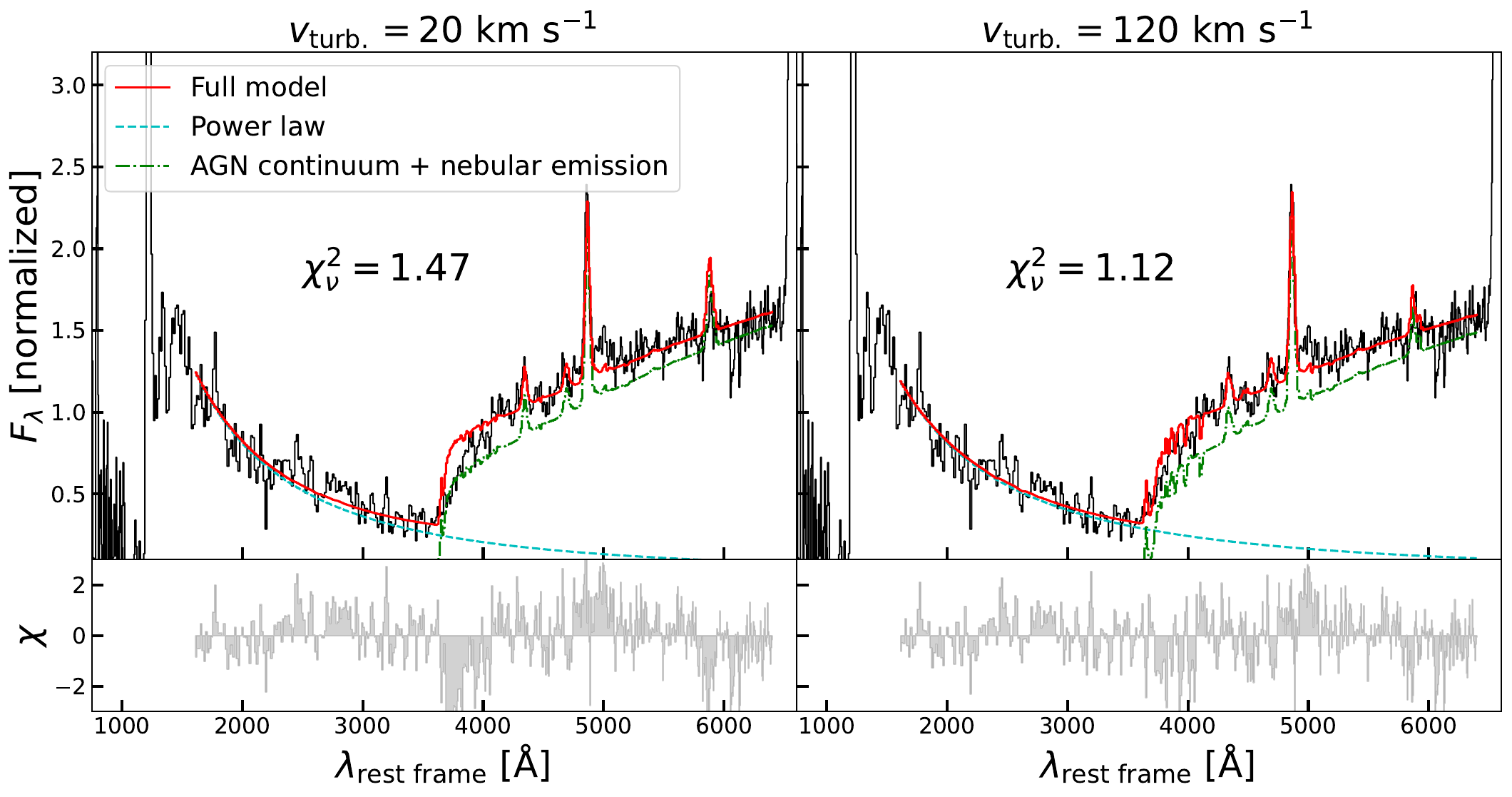}

    \caption{Comparison between model fits with low turbulence and high turbulence.
    \textit{Top:} \cloudy\ models for attenuated AGN continua with $v_{\rm turb}=20$ \kms\ and 120 \kms.
    Besides $v_{\rm turb}$, the rest of the model parameters are fixed to the fiducial values in Table~\ref{tab:models}.
    The models have been convolved to the LSF of the PRISM spectrum and resampled to the wavelength grid of the PRISM spectrum.
    The left panel shows the model continua with no dust attenuation.
    The right panel shows the model continua attenuated by a foreground dust screen with $A_{\rm V}=2$. The model with a higher $v_{\rm turb}$ produces deeper Balmer absorption and thus a more redshifted wing redward to the Balmer break.
    \textit{Bottom:} Best-fit models for the PRISM spectrum of \target\ (images A+C with the GTO reduction) with $v_{\rm turb}=20$ \kms\ and 120 \kms\ with a best-fit $A_{\rm V}\approx 2.1$ mag.
    The high-turbulence fit yields a smaller $\chi^2_{\nu}$ due to a better match between the model and the observed spectrum close to the Balmer break.
    }
    \label{fig:fit_turb}
\end{figure*}

\begin{figure*}
    \centering
    \includegraphics[width=\textwidth]{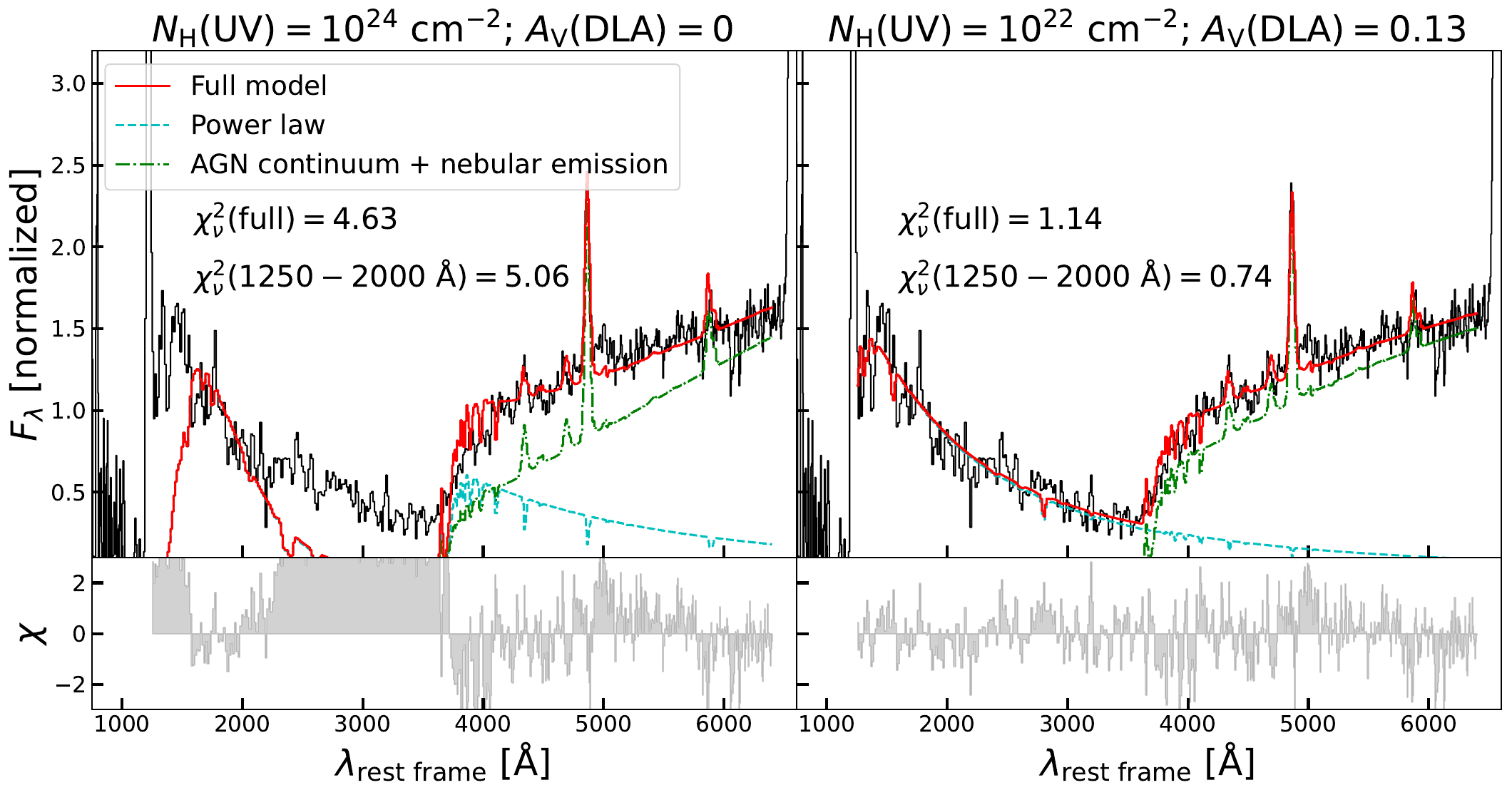}
    
    \caption{Comparison between two fits to the PRISM spectrum (extracted from the combined A+C images with the GTO reduction) with different UV continuum models.
    The UV model in the left panel is the AGN continuum emission attenuated by gas with $N_{\rm H}=10^{24}~{\rm cm^{-2}}$ (i.e., the same gas that attenuates the optical continuum).
    The UV model in the right panel is the AGN continuum emission attenuated by gas with $N_{\rm H}=10^{22}~{\rm cm^{-2}}$.
    Both UV continuum models are allowed to have dust attenuation characterized by $A_{\rm V}$(DLA) during the fit.
    The rest of the photoionization model parameters are the same as the fiducial values listed in Table~\ref{tab:models}.
    Different from previous fits, the fitted spectral range is extended to $1250-6400$ \AA\ in the rest frame of \target\ to capture part of the DLA-like feature and avoid Ly$\alpha$ emission.
    The UV component is better fitted by the model with a lower column density compared to that of the optical component.
    }
    \label{fig:fit_uvcol}
\end{figure*}

\begin{figure}
    \centering
    \includegraphics[width=\columnwidth]{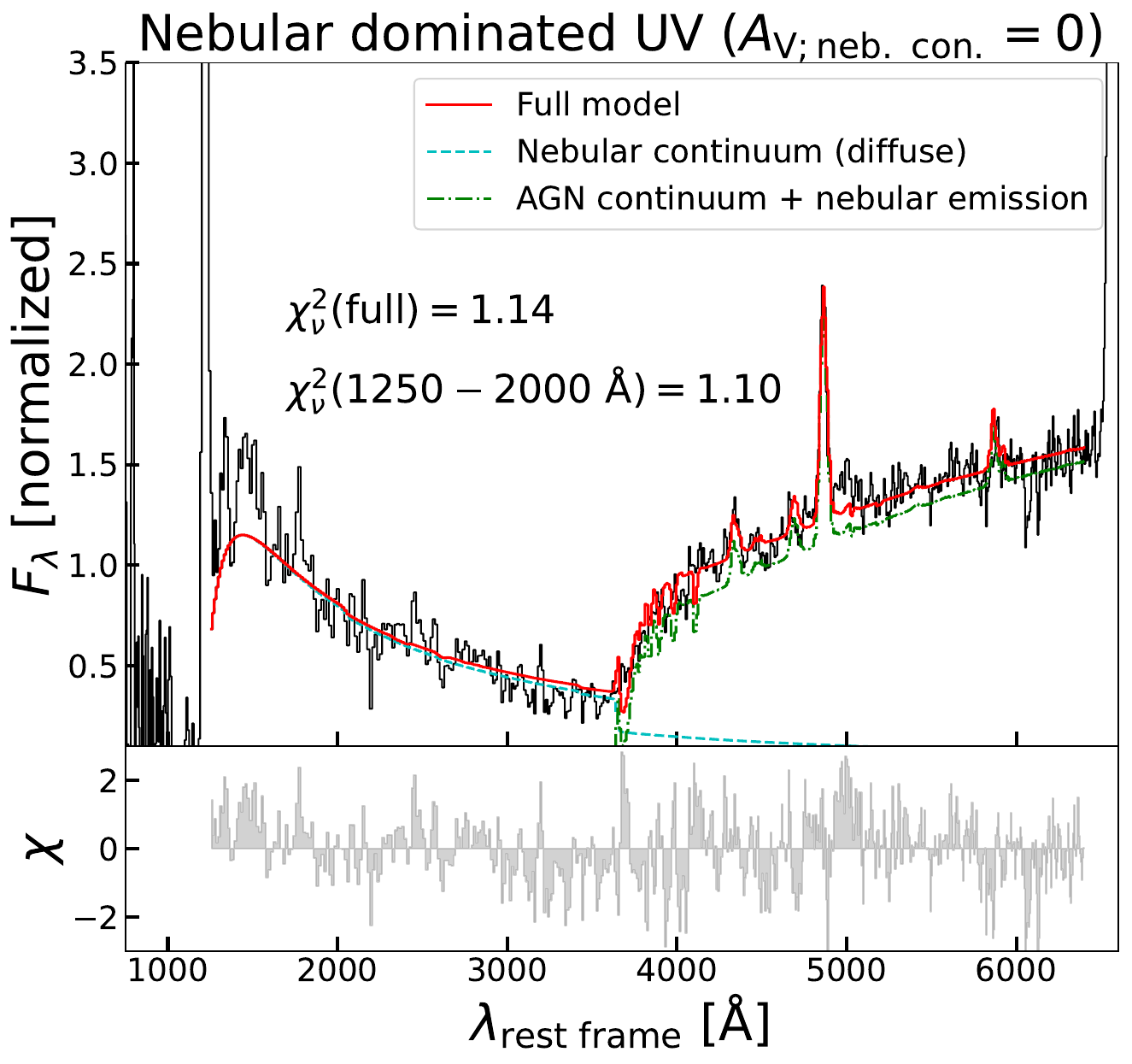}

    \caption{Fitting attempt that adopts a nebular dominated continuum for the UV component.
    The nebular continuum is assumed to come from low-density (diffuse) gas with no dust attenuation, and it includes hydrogen two-photon emission, the free-bound emission, and the free-free emission.
    The temperature and the normalization of the nebular continuum are allowed to vary.
    While the diffuse nebular continuum provides a equally good fit to that of the DLA model based on the overall $\chi^2_{\nu}$, the fit in the FUV region near Ly$\alpha$ is worse.
    %slightly worse fit near Ly$\alpha$ compared to the DLA model, the overall $\chi^2_{\nu}$ is the same as that of the DLA fit.
    }
    \label{fig:fit_2gamma}
\end{figure}

\subsubsection{Effects of different photoionization parameters}
\label{subsec:params_dependency}

%\redtxt{[TBD: change the fiducial value of the Eddington ratio to $\lambda _{\rm Edd}=0.01-0.1$?]}

In Table~\ref{tab:models}, we list a series of photoionization parameters to vary for \cloudy\ models, including the shape of the SED, the ionization parameter, $U\equiv\Phi_{0}/n_{\rm H}c$ ($\Phi_0$ is the hydrogen ionizing photon flux, $n_{\rm H}$ is the hydrogen density, and $c$ is the speed of light), the hydrogen density, $n_{\rm H}$, the hydrogen column density, $N_{\rm H}$, and the microturbulence velocity, $v_{\rm turb}$.
We set the metallicity to 0.1 solar, as no strong constraints from nebular emission lines are available \citep{furtak2024}.
Regardless, the metallicity has little impact on our results as we focus on spectral features produced by hydrogen.
Additionally, we considered no dust in our simulations, as expected for typical BLRs that lie within the dust sublimation radius \citep{netzer_1993,gaskell_2009}. The possible presence of dust surviving in the outer (possibly self-shielded) parts of the BLR \citep{shields2010}, is accounted for by the dusty-screen part of the model, whose location does not affect the results.

First of all, we explore the effects of varying parameters other than the SED.
Figure~\ref{fig:dependency} shows the model-predicted strength of the Balmer break as well as the \hb\ absorption as a function of different parameters, where the SED is fixed to the one with $M_{\rm BH}=10^7~M_\odot$ and $\lambda _{\rm Edd}=0.1$ that best match our measurements in Section~\ref{sec:mass_acc}.
Here we characterize the strength of the Balmer break using the ratio between the flux densities at 4050 \AA and {3640} \AA to avoid any line features {as well as the artificial spike seen in Figure~\ref{fig:model_demo}}.
The strength of the \hb\ absorption is simply represented by its equivalent width (EW) with respect to the continuum.
In each panel, we vary only one parameter while keeping all the other parameters fixed to the values shown on the top.

The top left panel of Figure~\ref{fig:dependency} shows the effects of $n_{\rm H}$ once the other parameters are fixed.
The Balmer break is much more sensitive to the variation in $n_{\rm H}$ compared to the \hb\ absorption.
Overall both the break and the absorption reach the maximum values around $n_{\rm H}=10^{10}~{\rm cm^{-3}}$.
The same effect was noted by \citet{im24} and is explained by the strength of the collisional excitation.
At very low densities, the collisional excitation of hydrogen is not strong enough to populate the atoms to their excited states.
At very high densities, the collisional excitation is so strong that a considerable fraction of hydrogen is populated at $n>2$ levels.
Both of the above tend to reduce the strength of the Balmer absorption, leading to the peak we see at $n_{\rm H}\approx 10^{10}~{\rm cm^{-3}}$ {While this density produces the strongest Balmer break, it should be noted that a Balmer break is expected at any density higher than $10^9~{\rm cm}^{-3}$}.

The top right panel of Figure~\ref{fig:dependency} shows the effect of $N_{\rm H}$.
Both the Balmer break and the \hb\ absorption are enhanced with increasing $N_{\rm H}$.
The only difference is that the Balmer break starts to increase with $N_{\rm H}$ faster at $N_{\rm H}\gtrsim 10^{23}~{\rm cm^{-2}}$, whereas the \hb\ absorption starts to saturate at $N_{\rm H}\gtrsim 10^{22-23}~{\rm cm^{-2}}$.
Increasing $N_{\rm H}$ strengthens Ly$\alpha$ trapping that helps to populate hydrogen to $n=2$, thereby leading to the observed trends  \citep{hall2007,juodzbalis_agnabs_2024}.

The bottom left panel of Figure~\ref{fig:dependency} shows the effect of the ionization parameter $U$.
Increasing $U$ significantly enhances the Balmer break and slightly enhances the Balmer absorption.
This can be understood as the fact that the bound-free absorption is enhanced with the increasing ionizing flux.

Finally, the bottom right panel shows the effect of the microturbulence velocity, $v_{\rm turb}$.
The microturbulence velocity describes the relative motions of gas within the line emitting/absorbing regions and is related to the observed velocity dispersion via $v_{\rm turb}=\sqrt{2}\sigma$.
While the physical origin of the microturbulence remains debated, mechanisms such as disk winds and magnetohydrodynamic (MHD) waves have been proposed to explain the formation of the turbulence \citep{bottorff_mhd_2000,bottorff2000}.
It has been shown by many previous modeling works that to reproduce the nebular emission of BLRs, a microturbulence velocity of $v_{\rm turb} \sim 100$ \kms\ is needed, which can enhance line escaping as well as continuum pumping \citep[e.g.,][]{baldwin2004,ferland2009,sarkar2021}.
In our case, $v_{\rm turb}$ clearly enhances the \hb\ absorption but only moderately enhances the Balmer break.
The enhancement of EW(\hb) can be understood as the strengthened continuum absorption due to a broader wing of the absorption profile boosted by the velocity dispersion \citep{juodzbalis_agnabs_2024}.
More specifically, the absorption profile around \hb\ can be written in the analytical form of
\begin{equation}
    f_\lambda \propto \exp[-\tau_0~e^{-(1 - \lambda/\lambda _0)^2c^2/(v_{\rm therm}^2+v_{\rm turb}^2)}],
    \label{eq:line_abs}
\end{equation}
where $\tau_0$ is the optical depth at the line center (which depends inversely on the line width, see \citealp{draine2011}), $\lambda _0$ is the central wavelength, $c$ is the speed of light, $v_{\rm therm}= \sqrt{2k_{\rm B}T_{\rm e}/m_{\rm H}}$ is the thermal broadening (typically 10-20 \kms), and $v_{\rm turb}$ is the turbulent broadening.
This equation indicates that the absorption profile is broadened at higher turbulence, leading to a larger equivalent width.
We note that \citet{im24} did not consider $v_{\rm turb}$ in their \cloudy\ simulations, which actually lead to an important implication on the smoothness of the Balmer break, as we discuss later in this section.

Besides the model dependencies, we also show the apparent strengths of the Balmer break and \hb\ absorption measured from the spectra of \target\ in Figure~\ref{fig:model_demo}.
These measurements are shown as lower or upper limits with uncertainties.
The Balmer break of \target\ is measured using the same definition as adopted in the model [i.e., $F(4050)/F(3640)$] and has a value of 2.28\footnote{As a comparison, \citet{wang_break_2024} reported break strengths for three LRDs in a range of 1.96-2.44, although they adopted a slightly different definition using spectral windows centered at wavelengths similar to those we adopted.}. However, since we have not separated the optical continuum from the UV continuum for \target\ at this stage, the unremoved UV continuum tends to lower this ratio.
Thus, we considered the measured $F(4050)/F(3640)$ as a lower limit, as shown in Figure~\ref{fig:model_demo}.
One can see this lower limit already sets some rough constraints on \cloudy\ model parameters, requiring $n_{\rm H}>10^{8.5}~{\rm cm^{-3}}$ and $N_{\rm H}>10^{22.5}~{\rm cm^{-2}}$, although this is subject to the choice of $U$.
%The effect of $U$ is mainly on the nebular emission.
\citet{juodzbalis_agnabs_2024} found that the ionization parameter of the Balmer-line absorbing gas in another \jwst-selected AGN at $z=2.26$ is $U\sim 10^{-1.8}-10^{-2.1}$ based on the detection of both Balmer and \hei\ absorption lines.
{We chose a slightly higher value of $U\sim 10^{-1.5}$ to start with.}
%For the moment, we take $U=10^{-1.8}$ as obtained from another \jwst-selected AGN at $z=2.26$ with much more stringent constraints on the ionization conditions of the Balmer-line absorbing gas \citep{juodzbalis_agnabs_2024}.
Overall the choice of $U$ within a range of $10^{-2}-10^{-1.5}$ does not impact our conclusions, as we show in Appendix~\ref{appendix:other_params}.
Furthermore, we show the measurement of EW(\hb) in absorption from the \blackthunder\ spectrum together with the $1\sigma$ uncertainty, which is $5.5^{+2.2}_{-1.7}$ \AA.
Again, we interpret this measurement as a lower limit, since the optical tail of the UV continuum should lower the apparent EW(\hb).
The lower limit on EW(\hb) sets a rough constraint on the turbulence, requiring $v_{\rm turb}\gtrsim 60$ \kms. Meanwhile, the measured width of the absorption sets an upper limit on the turbulence velocity, $v_{\rm turb}\lesssim \sqrt{2} \sigma _{\rm H\beta;abs} \approx 110$ \kms.
However, due to the large uncertainty in the measured width of the \hb\ absorption (see Table~\ref{tab:measurements}), the highest allowable value for the turbulence velocity is actually $110\pm 40$ \kms. We considered $v_{\rm turb}$ up to 120 \kms, which is still well within the $1\sigma$ uncertainty.
As we mention later in this section, $v_{\rm turb}=120$ \kms\ produces the smallest $\chi^2$ compared to other values by fitting the observed Balmer break best.
This motivates us to choose $v_{\rm turb}=120$ \kms\ as the fiducial value, although we note that making a more conservative choice of $v_{\rm turb}=100$ \kms\ only slightly increases the reduced $\chi^2$ of the fit.

Combining all the above analyses, to model the attenuated AGN emission with spectral features compatible with the observational limits and a deep Balmer break, we chose $n_{\rm H} = 10^{10}~{\rm cm^{-3}}$, $N_{\rm H} = 10^{24}~{\rm cm^{-2}}$, {$U=10^{-1.5}$}, and $v_{\rm turb}= 120$ \kms\ as the fiducial model parameters.
We further discuss the impacts of these parameters later in this section as well as in Appendix~\ref{appendix:other_params}.
%on the fitting results later in this section as well as in Appendix~\ref{appendix:other_params}. \redtxt{[note]}

%\redtxt{[this can also be moved later or to the appendix if needed?]} 
The remaining parameter is the shape of the AGN SED, for which we list a range of Eddington ratios in Table~\ref{tab:models}.
To see the impact of the AGN SED, we performed three fits with different $\lambda _{\rm Edd}$ and fixed the other \cloudy\ parameters to the fiducial values.
The fitting results for each of these cases are shown in {Figure~\ref{fig:fit_ledd} of Appendix~\ref{appendix:other_params}}.
The overall continuum shapes produced by different models are very similar.
\begin{comment}
    
The most notable difference is in the nebular emission.
When $\lambda _{\rm Edd}=0.1$, the strengths of \heii$\lambda 4686$ and \hei$\lambda 5876$ from the BLR is slightly overpredicted.
When $\lambda _{\rm Edd}=1$, \heii$\lambda 4686$ and \hei$\lambda 5876$ are further overpredicted but H$\gamma$ is underpredicted.
The smallest reduced $\chi^2$ is achieved at $\lambda _{\rm Edd}=10$.
The above difference is mainly driven by the hardness of the ionizing radiation, which is generally too hard for \target\ at the sub-Eddington regime.
This is consistent with the recent finding of \citet{lambrides2024}, who claimed that \jwst-selected Type 1 AGN generally show softer SEDs compared to those expected at the sub-Eddington regime.
Still, the difference in the reduced $\chi^2$ between the super-Eddington case and the sub-Eddington case is only 0.05. Also, based on our derivation in Section~\ref{sec:mass_acc}, \target\ is more likely to host a sub-Eddington AGN.
In addition, according to \citet{inayoshi_spedd_2024}, the UV spectrum of a super-Eddington AGN should actually become harder due to the emission from the hotter part of the accretion disk within the photon-trapping radius.
This is in contrast to the SED model we took from \citet{pezzulli2017}, which does not include the emission within the photon-trapping radius.
\end{comment}
Therefore, we choose $\lambda _{\rm Edd}=0.1$ in our following analysis, also because it is more aligned with the accretion rate inferred by us in Section~\ref{sec:mass_acc} ($\lambda _{\rm Edd}\sim 0.05-0.24$).
We caution that we are not trying to set a stringent constraint on the Eddington ratio, but rather select a plausible value.
{As shown in Figure~\ref{fig:fit_ledd}, the main effect of different Eddington ratios is the strength of the nebular emission, which also leads to a change in the best-fit covering fraction.
In Section~\ref{sec:discussion}, we further discuss the scenario where \target\ actually hosts an AGN accreting at a super-Eddington rate.
}

Finally, we can use the fiducial parameters to infer the mass of the dense gas. If we assume that the gas responsible for the Balmer absorption is the same gas emitting the broad H$\beta$ then gas mass can be simply inferred from the equation adopted by \citet{Carniani_2015} assuming $T_{\rm e}\sim 10^4$ K

%$$
%\rm M_{dense~gas} = 0.85 \left( \frac{L_{H\beta}}{10^{42}~\rm erg/s}\right)
%\left( \frac{n}{10^{10}~\rm cm^{-3}}
%\right)^{-1} ~\rm M_\odot
%$$

\begin{equation}
    M_{\rm dense~gas} = 0.85 \left(\frac{L_{\rm H\beta}}{10^{42}~{\rm erg~s^{-1}}}\right)\left(\frac{n_{\rm H}}{10^{10}~{\rm cm^{-3}}} \right)^{-1}~M_\odot.
\end{equation}
Using the observed H$\beta$ luminosity corrected for extinction as discussed in Section~\ref{sec:mass_acc}, and the gas density derived by our fitting of the absorber in this section, we obtain a gas mass of only $\sim 4~M_\odot$. 
\redtxt{We caution that the above calculation assumes typical emissivity coefficient, $\epsilon _{\rm H\beta}\equiv j_{\rm H\beta}/(n_{\rm e}n_{\rm p})$ (where $j_{\rm H\beta}$ is the \hb\ emissivity), for \hb\ in the NLR under the Case B assumption at $T_{\rm e}=10^{4}$ K, whereas in the BLR, $\epsilon _{\rm H\beta}$ can be different \citep{agn3_2006}.
With \cloudy, we checked this effect by computing the average $\epsilon _{\rm H\beta}$ of two AGN-ionized clouds with $n_{\rm H}=10^{10}~{\rm cm^{-3}}$ (BLR-like) and $n_{\rm H}=500~{\rm cm^{-3}}$ (NLR-like), and we found BLR $\epsilon _{\rm H\beta}$ is roughly 12\% higher that of the NLR $\epsilon _{\rm H\beta}$ at a fixed temperature of $T_{\rm e}=10^4$ K, slightly decreasing the gas mass estimate to $3.6~M_\odot$.
Compared to the density, the temperature has a stronger effect, and we found that by increasing $T_{\rm e}$ from $10^4$ K to $3\times 10^4$ K, $\epsilon _{\rm H\beta}$ decreases by roughly a factor of 2, increasing the gas mass estimate to $8~M_\odot$, which is still a relatively small amount of mass.}
Such a small gas mass is quite typical of the BLR of AGN, which can be very luminous in the recombination lines despite the small mass involved, as a consequence of the high densities \citep{maiolino2023b}.

\subsubsection{Decomposing the optical spectrum of \target}
\label{subsubsec:decompose}

\begin{table*}
        \centering
        \caption{Best-fit model parameters for different PRISM spectra of \target. All spectra are normalized to the flux densities at $\lambda =4260$ \AA\ before fitting. The model is given by $F_{\lambda;\rm model}=b\lambda ^m + 10^{-0.4A_\lambda}f_{\rm BLR}F_{\lambda; \rm BLR}+F_{\lambda;\rm NLR}$ (see Equation~\ref{eq:full_model}; here $f_{\rm BLR}$ is a normalization factor and is absorbed in the BLR terms in Equation~\ref{eq:full_model}). The fiducial photonionization model parameters we used to compute $F_{\lambda;\rm BLR}$ are listed in Table~\ref{tab:models}.
        The fitting results for the spectra from \citet{furtak2024}, \blackthunder, and the UNCOVER DR4 (the same one used by \citealp{mayilun_lrd_2024}) are shown in Appendix~\ref{appendix:other_reductions}.
        }
        \label{tab:best_fit}
        \begin{tabular}{l c c c c c}
                \hline
                \multicolumn{6}{|c|}{Fiducial model (images A+C); $\chi^2_{\nu} = 1.12$} \\
                \hline
                Parameter & $m$ & $b$ & $f_{\rm BLR}$ & $C_f$ & $A_{\rm V}$ \\
                Value & $-1.75\pm 0.07$ & $5.7\pm 0.2$ & $0.81\pm 0.01$ & {$0.24\pm 0.01$} & $2.13\pm 0.02$ \\
                \hline
                \multicolumn{6}{|c|}{Fiducial model (images A+B+C with \citealp{furtak2024}'s reduction); $\chi^2_\nu=1.86$} \\
                \hline
                Parameter & $m$ & $b$ & $f_{\rm BLR}$ & $C_f$ & $A_{\rm V}$ \\
                Value & $-1.61\pm 0.01$ & $5.20\pm 0.05$ & $0.723\pm 0.007$ & $0.22\pm 0.01$ & $2.39\pm 0.02$ \\
                \hline
                \multicolumn{6}{|c|}{Fiducial model (image A from \blackthunder); $\chi^2_\nu = 1.09$} \\
                \hline
                Parameter & $m$ & $b$ & $f_{\rm BLR}$ & $C_f$ & $A_{\rm V}$ \\
                Value & $-2.17\pm 0.11$ & $7.0\pm 0.4$ & $0.86\pm 0.01$ & $0.25\pm 0.01$ & $2.04\pm 0.03$ \\
                \hline
                \multicolumn{6}{|c|}
                {Fiducial model (image A from UNCOVER DR4, \citealp{uncover_dr4}); $\chi^2_\nu = 2.38$} \\
                \hline
                Parameter & $m$ & $b$ & $f_{\rm BLR}$ & $C_f$ & $A_{\rm V}$ \\
                Value & $-2.00\pm 0.05$ & $6.5\pm 0.2$ & $0.79\pm 0.01$ & $0.22\pm 0.01$ & $2.18\pm 0.02$ \\
                \hline
        \end{tabular}
\end{table*}

%The turbulent velocity is related to the velocity dispersion by $v_{\rm turb}=\sqrt{2}\sigma$.
%\redtxt{[update best-fit values]}

In Figure~\ref{fig:fit_demo} we show the fitting result of our fiducial model, including the spectral model from the rest-frame UV to the optical for \target, and the best-fit parameters, which are summarized in Table~\ref{tab:best_fit}.
In the top panel of Figure~\ref{fig:fit_demo}, we plot individually the component that dominates the rest-frame UV and the component that dominates the rest-frame optical.
For the moment, we concentrate our discussion on the optical part, which is the focus of this paper, and we resume the discussion on the UV part in Section~\ref{subsec:uv_fit}.

The most notable feature in the optical part of the model is the Balmer break.
%Compared to the model example shown in Figure~\ref{fig:model_demo}, the Balmer break is smoother due to the Balmer absorption on top.
Close to the Balmer limit, the Balmer absorption features are increasingly blended as the energy level $n$ increases.
At the resolution of the PRISM spectrum, the blended absorption ``erodes'' the red side of the Balmer break and leads to the slowly rising break in \target.
To make this smooth break, a key parameter is the turbulence velocity.
As we have seen in Figure~\ref{fig:dependency}, the depth of the \hb\ absorption increases with increasing $v_{\rm turb}$.
The same process applies to other Balmer absorption lines, making the erosion on the red side of the Balmer break stronger at high $v_{\rm turb}$.
We note that in the \cloudy\ models of \citet{im24}, which also predict a Balmer break, $v_{\rm turb}$ is effectively 0 and thus the Balmer break is always sharp.
To demonstrate the importance of the turbulence, we compare two models with $v_{\rm turb} = 20$ \kms\ and 120 \kms, respectively, in Figure~\ref{fig:fit_turb}.
In the top panel, we show the attenuated AGN continua of the two models normalized at $\lambda =4260$ \AA.
The attenuated continuum with $v_{\rm turb} = 120$ \kms\ exhibits significantly deeper Balmer absorption and a redder Balmer jump due to the deepened and blended higher-order Balmer absorption lines.
After applying the dust attenuation, the more turbulent model shows a more gradual break.
The bottom panel of Figure~\ref{fig:fit_turb} compares the fitting results based on these two models.
Notably, without considering a strong turbulence, the fit is much worse around the Balmer break, resulting in $\chi^2_{\nu} = 1.35$, much larger than $\chi^2_{\nu} = 1.10$ arising from a strong turbulence.

%\subsection{Origin of the Balmer break}
The strongest turbulence we adopted is compatible with the upper limit set by the width of the \hb\ absorption measured from the \blackthunder\ spectrum, which is $\rm FWHM _{\rm H\beta;abs}=185^{+69}_{-72}$ \kms\ (corresponding to $v _{\rm turb}<110\pm 40$ \kms).
In addition, a consistency check can be performed using $\rm EW(H\beta)$, which is strongly dependent on $v_{\rm turb}$ as shown in Figure~\ref{fig:dependency}.
The best-fit model with $v_{\rm turb} = 120$ \kms\ gives $\rm EW(H\beta)=8.0$ \AA with respect to the co-added continuum (i.e., attenuated AGN continuum + nebular continuum + UV continuum), which is consistent with the value measured from the high-resolution spectrum, {$\rm EW(H\beta)_{R2700}=5.5^{+2.2}_{-1.7}$ \AA} within {$1.1\sigma$}.
If we lower the turbulence velocity to $v_{\rm turb} = 100$ \kms, the fit is slightly worse with $\chi^2_{\nu} = 1.15$. The resulting strength of the \hb\ absorption becomes $\rm EW(H\beta)=6.8$ \AA, consistent with the observed value within $1\sigma$.
Overall, our fitting results suggest that under the assumption of an AGN dominated Balmer break, the best-fit model indicates a level of turbulence with $v_{\rm turb} \sim 100$ \kms.
This value is consistent with the turbulence velocity adopted in previous photoionization modeling of BLR clouds \citep{baldwin2004,ferland2009,sarkar2021}.

Next, we examine other spectral features in the optical.
The optical spectrum has a rising slope towards longer wavelengths, which can be explained by the dust reddened AGN continuum.
The best-fit value for the visual extinction is $A_{\rm V}=2.13\pm0.02$ mag ($A_{\rm V}=2.18\pm0.02$ mag for the image A used by \citealp{mayilun_lrd_2024}).
This value is slightly higher than the best-fit magnitude of the dust attenuation estimated by \citet{mayilun_lrd_2024} for their AGN-only model ($A_{\rm V}=2.08\pm 0.01$ mag) and Galaxy-only model ($A_{\rm V}=2.12\pm 0.02$ mag), and significantly higher than the best-fit $A_{\rm V}$ of their fiducial model (AGN + Galaxy; $A_{\rm V}=0.50\pm 0.09$ mag for the AGN component and $A_{\rm V}=2.02\pm 0.02$ mag for the Galaxy component).
Compared to the fiducial fit of \citet{mayilun_lrd_2024} to the image A of \target\ from the UNCOVER DR4, beside obtaining an improved $\chi^2_{\nu}$ (2.38 vs. 2.85; see Table~\ref{tab:best_fit}), our fiducial fit does not require an unusually steep dust attenuation law.
The implication of the dust attenuation is further discussed in Section~\ref{sec:discussion}.
Finally, we note that the best-fit nebular emission model, which includes nebular continuum as well as emission lines, is best described by a covering fraction of $C_f=0.24\pm0.01$.
The best-fit value of $C_f$ depends on the choice of $\lambda _{\rm Edd}$ and $U$, both of which impact the relative strength of the nebular emission with respect to the strength of the continuum.
For $\lambda _{\rm Edd}=0.1-10$ and $U=10^{-1.5}$, the best-fit covering fraction is $C_f=0.22-0.46$; for $\lambda _{\rm Edd}=0.1$ and $U=10^{-2}-10^{-1}$, the best-fit covering fraction is $C_f=0.23-0.34$ (see Appendix~\ref{appendix:other_params}).
For AGN at lower redshift, observations imply a wide range of $C_f$ from 0.05 to 0.5 and might depend on the luminosities and Eddington ratios \citep[e.g.,][]{ferland2020}.
The best-fit $C_f$ for \target\ is still within the range found in the local Universe.
We note that the models of \citet{im24} assume $C_f=1$, which leads to a stronger reduction in the strength of the Balmer break at high densities due to the contribution from a Balmer jump in the nebula.
We further discuss the implication of the covering fraction in Section~\ref{sec:discussion}.

\subsubsection{Origin of the rest-frame UV emission}
\label{subsec:uv_fit}

Thus far, we have limited the spectral range of the fit away from the proximity of the Ly$\alpha$ damping wing.
As already noted by \citet{mayilun_lrd_2024}, the PRISM spectrum of \target\ might have damped Ly$\alpha$ (DLA) absorption, which can be seen in the top panel of Figure~\ref{fig:spec_qso1}.
If this feature close to Ly$\alpha$ emission is indeed a DLA, it might help to constrain the physical origin of the UV continuum.

In previous subsections, we have fitted the UV continuum of \target\ with a featureless power law without assuming its physical nature.
Previous studies on LRDs have suggested their blue UV continua could come from less attenuated AGN continua or stellar continua \citep[e.g.,][]{greene2024,li_lrdatt_2024,mayilun_lrd_2024,Volonteri_2024}.
Given the potential presence of the DLA in \target, we aim to test the following two scenarios for the UV continuum.
First, the UV continuum comes from the AGN continuum emission attenuated by the same dense gas as the optical continuum, but without being strongly attenuated by a foreground dust screen to preserve the blue UV slope.
Second, the UV continuum comes from either the AGN continuum emission or a stellar continuum with the DLA absorber located outside the BLR and without being strongly attenuated by a foreground dust screen.
The AGN scenarios above would require either that the dusty absorber has partial covering towards the UV continuum \citep[e.g.,][]{Finn2014} or that the AGN continuum is scattered from a sightline that does not intercept the dusty medium.

% the geometry of the BLR might require some fine tuning as one needs to have the dust screen only affecting part of the continuum emission - this may happen be
% \footnote{This might be achievable if there is a distribution of gas column densities and only clouds with large column densities reach the dust sublimation radius. Or, the blue UV continuum might be scattered light from the AGN. However, detailed modeling of such geometry is beyond the scope of the current work.}.

In Figure~\ref{fig:fit_uvcol}, we compare two fits with different UV continuum models and extend the fitted spectral range from $1600-6400$ \AA\ to $1250-6400$ \AA.
In the left panel, the UV continuum is assumed to be the AGN continuum emission attenuated by a slab of gas with $N_{\rm H}=10^{24}~{\rm cm^{-2}}$ (i.e., the same model as the optical continuum).
{The corresponding neutral hydrogen column density is $N_{\rm HI}=9\times10^{23}~{\rm cm^{-2}}$.}
{We also allowed the UV continuum to have additional dust attenuation (independent of the best-fit magnitude for the optical attenuation) during the fit, which is characterized by a free parameter $A_{\rm V}$(DLA) and an extinction curve with the same shape as adopted for that of the optical continuum.
Still, the best-fit model yields $A_{\rm V}{\rm (DLA)}=0$.
}
It is clear that such a model cannot provide a proper fit to the spectrum of \target.
This model not only fails at the location of the Balmer break by pushing up the break too much, but also fails in the whole UV regime due to the significantly more pronounced DLA feature compared to that observed in \target.
In the right panel, we reduce the gas column density that attenuates the AGN emission to $N_{\rm H}=10^{22}~{\rm cm^{-2}}$ {(with a neutral hydrogen column density of $N_{\rm HI}=2.7\times10^{21}~{\rm cm^{-2}}$)} for the UV continuum but keep other model parameters the same.
In this case, the overall fit is significantly improved and the DLA feature of the model roughly fits that in the observed spectrum.
{The best-fit V-band attenuation is $A_{\rm V}{\rm (DLA)}\approx 0.13$ mag, which is significantly lower than that of the optical attenuation.}
An immediate question is whether the DLA absorber is close to the BLR under the AGN scenario.
As shown by \citet{fabian_2008}, considering the effect of radiation pressure from the AGN emission on the dusty gas in the nuclear region, the gas is only long-lived when $N_{\rm H}>5\times 10^{23}\lambda_{\rm Edd}~{\rm cm^{-2}}$.
If the DLA absorber in \target\ does originate in the nuclear region, the column density adopted in our \cloudy\ model ($N_{\rm H}=10^{22}~{\rm cm^{-2}}$) is close to the critical value when the gas becomes dynamically unstable at $\lambda _{\rm Edd}\sim 0.1$.
We note that if we reduce the gas column density of the optical continuum model to the same value of $N_{\rm H}=10^{22}~{\rm cm^{-2}}$, the Balmer break in the model would become too shallow to fit the observed Balmer break.

Although here we used the AGN continuum to test the fit of the UV spectrum, the same should be applicable to the stellar continuum as the physical mechanism for creating the DLA feature is most sensitive to the gas column density.
Also, an advantage of a UV stellar continuum (while the optical continuum being still AGN-dominated) is that it can alleviate the problem where the UV and optical continua need to have different column densities and dust attenuation, if one considers the stellar populations providing the UV light originate from a much larger physical scale \citep[see e.g.,][]{volonteri_2017,Volonteri_2024}.
This simple practice shows that the UV continuum needs to be attenuated not only by a significantly less amount of dust but also by a significantly less amount of gas, if there is indeed a DLA absorber in \target.
\redtxt{If the UV stellar continuum has a stellar origin, one can estimate the stellar mass associated with the UV. This has been done by \citet{juodzbalis_qso1_2025} following the $M_{\rm UV}$\,-\,$M_\star$ relation derived by \citet{Simmonds_jadesphoto_2024} using photometric observations of galaxies at $3\leq z_{\rm phot} \leq 9$ in JADES. The UV-based stellar mass estimate is $M_\star \approx 10^6~M_\odot$, compatible with the dynamical mass upper limit of $10^{7.3-8.6}~M_\odot$ (see Section \ref{sec:mdyn}).}

%However, we caution that this analysis depends on the spectrum we adopted.
%As shown in Appendix~\ref{appendix:other_reductions}, for the PRISM spectrum adopted by \citet{furtak2023}, the AGN+DLA model cannot provide an equally good fit compared to the other modeling approach we introduce below, although this might be related to the inclusion of image B that has a background subtraction issue in the combined spectrum.
%For the PRISM spectrum reduced from the observations by \blackthunder\ as well as the PRISM spectrum extracted from image A from UNCOVER DR4, the AGN+DLA model provides a sensible fit and we reach the same conclusion as that based on the fiducial reduction.

One might wonder how the presence of the Ly$\alpha$ emission \redtxt{(see Table~\ref{tab:measurements} for the measurements)} can be explained if there is DLA at the same time.
One possibility is that the Ly$\alpha$ emission comes from the more diffuse gas on a larger scale (e.g., ionized or reflecting gas in the Circum-Galactic Medium, CGM), whereas the DLA is produced in gas blocking the UV light emitters on a smaller scale along the LOS.
Such a geometry has been proposed for several sources with strong Ly$\alpha$ emission and DLA-like feature \citep[e.g.,][]{hu_lya_2023,tacchella2024,wu_lya_2024,Witstok2025}. The BlackTHUNDER PRISM IFU data do show that the Ly$\alpha$ emission is offset (by $0.\!\!^{\prime\prime}05=270$ pc \redtxt{in the image plane and roughly 160 pc in the source plane; \citealp{furtak2024}}) relative to the optical continuum, possibly supporting this scenario; however, the analysis of the Ly$\alpha$ emission goes beyond the scope of this work and will be discussed in a separate paper.
An alternative explanation is that the UV continuum is actually dominated by a nebular continuum, where the DLA-like feature is actually the signature of a two-photon continuum \citep{cameron_gs9422_2024}.
Indeed, given the current S/N of the spectrum, we cannot rule out the case where the flux density of the UV continuum peaks at 1430 \AA\ as the two-photon continuum \citep{gaskell1980}.
This nebular continuum cannot originate in the BLR or its proximity as the two-photon continuum is significantly suppressed at $n_{\rm H}\gtrsim 10^4~{\rm cm^{-3}}$ \citep{bottorff2006}.
It can be created either in the diffuse gas close to/outside the BLR or in the gas surrounding very hot and massive stellar populations \citep[e.g.,][]{grandi1982,schaerer_pop3_2002,raiter_2010,Zackrisson_2011,Inoue_2011}.

The existence of the two-photon continuum must be accompanied by the bound-free continuum of hydrogen as well as Balmer emission lines.
We can thus perform another fit by including the above components to see whether the nebular continuum scenario works.
We set up a simple configuration by using the nebular continuum computed with \textsc{PyNeb} \citep{luridiana2015}.
We used the \textsc{get\_continuum} function and included two-photon emission, bound-free emission, and free-free emission from hydrogen.
The density is set to $n_{\rm H}=100~{\rm cm^{-3}}$, below the low-density limit of the two-photon emission.
The temperature and the normalization of the nebular continuum is set to vary freely during the fit.
Specifically, we allow the temperature to vary within $5\times 10^3~{\rm K}<T_{\rm e}<3\times 10^{4}~{\rm K}$.
In principle, the normalization of the nebular continuum should be fixed to the fluxes of Balmer lines emitted by the same cloud \citep{peimbert1967}.
In the case of \target, a natural assumption might be that the nebular continuum originates in the gas emitting the narrow \hb.
Regardless, we set the normalization as a free parameter to see whether the nebular continuum can provide a sensible fit without this physical constraint.
Also, we included additional dust attenuation characterized by $A_{\rm V;neb.~con}$ as a free parameter.
Figure~\ref{fig:fit_2gamma} shows the best fit model for \target\ with a nebular dominated UV component and an optical component dominated by attenuated AGN emission.
This fit results in the same $\chi^2_{\nu}$ compared to the DLA fit with $N_{\rm H}=10^{22}~{\rm cm^{-2}}$.
{However, the fit is worse in the FUV region as reflected by a significantly larger $\chi^{2}_\nu$ at $1250-2000$ \AA\
mainly due to the underestimation of the flux densities.} Clearly, the nebular continuum is not steep enough to describe the UV continuum of \target\ especially at shorter wavelengths, even if we included an extra freedom (i.e., the normalization of the nebular continuum) during the fit.
The fit also prefers the temperature of the diffuse gas to be $T_{\rm e} \approx 30,000$ K, which would be among the highest temperatures currently found by \jwst\ at $z\gtrsim 4$ \citep{laseter2023}.
In Appendix~\ref{appendix:other_reductions} we show the results for PRISM spectra with other reductions and we found a general agreement between most reductions.

%Again, we note that the above results are subject to data reduction in the rest-frame FUV regime.
%In fact, as we show in the Appendix~\ref{appendix:other_reductions}, the nebular continuum model fits the UV part better compared to the DLA model for the reduced spectrum adopted by \citet{furtak2023}.
%For the spectra from \blackthunder\ and UNCOVER DR4, the DLA fit is still better than the nebular continuum fit. Once again, the difference between the two results may be a consequence of including the spectrum of image B in \citet{furtak2023}'s spectrum, which is affected by background subtraction issues.

To conclude, while the nature of the UV continuum of \target\ remains unclear, it is most likely explained by either stellar populations or AGN continuum emission with little dust attenuation and a DLA produced by gas with $N_{\rm H}\sim 10^{22}~{\rm cm^{-2}}$.
In the case of stellar light, the UV component is more spatially extended compared to the optical component.
In the case of AGN light, the UV component might be as compact as the optical component but a specific geometry is needed to explain the differential attenuation.
Still, without better observational constraints in the FUV, we cannot fully rule out the nebular dominated scenario, which is further discussed in Appendix~\ref{appendix:other_reductions}.
%as the fit of the DLA-like feature apparently depends on the data reduction.

Thus far, we have explored models where the optical spectrum of \target\ is dominated by AGN light.
An immediate question related to this model is whether there is any sign of variability associated with the AGN emission as typically found at lower redshift \citep[e.g.,][]{burke2021}, which we discuss next.

% \redtxt{UV must be a separate component. two explanations: less attenuated AGN continuum or SF continuum with a high escape fraction (Francesco's test of UV-\oiii?)}

\begin{figure}
    \centering
    \includegraphics[width=\columnwidth]{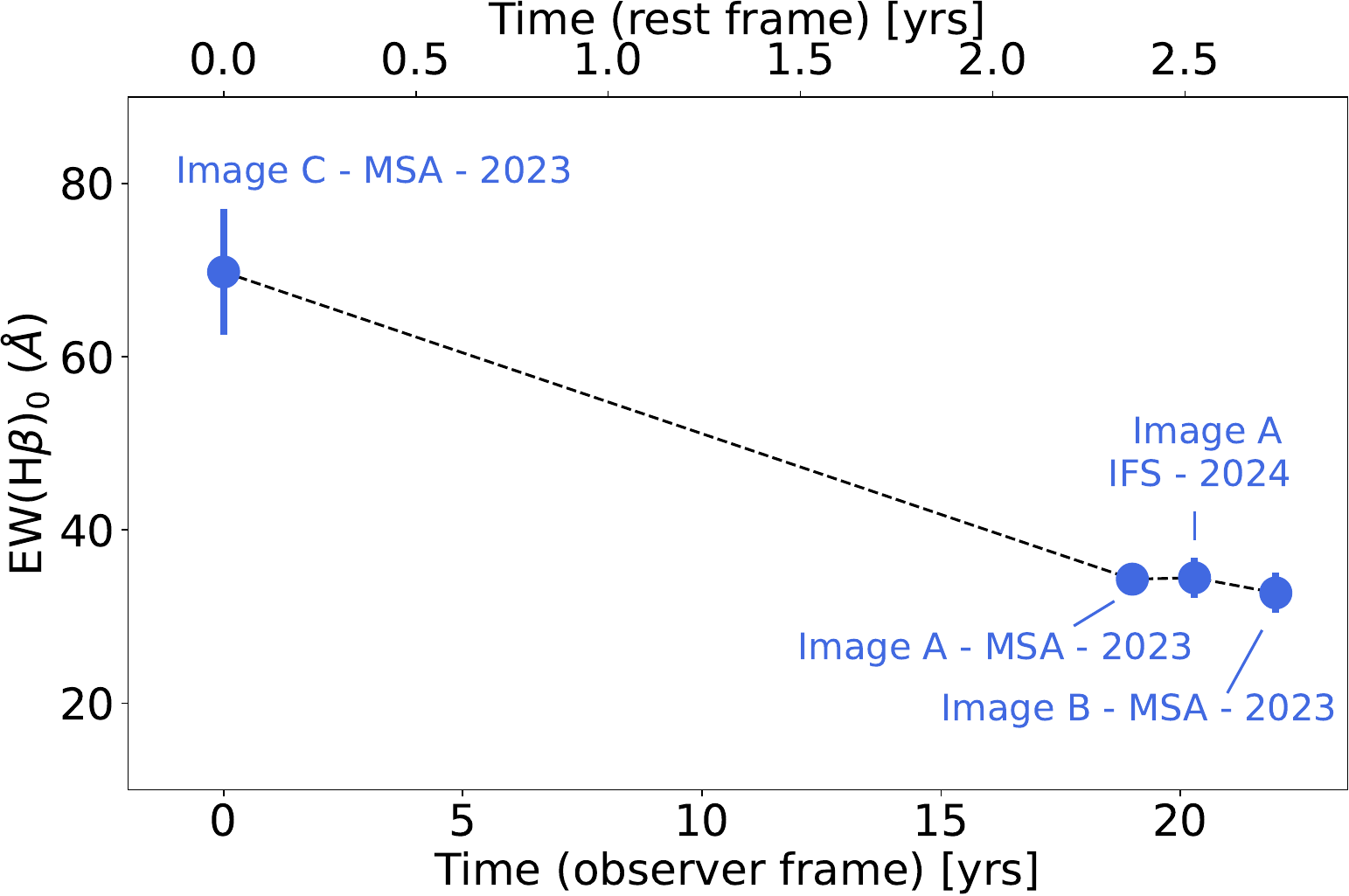}
    \caption{
    Variation (or lack thereof) of the rest-frame equivalent width of H$\beta$ measured in the four spectra of Abell2744-QSO1: the three MSA spectra of the three images obtained in 2023 and BlackTHUNDER IFU spectrum obtained in 2024. The spectra are plotted as a function of arrival time since the observation of image C, taking into account the lensing time delay. While images A and B do not show evidence for variation of the EW, image C clearly shows an EW that is about two times higher than in the other images, at high significance.
    }
    \label{fig:EWs}
\end{figure}

\begin{figure}
    \centering
    \includegraphics[width=\columnwidth]{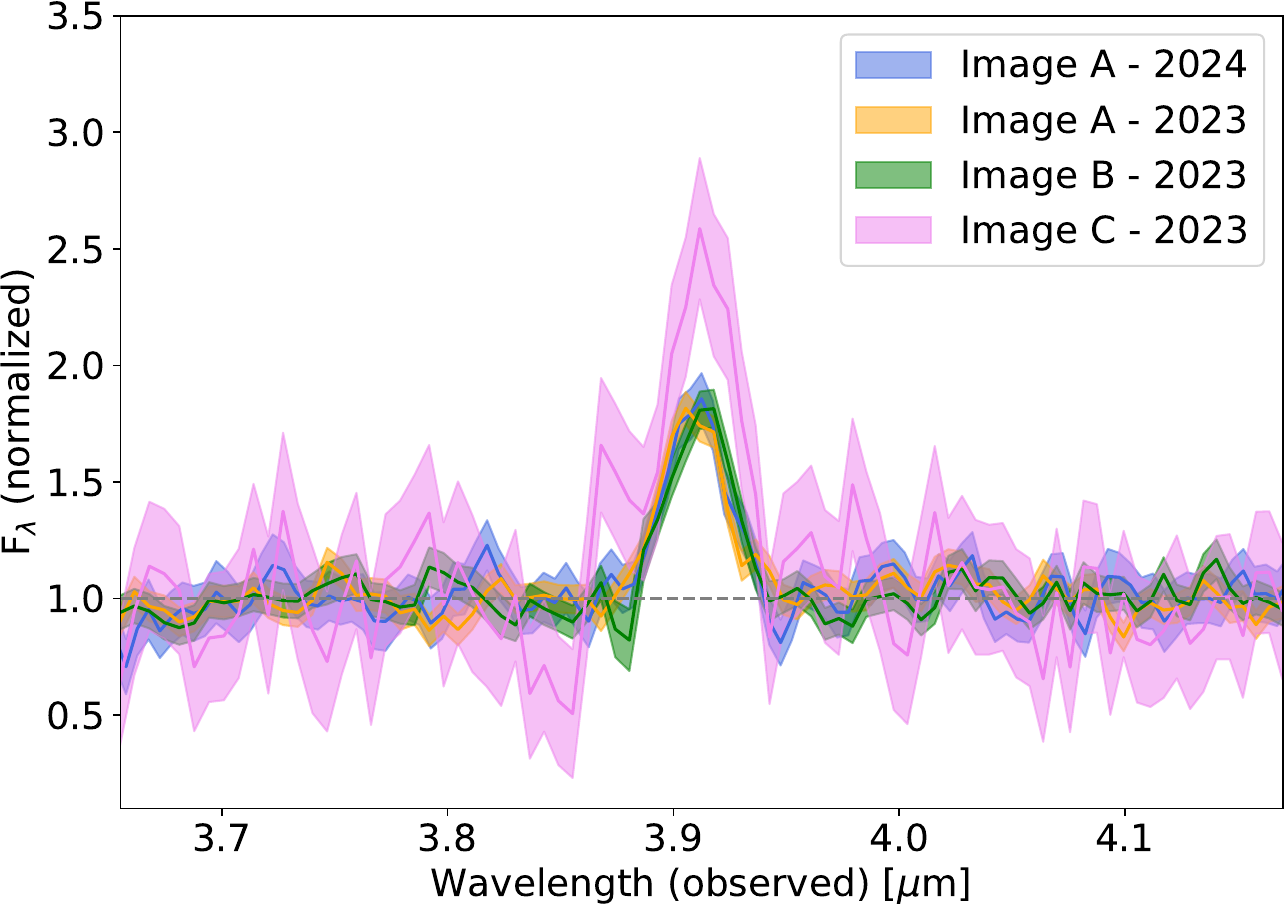}
    \includegraphics[width=\columnwidth]{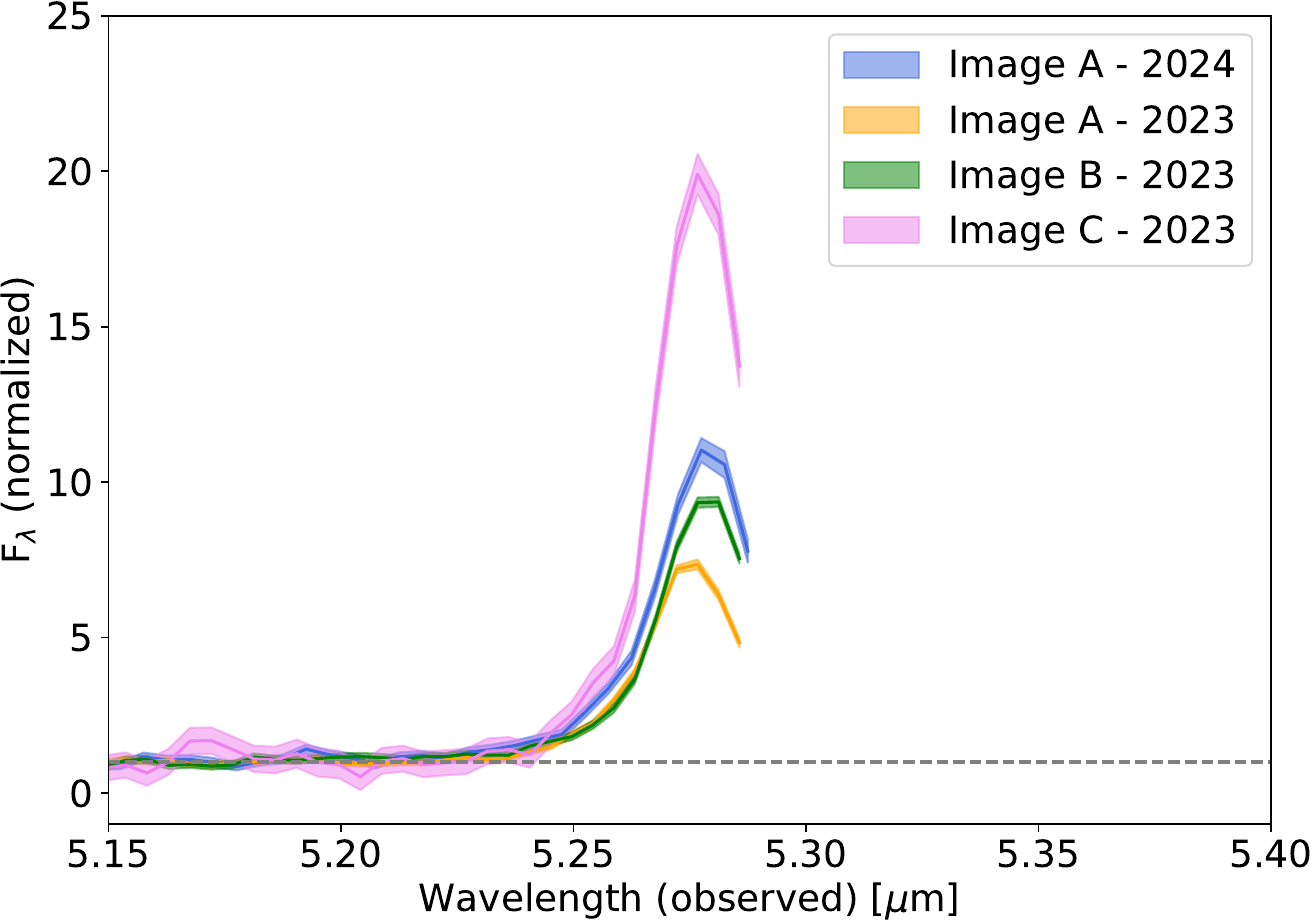}
    \caption{
    Top: Comparison of the four spectra of Abell2744-QSO1 around H$\beta$, normalized to the continuum level. Although the spectrum of image C is noisier than the others, its H$\beta$ is clearly more prominent than in the other images (i.e. has a higher equivalent width). 
    Bottom: Same comparison for H$\alpha$. The H$\alpha$ line is at the edge of the available wavelength range, hence its red wing is partly chopped. Additionally, the calibrations at such long wavelengths are more uncertain, hence variations of the order of 15-30\% may not be real. However, the spectrum of image C shows that H$\alpha$ is clearly much stronger (relative to the continuum) when compared with the other two images by about a factor of two, consistent with what observed for H$\beta$.
    }   \label{fig:Comp_norm_spec}
\end{figure}

\section{Variability and reverberation mapping}
\label{sec:variability}

The triply imaged system, \target, is in principle an excellent tool to explore variability of AGN at high redshift. Indeed, the arrival times of the three images are subject to delays associated with the image-lens configuration. Specifically, as pointed out in Appendix~\ref{sec:lens_model}, the arrival of image C is followed by image A 18--19 years later ($=2.2-2.4$ years rest-frame),  which is then followed by image B another $\sim 2.2-3$ years later (3.2--4.5 months rest-frame), depending on the lens model.
Unfortunately, as mentioned in Appendix~\ref{sec:lens_model},
the uncertainties in the lensing magnifications of the three images do not allow exploring variability by simply comparing the fluxes of the three images.

However, the BlackTHUNDER IFU spectrum provides an additional, new epoch of image A, taken one year after the MSA spectrum. 
In Appendix~\ref{appendix:var}, we show that the spectra taken at the two epochs (July 2023 and November 2024) are consistent with each other within the uncertainties,
implying that the source has probably been stable within the $\sim 2$ months in the rest frame covered by the two observations (although we cannot exclude variability between the two epochs).

While comparing the absolute fluxes is not really possible between the different images, because of the magnification factor uncertainties, it is possible to compare the equivalent widths of the lines, as these are calculated relative to the continuum at similar wavelengths and thus are insensitive to lensing magnification.
Figure~\ref{fig:EWs} shows the variation of the equivalent width (rest frame) of H$\beta$ as a function of time relative to the arrival time of image C. While the EWs of image A in the two epochs (MSA and IFU) and image B are fully consistent with each other, the EW of image C is clearly higher, by about a factor of two and with high significance relative to the other two images.

We note that when undertaking this kind of variability analysis it is important to use the full-shutter, pipeline extracted 1D spectra, as these fully and rigorously take into account path losses and diffraction losses. Using 1D spectra extracted from the 2D spectra with any kind of custom aperture is deprecated in this case, as the extraction from the 2D spectra loses information on the path- and diffraction-losses. In Appendix~\ref{appendix:var}, we illustrate this issue by showing that the analysis performed on spectra extracted with the latter method provides results that are quantitatively different because of such issues, although the qualitative trends remain.

The higher equivalent width of H$\beta$ in image C can also be seen visually in the top panel of Figure~\ref{fig:Comp_norm_spec}, where the four PRISM spectra (three MSA and one IFU) are compared with each other after being normalized to the continuum level in the vicinity of H$\beta$. Although the spectrum of image C is more noisy, the relative flux of H$\beta$ is clearly higher, even taking into account the noise.
The same comparison for H$\alpha$ is shown in the bottom panel of Figure~\ref{fig:Comp_norm_spec}. Unfortunately, H$\alpha$ is at the edge of the wavelength range, and the red wing of its profile is chopped. Additionally, at such long wavelengths the calibrations are more uncertain; therefore, the 10\%-30\% variation seen in images A and B should be considered with caution and may not be real. However, the H$\alpha$ line of image C is much stronger (relative to the continuum) than observed in images A and B, by about a factor of two (i.e., at the same level seen in H$\beta$).

Clearly, the AGN in Abell2744-QSO1 must have undergone a short phase of enhanced accretion before the spectroscopic observation of image C or, equivalently, the accretion rate (hence continuum luminosity) must have dropped during the spectroscopic observation of image C. Then, during the spectroscopic observation of image C we are observing the BLR reverberation traced by the H$\beta$, which has not yet adjusted its flux relative to the continuum variation, due to the extension of the BLR.

\begin{figure}
    \centering
\includegraphics[width=\columnwidth]{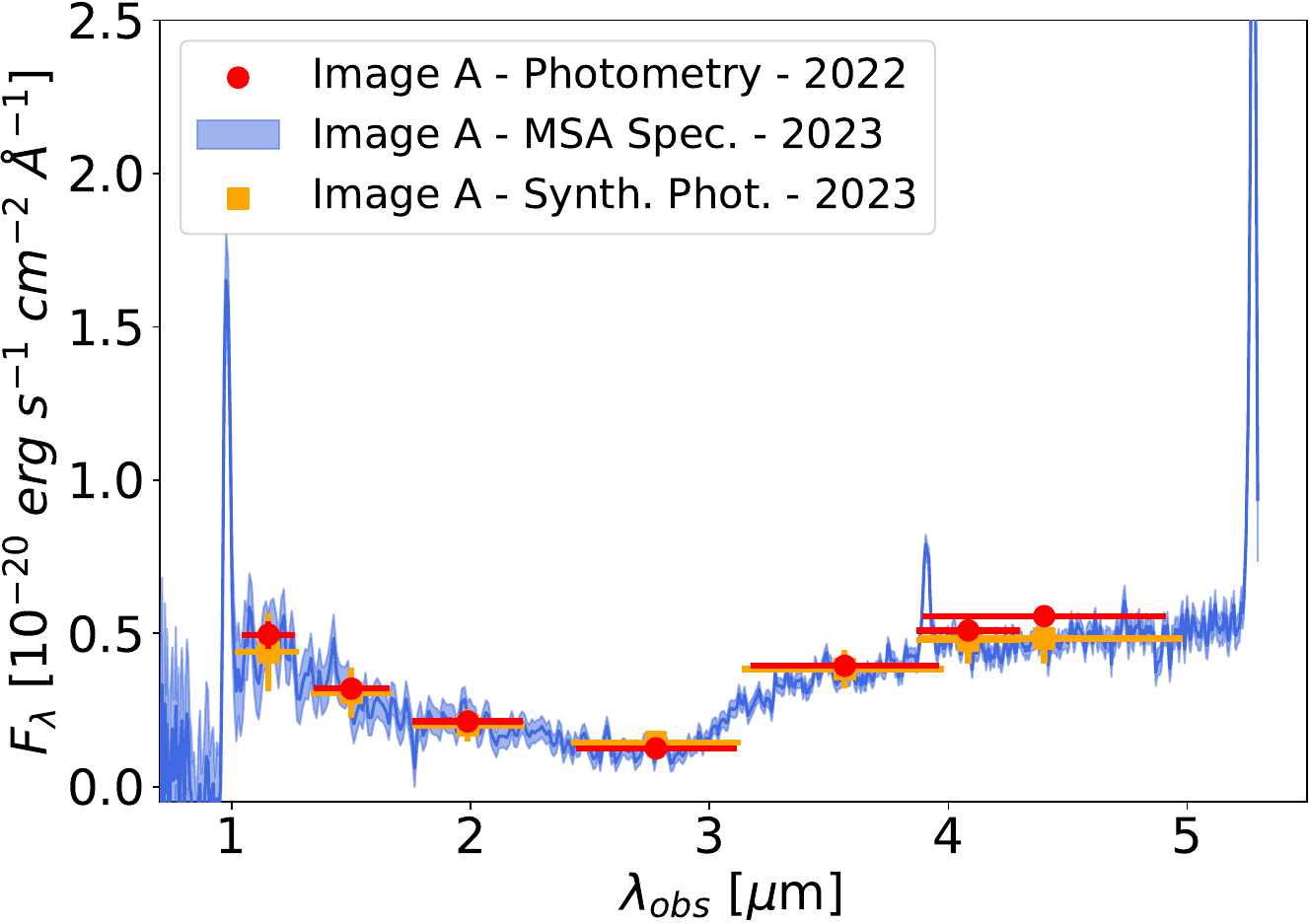}
\includegraphics[width=\columnwidth]
    {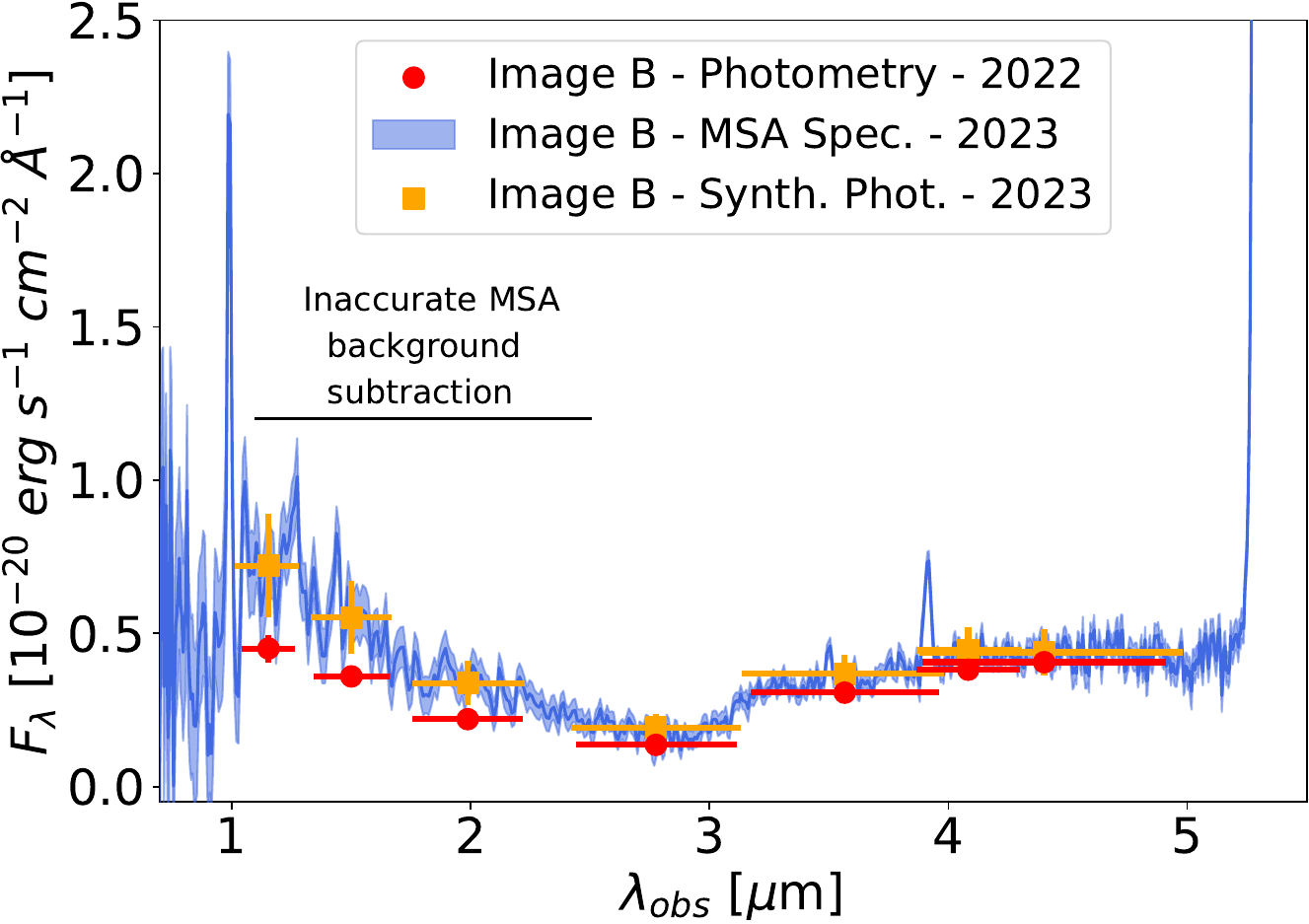}
\includegraphics[width=\columnwidth]   
    {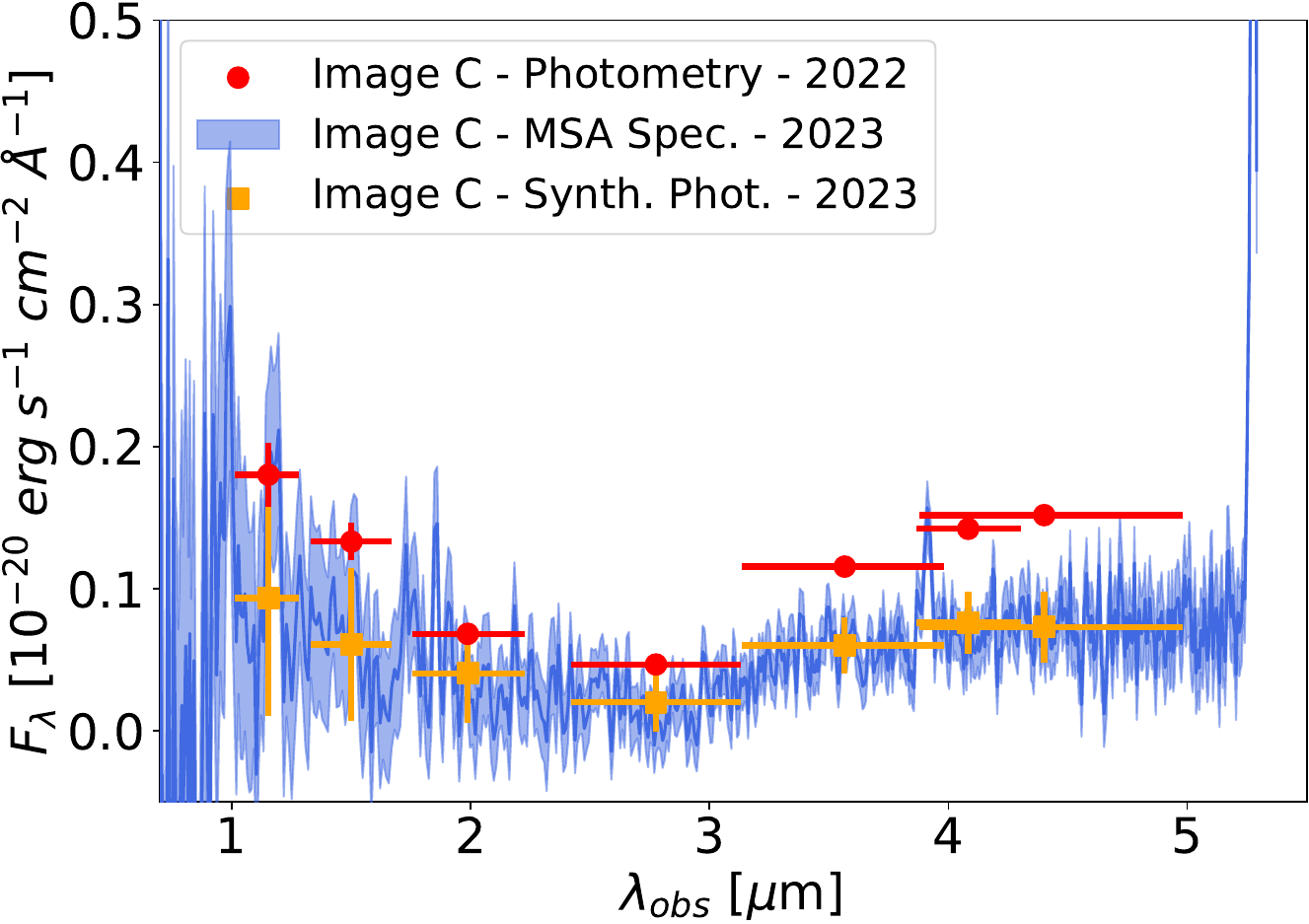}
    \caption{
    Comparison of the MSA spectra, and associated synthetic photometry (red points), taken in July-August 2023, with the photometry from the images take in November 2022 (orange symbols), for image A (top), B (middle) and C (bottom). Clearly the continuum flux in the optical rest-frame of image C was higher in 2022, by about a factor of two, relative to the spectrum taken in 2023, showing a flux drop, at least in the optical, by a factor of about two between 2022 and 2023. On the contrary, 2022 photometry and 2023 spectroscopy of image A are in agreement within errors, and the same is for the optical part of image B. The UV part of image B shows a variation, but this is likely due to the background subtraction issues in image B due to the blue foreground galaxy, $0\farcs9$ from image B 
    }
    \label{fig:spec_photo_var}
\end{figure}

\begin{figure}
    \centering
    \includegraphics[width=\columnwidth]{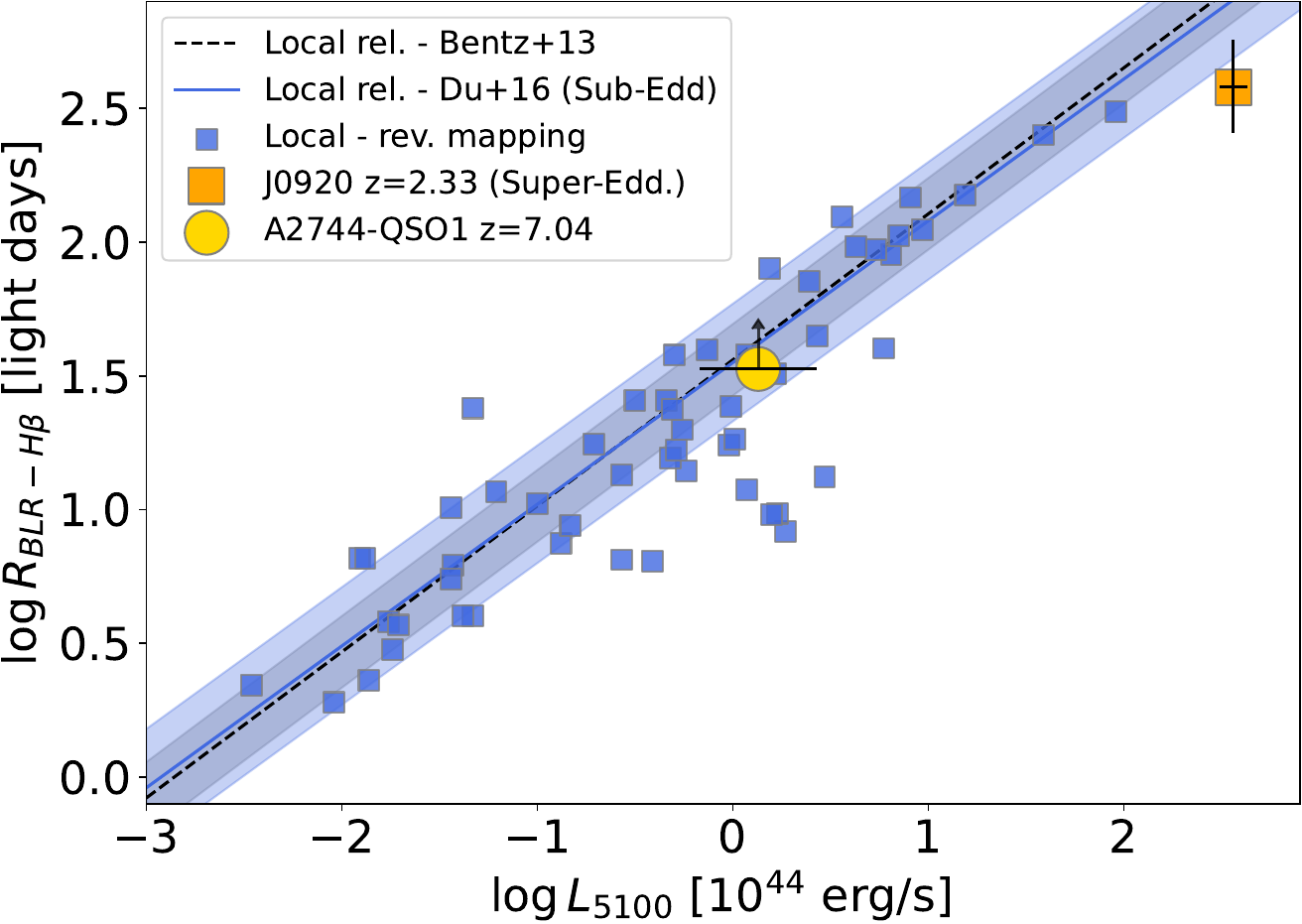}
    \caption{Relation between radius of the BLR and optical continuum luminosity (L$_{5100}$) from H$\beta$ reverberation mapping in local AGN (blue points), where the blue solid line and black dashed lines show the best fit of the local relations from 
    \citet{Du2016_refrel} (for their sub-Eddington sample) and \citet{Bentz2013_rev}, 
    and the shaded region show the scatter of the relations. The golden circle shows the lower limit on the radius of the BLR inferred for Abell2744-QSO1 based on the variability of image C. 
    The orange square shows the interferometric measurements at \redtxt{$z=2.33$} obtained by \citet{Abuter2024} for a QSO accreting at a highly super-Eddington rate.
}
    \label{fig:rev_map}
\end{figure}

It is possible, in principle, to test this by looking at the flux difference in image C between the photometry in 2022 and its MSA spectrum in 2023. One should always be cautious about comparing photometry and slit spectroscopic measurements because of slit losses plaguing the latter. However, given that Abell2744-QSO1 is point like, and that its location in the shutter is known accurately, the slit losses provided by the pipeline are quite accurate (better than 15\%). The comparison of photometry and spectroscopy of image C is shown in bottom panel of Figure~\ref{fig:spec_photo_var}, where the red circles show the photometry from 2022 and the orange squares show the synthetic photometry obtained from the spectrum (where the errorbars include both the Poissonian error and calibration uncertainties).
% While in the UV rest frame the spectrum is too noisy to draw conclusions,
In the optical rest frame the photometry from 2022 is clearly higher than the spectrum in 2023 by a factor of about two. This is exactly the flux difference expected by the higher EW(H$\beta$) observed in 2023 for image C. In other words, the MSA spectrum of image C shows either the increase of EW(H$\beta$) is due to a drop in continuum ionizing emission between 2022 and 2023, or the H$\beta$ reverberation of an increase in continuum flux in 2022.

The UV rest-frame spectrum of image C is too noisy to draw any conclusions, and it is consistent with no variability within 1--2$\sigma$. However, should variability be confirmed also in the UV, this would favour an AGN contribution also in this spectral region, possibly within the partial covering scenario discussed in Section~\ref{subsec:uv_fit}.
\redtxt{
If the UV is truly varying, following the variability pattern of SDSS quasars measured by \citet{MacLeod_2010}, one expects the RMS flux variation in the UV is stronger than that in the optical.
For image C, given its optical variation of 50\%, the expected UV variation should be roughly 67\%, compatible with the measured UV variation of $50\pm 50$\%.
However, we cannot draw any further conclusions due to the large uncertainty and the fact that the RMS variation is not well constrained by two epochs.
}

Unfortunately, there is no photometry available at the epoch of the spectroscopic observation. As discussed in Section~\ref{sec:photometry}, images in F356W and F444W are available about 8 months before and 4 months after the spectroscopic observation. The photometry does not show evidence for variability at those two epochs, at a level higher than 3$\sigma$, although for image C variability of up to 25\% would be consistent with the photometric observations and associated uncertainties.

We notice that, in contrast to image C, the comparison of photometry and MSA spectroscopy for images A and B does not show evidence for significant flux variation among the two epochs (a variation in the UV for image B is likely due to the problematic background subtraction discussed above and associated with a bright foreground galaxy). This is shown in the top and middle panels of Figure~\ref{fig:spec_photo_var}. This means that about 17 years after the arrival of image C ($\sim$2 years rest frame) the source was relatively stable in flux for a few rest-frame months, as also evident from the stable EW(H$\beta$).
\redtxt{We note that if the UV of image B is truly varying, taking the small optical variation of $\sim8$\% at $\lambda _{\rm obs}\approx 4.4~{\rm \mu m}$, one expects a UV variation of 13\% at $\lambda _{\rm obs}\approx 1.5~{\rm \mu m}$ following the result of \citet{MacLeod_2010}.
While this is marginally compatible with the measured UV variation of $35\pm 16$\%, we again caution on the large uncertainty and the two-epoch measurement.
}

Unfortunately, with only two data points for image C it is very difficult to attempt a reverberation mapping. However, the finding that the EW(H$\beta$) is two times higher than its regime value at later epochs, and as expected by the higher continuum flux in 2022, indicates that the BLR has not yet adjusted its flux to the flux drop between November 2022 and July 2023. This essentially means that the radius of the BLR must be larger than about 9-light-months/$(1+z)$, which is about 45 light-days. This constraint is shown in Figure~\ref{fig:rev_map} and compared with the scaling relation between the BLR radius (for H$\beta$) and $L_{5100}$, as inferred from reverberation mapping in local/low-$z$ galaxies, specifically data from Table 7 of \cite{Du2015_revdata} and the relations for sub-Eddington AGN taken from \cite{Du2016_refrel} (blue solid line and blue shaded region), as well as the relation (from `Clean2' parameters) of \cite{Bentz2013_rev}. We also show the interferometric image obtained by \cite{Abuter2024} at \redtxt{$z=2.33$} (orange square) for a super-Eddington quasar. The lower limit on the size of the BLR inferred for \target is consistent with the scatter of the local relations. Additional spectroscopy of image C in the future will be able to verify whether the BLR in this early AGN at $z=7.04$ behaves as the BLRs of local AGN.

Finally, the simple fact that the continuum of image C has been varying by a factor of two between 2022 and 2024 indicates that at least the optical continuum cannot be dominated by stellar light.
This is consistent with our analyses in previous sections, based both on the dynamical mass analysis and on the modeling of the Balmer break and SED.

\section{Discussion}
\label{sec:discussion}

In previous sections, we present our physical model for \target\ where the optical spectrum is dominated by AGN light based on observational data.
In this section, we discuss whether the physical model we adopted for \target\ can in principle be generalized to other \jwst-selected LRDs that host AGN.
In addition, we discuss the implications of our fitting results in the context of the peculiar observational features of \jwst-selected Type 1 AGN.

\subsection{Applicability to other LRD AGN}
\begin{comment}
\begin{itemize}
    \item The shape is controlled by turbulence and dust; thus sharper break is expected at lower turbulence and weaker absorption
    \item The DLA requires more statistics
    \item the breaks weakens at high covering fractions and low/high densities; this might have implications on the fraction of AGN as LRDs
\end{itemize}
\end{comment}

Thus far we have focused our analyses on \target, an LRD and a Type 1 AGN identified by \jwst.
An immediate question is whether the V-shape spectra and Balmer break observed in other LRD AGN can be explained by the same modeling approach.

Based on our analyses shown in Section~\ref{sec:results_BB}, the V-shape can be explained by an unobscured UV component of a stellar or an AGN origin, and an optical component corresponding to an AGN continuum attenuated first by high-density and dustless gas and then by a dust screen (which can potentially be also the outer regions of the high-density gas).
The shape of the Balmer break in such a model is mainly determined by the physical conditions of the dense gas that absorbs the AGN continuum, as well as the amount of external dust attenuation.
It is worth noting that the shape of the Balmer break varies among LRDs.
For example, \citet{labbe_monster_2024} recently reported an LRD at $z=4.47$, which has both H$\alpha$ absorption and a ``sharp'' Balmer break that rises rapidly at the wavelengths longward of the Balmer limit, unlike the gradual break we see in \target.
One might thus wonder whether our modeling approach can explain the diversity of the Balmer breaks in LRDs.
One potential clue to resolve this question comes from the variation in the turbulence.
As shown in Figure~\ref{fig:fit_turb}, at low turbulence, the Balmer break becomes sharper compared to the case of high turbulence.
This is due to the weakening of the Balmer line absorption at low turbulence shown in Figure~\ref{fig:dependency}, which results in less erosion of the break.
Therefore, in the context of our models, the diversity of the Balmer break could be partly related to the diversity of the turbulence in the dense absorbing gas \citep[see e.g.,][]{bottorff2000}.
Additionally, superpositions (and blending) of strong 
% narrow 
emission lines might change the shape of the break.
While in the case of \target\ the emission lines are not very strong, in the case discussed by \citet{labbe_monster_2024}, an abundant set of strong emission lines are present.
Furthermore, we have not fully excluded the contribution from a stellar Balmer break, which would add more complexity to the shape of the break.
We will present our analyses on other LRDs including the one reported in \citet{labbe_monster_2024} in a future paper.

One might wonder what observational evidence is required to verify our models.
One piece of evidence comes from the connection between the Balmer break and the Balmer absorption.
As already shown by \citet{im24} and in Figure~\ref{fig:dependency}, when there is a strong Balmer break, there is likely also strong Balmer absorption produced in the same gas.
Despite the complex dependencies of the break and absorption strengths on various parameters, statistically, stronger Balmer breaks are accompanied by stronger Balmer-line absorption. The fact that at least 20\% of the LRDs and \jwst-discovered AGN show evidence of H$\alpha$ absorption suggests that the observed Balmer breaks likely have, at least partly, a dense-gas absorption component. Still, a detailed fit and assessment is required for each individual LRD/AGN, which is beyond the scope of this paper.
In addition, constraining the shape of the Balmer break in observations will be useful, 
since statistically, narrower and weaker absorption would correspond to sharper breaks.
We do emphasize that there could be large stochasticity within small samples.
Thus, it is vital to accumulate more high spectral resolution observations aiming to find Balmer lines absorption and combine them with PRISM observations.
Another interesting observable is the potential DLA-like features in the rest-frame FUV of LRDs.
Verifications and detailed modeling of these features can help us understand the origin of the UV continuum.
This requires more PRISM observations with high S/N to constrain the shapes of the DLA-like features and compare them with the shapes of the Balmer breaks.

Finally, we would like to comment on the occurrence rate of LRDs given our physical model.
It is known that not all \jwst-selected AGN are LRDs.
As shown by \citet{hainline2024}, about 30\% of the \jwst-selected Type 1 AGN are LRDs.
If the case of \target\ is typical, given a covering fraction of $C_f\approx 0.24$ for BLR clouds (see Section~\ref{subsubsec:decompose}), one expects about 24\% of the time our LOS is blocked by dense BLR gas, which implies an LRD fraction of 24\%.
This is certainly a very rough estimation and does not take into account the variations in the AGN and BLR properties.
Also, as we have shown in Section~\ref{sec:balmer_break} and in Appendix~\ref{appendix:other_params}, the derived value for $C_f$ depends on the assumption of the Eddington ratio and the ionization parameter and can vary in a range of $0.22-0.46$ for \target.
Furthermore, we caution that it remains debated whether all LRDs are hosting AGN since the majority of these sources are color-selected.
Still, it would be interesting to investigate the properties of the absorbing gas over a larger sample and see if they match the statistics.

\subsection{Implications of a non-stellar origin of the Balmer break}
\label{sec:disc-non-st-BB}

Past attempts to fit the Balmer break observed in \target and in other LRDs have faced difficulties with scenarios advanced that are extreme and difficult to reconcile with other findings.
The stellar interpretation of the Balmer break typically leads to high stellar mass. Combined with the large number of LRDs, this results in extremely high stellar-mass density per cosmic volume, possibly in tension with the standard cosmological framework \citep{wangbingjie_lrd_2024,inayoshi_ombh_2024,akins_lrd_2024}.

Additionally, when combined with the very compact sizes of the LRDs (often unresolved), a stellar interpretation of the Balmer break and optical continuum results in extremely high stellar densities of $10^6-10^7~M_\odot~{\rm pc}^{-2}$ \citep{baggen_2024,labbe_monster_2024}, which have never been seen (with such high masses) at lower redshifts or in the local Universe, hence posing the question of how these hyper-dense and massive stellar system would dissolve across the cosmic epochs. The same issue also applies specifically to \target, for which \cite{mayilun_lrd_2024} reported a lower limit on the stellar density of $10^6~M_\odot~{\rm pc}^{-2}$, if one associates the Balmer break and optical continuum with a stellar origin. However,
the same authors recognized that any stellar fit is unsatisfactory for this object, and an unusually steep extinction curve (steeper than both the \citealp{calzetti2000} and SMC extinction curves of \citealp{gordon2003}) is required.

Our work strongly disfavours a stellar origin of the optical continuum and of the Balmer break on multiple grounds. The conservative upper limit on the dynamical mass is at least an order of magnitude below the stellar mass obtained when interpreting the Balmer break and optical continuum in terms of stellar emission, which implies a very conservative upper limit of 10\% on the contribution of the stellar light to the optical continuum and Balmer break. Our dense-gas absorption scenario provides a good fit ($\chi^2_{\nu}=1.12$) to the spectrum and predicts the presence of \hb\ absorption which is roughly consistent with the high-resolution spectrum. The indications of variability provide further support for the non-stellar origin of the optical continuum. 

The scenario of an optical continuum dominated by AGN light and the Balmer break originating from gas absorption naturally avoids all extreme physical scenarios faced when interpreting the Balmer break and optical continuum as stellar, as discussed above. AGN are intrinsically very compact; hence, if they dominate the optical continuum, they very naturally explain the very compact (often unresolved) emission. It has been known for decades that AGN are surrounded by extremely dense gas (the BLR; \citealp{netzer1990}), hence ascribing the Balmer break to dense gas absorption does not require invoking new, exotic scenarios.

Clearly,  AGN dominating the optical continuum of LRDs and dense gas absorption being responsible for the Balmer break, naturally explain most of their properties. In this paper, we have verified this model for the specific case of \target. However, as discussed in the previous section, the same scenario may explain the properties observed also in other LRDs. This would more broadly alleviate the issues faced by previous work when attempting to explain the Balmer break with a stellar origin. An analysis of other LRDs with a methodology similar to that adopted here will be presented in a future work.

\subsection{Insights into the X-ray weakness of \jwst-selected AGN}

\begin{comment}
\begin{itemize}
    \item deep break requires nearly Compton-thick gas; this might explain the X-ray weakness
    \item on the other hand, the AGN might also be intrinsically weak in the hard X-ray due to super-Eddington accretion; evidence from weak optical high-ionization lines
    \item the covering fraction might also be related to the fraction of AGN as LRDs (as mentioned in the previous subsection); in this case the super-Eddington accretion needs to be ubiquitous
\end{itemize}    
\end{comment}

In the local Universe, it is observed that AGN are frequently accompanied by strong hard X-ray emission, which has been proposed to be associated with ``hot coronae'' above accretion disks.
Photons from the accretion disks undergo Compton up-scattering by electrons in the hot coronae, leading to non-thermal X-ray emission.
The non-thermal hard X-ray emission is generally considered as a ubiquitous feature of AGN, whose contributions to the AGN SED can be characterized by a spectral slope from the optical to the X-ray, $\alpha_{\rm ox}=-0.384\times\log(L_{\rm 2~keV}/L_{\rm 2500 ~\AA})$, which has a typical value of $-1.2$ to $-1.4$ for Type 1 AGN and QSOs at $z\lesssim 3$ \citep{zamorani1981,lusso_2017}.

However, one of the most puzzling observational results obtained by combining data from \jwst\ and from X-ray missions, such as \textit{Chandra}, is the lack of hard X-ray emission in most \jwst-selected AGN \citep{yue_agn_2024,Maiolino2024_Xrays,ananna_agnxray_2024,kokubo_harikane2024,lambrides2024,lyu_miriagn_2024}.
For example, as shown by \citet{Maiolino2024_Xrays}, compared to quasars selected in the UV/optical at low redshift, \jwst-selected AGN at $z\gtrsim 4$ with similar bolometric luminosities, no matter whether they are Type 1 or Type 2, are 1-2 orders of magnitude weaker in their X-ray luminosities.
Similarly, \citet{lambrides2024} found that the broad-line AGN selected by \jwst\ must have their $\alpha _{\rm ox}$ offset from the lower redshift broad-line AGN at similar UV luminosities by more than 0.6 dex.

The explanations for the X-ray weakness of the \jwst-selected AGN (apart from the non-AGN explanations) can be broadly divided into external processes and internal processes.
For external processes, \citet{Maiolino2024_Xrays} suggest Compton-thick obscuration by dust-free clouds within BLRs leads to the X-ray weakness.
This picture is further investigated by \citet{juodzbalis_agnabs_2024} using the observation of an extremely X-ray weak AGN at $z=2.26$, suggesting that the Compton-thick clouds along the LOS also lead to strong Balmer and \hei\ absorption detected in the \jwst/NIRSpec grating spectra of this source.
Furthermore, \citet{trefoloni_feii_2024} suggest the metal-poor nature of the BLRs of the \jwst-selected AGN might lead to more compact BLRs compared to metal-rich AGN at low redshift as well as metal-rich QSOs at high redshift, increasing the chance of obscuration.
For internal processes, various authors have suggested super-Eddington accretion as a mechanism to reduce hard X-ray luminosities \citep{Dai_spedd_2018,inayoshi_spedd_2024,king_2024,lambrides2024,madau_spedd_2024,pacucci_spedd_2024}.
For example, \citet{lambrides2024} argue the super-Eddington accretion leads to much steeper optical-to-X-ray slopes and thus reduces the X-ray luminosities.
\citet{lambrides2024} also show that the high-ionization emission lines are suppressed under super-Eddington accretion due to photon trapping.
Meanwhile, super-Eddington accretion can also explain the lack of UV/optical variability in \jwst-selected AGN due to photon trapping \citep{inayoshi_spedd_2024}.

For \target\ \citet{furtak2024} report that the hard X-ray emission constrained by \textit{Chandra} in the rest-frame 40 keV only has a $3\sigma$ upper limit of $L_{40~\rm keV}<3\times 10^{43}~{\rm erg~s^{-1}}$, which is at least 10 times weaker than the value expected from UV/optical luminosities given a typical AGN SED.
Given the confirmation of the X-ray weakness, we discuss below whether any of the physical conditions in \target\ we derive would imply reduction in the X-ray emission.

First, we note that the column density for the dust-less gas in our fiducial model is $N_{\rm H}=10^{24}~{\rm cm^{-2}}$, and the Compton-thick column density is $N_{\rm H; thick}\approx1.5\times10^{24}~{\rm cm^{-2}}$.
Thus, our fiducial model of a slab of gas along the LOS is capable of reducing the hard X-ray emission by a factor of 2 (or 0.3 dex) at 40 keV, whereas a cloud that is twice as thick as the cloud in our fiducial model can reduce the hard X-ray emission by roughly a factor of 4 (0.6 dex).
We emphasize that the above estimations are only tentative as we are not constraining the total $N_{\rm H}$ accurately in our analyses, and the photoelectric absorption of X-ray is not quantitatively calculated \citep[see][]{hep_longair}. Additionally, the column density estimated from our models is only sensitive to the fraction of the gas responsible for the Balmer absorption and Balmer break. It is possible that there is additional gas (on the opposite side of the illuminated side of the clouds; see discussion in \citealp{juodzbalis_agnabs_2024}) that is colder and, while not contributing to the Balmer absorption, its column contributes to the X-ray absorption. In other words, the column density estimated in our model is only a lower limit of the column density absorbing the X-rays.

An immediate question is how often one expects to see X-ray weak AGN if it is due to Compton-thick absorption.
In principle, the occurrence rate of the X-ray weak AGN should depend on the covering fraction of Compton-thick clouds.
If we took the covering fraction from our best-fit fiducial model, which is $C_f\approx 0.24$, the occurrence rate would be 24\%, which is significantly lower than the nearly 90\% of the X-ray weakness in \jwst-selected Type 1 AGN (\citealp{Maiolino2024_Xrays}; Ji et al. in prep) but close to the occurrence rate of LRDs among the sample of \jwst-selected Type 1 AGN \citep[$\sim 30\%$,][]{hainline2024} as discussed in the last subsection.
%On the other hand, if we considered a strong Balmer break only occurs in such an LOS obscuration, the occurrence rate for a V-shaped LRD in AGN is also 24\%, which is close with the estimation by \citet{hainline2024} ($\sim 30\%$ depending on redshift and other quantities).
Therefore, if \target\ is representative for \jwst-selected AGN, the (near) Compton-thick obscuration cannot explain the ubiquity of X-ray weakness but might explain the occurrence rate of LRDs among AGN.
\redtxt{To check whether a high-$C_f$ model can provide a comparably good fit, we performed a test by fixing $C_f=0.9$. This produced a significantly worse fit with $\chi _{\nu}^2=3.2$, mainly due to the over prediction of the strength of the nebular emission with respect to the strength of the Balmer break.
The fit can be improved by increasing the column density, thereby increasing the optical depth of the nebular emission. At $N_{\rm H}=10^{25}~{\rm cm^{-2}}$, the fit is improved to have $\chi _{\nu}^2=2.1$, but still significantly worse than the fiducial fit.
At $N_{\rm H}\gtrsim 10^{25}~{\rm cm^{-2}}$, the improvement in the fit becomes negligible.
While such a high column density possibly exists in some LRDs \citep[e.g.,][]{degraaff_lrd_2025,naidu_lrd_2025}, one needs to start considering the electron scattering optical depth for the optical emission \citep[e.g.,][]{Panda_2020,Rusakov_2025}, and we defer the investigation on \cloudy's calculations in this regime in future work.
Overall, while the best-fit $C_f$ can be affected by other model parameters, whether a high $C_f$ is preferred requires more statistical evidence.
}
Certainly, one should be cautious on the interpretations based on a single target. Also, the above analyses are based on another idealization that all BLR clouds have the same column density.

In addition to the dense gas obscuration, the intrinsic X-ray weakness might occur in \target\ due to super-Eddington accretion.
\citet{furtak2024} estimated a sub-Eddington accretion ratio of $\sim 0.3$ for \target, and our estimations in Section~\ref{sec:mass_acc} gives $\lambda _{\rm Edd}\sim 0.05-0.24$, which put \target\ into the sub-Eddington regime.
Still, if we assume the scenario proposed by \citet{lambrides2024} and \citet{Lupi2024} where the black hole masses of essentially all \jwst-selected AGN might be overestimated (in our case by a factor of $\gtrsim3-20$ to be compatible with super-Eddington accretion) and thus the Eddington ratios are underestimated, some of the features seen in the observations of \target\ including the X-ray weakness might be explained with super-Eddington accretion.
The underestimated factor for the mass accretion rate, $\dot m/\dot m_{\rm Edd}$, can be even larger due to the photon trapping effect at super-Eddington accretion \citep{madau_2014,inayoshi_spedd_2024}.
{Since $\lambda _{\rm Edd}\propto L_{\rm bol}/M_{\rm BH}$, if $L_{\rm bol}$ is not biased, an underestimation of $\lambda _{\rm Edd}$ by a factor of 20 would mean an overestimation of $M_{\rm BH}$ by the same factor. In Figure~\ref{fig:Mdyn_Mstar}, the black hole in \target\ would still be overmassive compared to the stellar mass. 
In terms of dynamical mass, the system will get closer to the \citet{Kormendy13} relation, approaching the $1\sigma$ dispersion, although one should take into account that the dynamical mass is still a very conservative upper limit.
We caution that the potential bias associated with the bolometric conversion to get $L_{\rm bol}$ at the super-Eddington regime is not taken into account in the above estimation.
}
%this value could be significantly underestimated if the AGN is actually accreting at a super-Eddington rate \citet{lambrides2024}.
%Supporting evidence for the bias in the Eddington ratio comes from the predicted strength for high-ionization lines such as \heii$\lambda 4686$.

An additional piece of evidence may potentially come from the strength for high-ionization lines such as \heii$\lambda 4686$.
As shown in Figure~\ref{fig:fit_ledd}, to fit the observed \heii$\lambda 4686$ together with Balmer emission lines, one can use an AGN SED with super-Eddington accretion, which performs slightly better than the sub-Eddington model.
{However, as we mentioned in Section~\ref{sec:balmer_break}, the softening of the EUV photons at the super-Eddington regime might not be realistic due to the complete exclusion of the emission within the photon-trapping radius \citep{inayoshi_spedd_2024}.
As a result, further understanding of the effect of the super-Eddington SED is still needed and we leave this investigation to future work.
}
%Still, we cannot completely exclude the case of low Eddington ratios at $\lambda _{\rm Edd}\lesssim 0.1$, where the relative strength of \heii$\lambda 4686$ is also reduced due to the overall decrease in the temperature of the accretion disk.

To distinguish the super-Eddington case from the sub-Eddington case, evidence from the UV/optical variability can also be useful, as the super-Eddington accretion should in principle result in less continuum variability due to photon trapping \citep{inayoshi_spedd_2024}.
As shown in Section~\ref{sec:variability}, there is evidence for the variation in EW(\hb) over a rest-frame time interval of $\sim 2.5$ years, which might be tracing an earlier variation in the continuum if this is real.
Still, we note that based on multi-epoch \jwst\ and \textit{HST} photometry, most of the \jwst-selected AGN and LRDs show no evidence for continuum variability \citep{kokubo_harikane2024,Maiolino2024_Xrays,ubler2023b,zhang_lrdvar_2024}.
More photometric/spectroscopic follow-up is needed to understand the variability or the lack thereof in \jwst-selected AGN and LRDs.

To conclude, \target\ serves as an interesting case to test theories of X-ray weakness in \jwst-selected AGN.
The observed Balmer break, if indeed originating in a dense and warm absorber, might also lead to strong absorption of the hard X-ray emission due to Compton down-scattering.
However, the best-fit covering fraction is not high enough to explain the ubiquitous X-ray weakness in \jwst-selected AGN, if \target\ is a representative case.
Alternatively, there could be intrinsic X-ray weakness caused by super-Eddington accretion, which is supported by the lack of high-energy photons.
In the case of intrinsic X-ray weakness, the covering fraction of \target\ might instead imply the fraction of LRDs among \jwst-selected AGN is $\sim 24\%$, if the physical conditions of its BLR are representative.

\subsection{Implications of an overmassive black hole}
\label{sec:disc_overm}

The finding that the AGN discovered by \jwst\ tend to be overmassive relative to the stellar mass of their host galaxies has been reported by several studies
\citep[e.g.,][]{harikane2023, ubler2023a, maiolino2023b, furtak2024,
juodzbalis_dormantagn_2024, Marshall24,Bogdan23,Pacucci2023_Overmassive}.
It has been claimed that part of this offset is due to a much larger dispersion of the $M_{\rm BH}-M_{\star}$ and selection effects that make overmassive black holes preferentially observed, since they are on average more luminous \citep{Li2024_overmassive,Zhang2023_trinity}. Yet, the discovery of highly overmassive and dormant (hence low-luminosity) black holes suggests that the bias and putative high-dispersion effects cannot entirely explain the offset on this relation \citep{juodzbalis_dormantagn_2024}. Regardless of the role of selection effects, the simple finding that even a small fraction of the black holes found by \jwst\ are as massive as 10\% of the host's stellar mass (or even more), is very important, as it indicates that nature somehow manages to produce such overmassive black holes in the early Universe.
Within this context, \target\
is a remarkable object as it is not only one of the most overmassive black holes confirmed to date, but also one of the most distant overmassive black holes.

Based on high-resolution spectroscopic data,
\cite{maiolino2023b} and \cite{ubler2023a} found that, in terms of dynamical masses, these early black holes are closer to the local relation. One implication could be that the host galaxies of these black holes have about the right amount of baryonic mass, but that the formation of stars has not been as efficient as the black hole growth. This could be due to the feedback of the black hole during its accretion, although mostly in the form of heating/photo-dissociation feedback, rather than ejective feedback. However, \target\ is an interesting object in this context in the sense that its black hole is highly overmassive also relative to the dynamical mass of the host galaxy. Specifically, the conservative upper limit on the dynamical mass indicates that the black hole must be more massive than 10\% of the dynamical mass of the host galaxy. If one takes the tentative measurement of the velocity dispersion, then the implied dynamical mass of the host galaxy would be just a factor of about two higher than the black hole mass, meaning that the black hole potentially contributes to the observed dynamical mass.

It is beyond the scope of this paper to explore the possible theoretical implications of this finding. However, we note that highly overmassive black holes are predicted both in scenarios where black holes originate from heavy seeds, such as the so-called Direct Collapse Black Holes \citep[e.g.,][]{natarajan_2024,Zhang2023_trinity,Bhowmick2024}, remaining outliers in the population through sustained growth across cosmic times \citep{Hu2022,scoggins2024}, and/or scenarios in which black holes experience episodes of super-Eddington accretion \citep[e.g.,][]{volonteri_2015,Inayoshi2022,schneider_2023,Trinca2023,trinca2024,bennett_2024,Li2024}.
%\RMcomm{More references needed from the theoreticians in our groups}
These two scenarios are not in conflict, and both of them could apply to different classes of black holes, or even the same objects. Primordial black holes, formed from fluctuation in the early phases after inflation, are also a possible alternative to explain overmassive black holes in the early Universe \citep{Dayal2024}. Many of these studies manage to reproduce the overmassive nature of early black holes on the $M_{\rm BH}-M_{\star}$ relation. However, there has been little effort in exploring the offset on the $M_{\rm BH}-M_{\rm dyn}$ relation. 
Thus, it is vital in the near future to explore this relation with models and simulations.
The extreme properties of objects such as \target\ may help to break degeneracies between the models.
%Hopefully, in the near futures simulations and models will explore also this relation, and the extreme properties of \target\ may help to break degeneracies between models.

\subsection{Low metallicity and possible feedback}
\label{sec:disc-feedback}

As already illustrated by \citet{furtak2024}, the MSA PRISM spectrum of \target\ shows peculiar narrow emission lines with very weak \oiii$\lambda 5008$.
%Already the MSA prism spectrum obtained by \cite{furtak2023} illustrated the peculiarity of \target in terms of narrow emission lines, by revealing a weak emission of [OIII]5007.
Our high-resolution spectrum confirms the weakness of \oiii$\lambda 5008$, with EW $<$ 3.7 \AA. Such a low EW of the \oiii$\lambda 5008$ has been observed also in many lower redshift quasars. 
Potential explanations for the weak \oiii$\lambda 5008$ include
very extended NLRs (with sizes larger than the host galaxies, hence running out of gas to ionize; \citealp{Netzer2004}), and the so-called Baldwin effect for \oiii$\lambda 5008$ caused by the inclinations of the accretion disks \citep{Risaliti2011}.
%and it has been ascribed to possibly the size Narrow Line Region extending beyond the ISM of the host galaxy, hence running out of gas to ionize \citep{Netzer2004}, and to effects of the inclination of the accretion disc, the so-called Baldwin effect for [OIII] \citep{Risaliti2011}. 
However, the above explanations were proposed in the context of very luminous quasars ($L_{\rm bol}\approx 10^{46}-10^{48}~{\rm erg~s^{-1}}$), while \target\ is a much lower luminosity AGN ($L_{\rm bol}\approx 0.2-1\times 10^{45}~{\rm erg~s^{-1}}$). Specifically, in terms of bolometric luminosity \target is similar to many other AGN discovered by \jwst\ at high-$z$ \citep[e.g.,][]{scholtz2023,Adamo2024_firstbillion,maiolino2023b,taylor24}, which are instead characterized by prominent \oiii\ emission. In principle, the weakness of \oiii\ could be due to a very low metallicity. However, our high-resolution spectrum reveals that the narrow component of H$\beta$ is relatively weak as well, in contrast to many other \jwst-discovered Type 1 AGN, whose H$\beta$ emission is dominated by the narrow component \citep{maiolino2023b,ubler2023a,ubler2023b}. The weakness of the narrow \hb\ cannot be explained by the low metallicity. 
%This suggests that the weakness of the narrow lines is not a metallicity effect given the weak H$\beta$. 
%Therefore, 
One possible explanation for the weak narrow lines is that the host galaxy has experienced a recent strong feedback effect by the AGN, which has removed a large fraction of the ISM in the host galaxy, hence leaving little gas to ionize in the host galaxy (i.e., a very weak NLR).

%This would be consistent with the scenario discussed in Sect.\ref{subsec:uv_fit}, whereby, in the scenario in which the UV is dominated by stellar emission, the weak narrow H$\beta$ implies that the star formation must have quenched in the last few Myr.
In Section~\ref{subsec:uv_fit}, based on the fit of the UV continuum, we discuss a scenario where the UV light of \target\ is dominated by stellar light.
The strong feedback scenario inferred from the weak narrow H$\beta$ implies that the star formation must have quenched in the last few Myr.
Under the hypothesis that the UV continuum is dominated by young stars, one additional characteristic of \target is the mismatch between M$_{\rm UV}$ and the luminosity of H$\beta$. Specifically, using the UV calibration of \citet{Kennicutt1998}, corrected for a \citet{Chabrier2003} initial mass function, we can estimate a star-formation rate (SFR) of 0.2~M$_\odot$~yr$^{-1}$
\citetext{we assumed a calibration for solar metallicity, $M_{\mathrm{UV}}=-16.98\pm0.09$~mag, \citealp{furtak2024}, and uncertainties on the calibration of 0.3~dex}. From the luminosity of the narrow H$\beta$ (Table~\ref{tab:measurements}), assuming Case~B recombination and the SFR law of \citet{Shapley2023}, we infer {an SFR of 0.06~M$_\odot$~yr$^{-1}$}. This is a discrepancy of a factor of three, without even considering that any dust attenuation would suppress the UV-inferred SFR more than the SFR based on H$\beta$, and that the latter estimate assumes no NLR to be present in the AGN. Since the UV and H$\beta$ SFRs trace star formation on timescales of 100~Myr and 10~Myr, respectively, this mismatch could point to a decreasing SFR trend, reminiscent of rapidly quenched galaxies \citep{strait+2023,looser+2024,baker+2025} and consistent with the strong-feedback interpretation.

On the other hand, it is also true that \target\ has an anomalously low \oiii$\lambda 5008$ to H$\beta$ flux ratio, which is \oiii/\hb$_{\rm narrow}=0.60\pm 0.18$ from the R2700 \blackthunder\ spectrum.
This \oiii/\hb$_{\rm narrow}$ is the lowest among all high-$z$ AGN to our best knowledge 
\cite[e.g.,][]{Kokorev2023,maiolino2023b,ubler2023a,ubler2023b,trefoloni_feii_2024,Marshall2023,Marshall24}.
%In fact, this is the AGN discovered by \jwst\ with the lowest ratio, with \oiii/\hb$_{\rm narrow}=0.60\pm 0.18$ from the R2700 spectrum. To our knowledge this is the lowest \oiii/\hb$_{\rm narrow}$ ever found among high-$z$ AGN \cite[e.g.][]{maiolino2023b,Kokorev2023,ubler2023a,ubler2023b,trefoloni_feii_2024,Marshall24,Marshall2023}. 
Such a low ratio could be due to a very low metallicity in the ISM of the host galaxy. 
While it is difficult to assess the metallicity without other nebular emission lines, based on the photoionization models presented by \citet{NakajimaMaiolino2022}, it is possible that the metallicity of the host is below $0.01~Z_{\odot}$. 
Such a low metallicity inferred from \oiii/\hb$_{\rm narrow}$ implies that we might be witnessing the formation of a primeval black hole in a primeval galaxy. Alternatively, the weakness of \oiii\ relative to H$\beta$ could be a consequence of very high gas density in the ISM of this system. Specifically, if the electron density is higher than the critical density of \oiii$\lambda 5008$, which is $n_{\rm e;ISM}>n_{\rm [OIII];critical}\sim10^6~{\rm cm}^{-3}$, the observed strength of \oiii\ would be significantly suppressed relative to H$\beta$ \citep{draine2011}.
However, this scenario would imply that there is no intermediate/low-density gas in the system and that most of the ISM has extremely high densities, which is atypical even in high-$z$ galaxies \citep[e.g.,][]{isobe_ne_2023,lisijia_ne_2024}.

\section{Conclusions}
\label{sec:conclude}

In this work, we have presented new NIRSpec-IFU data of a lensed Little Red Dot (LRD) at $z=7.04$, obtained within the context of the BlackTHUNDER \jwst\ Large Programme, using both high (G395H, R$\sim$2700) and low (PRISM, R$\sim$100) spectral resolution modes. We also re-analyzed archival, low-resolution MSA spectra of the three images of the lensed LRD.

The spectrum confirms a prominent, but smooth Balmer break and the presence of broad Balmer lines tracing the Broad Line Region (BLR) of an AGN. We also detect a very narrow component of H$\beta$ and \oiii$\lambda5008$ emission and tentative (3.2$\sigma$) \hb\ absorption. 
% In this work we present new high-resolution \jwst\ observations for \target, a broad-line AGN at $z=7.04$, and propose a new approach for fitting the rest-frame UV-to-optical NIRSpec/PRISM data of this target.
% Our new fitting approach includes an AGN continuum attenuated by first dense dust-less gas and then a dust screen dominating the rest-frame optical, and an unattenuated component coming from either young stellar populations or AGN continuum emission dominating the rest-frame UV.
By analyzing these spectral properties in detail, we reach the conclusions listed below.
% We summarize our main conclusions as follows.
% \RMcomm{To be written}

\begin{itemize}
    \item Assuming local virial relations, we infer a black hole mass of $4^{+4}_{-2}\times 10^7~M_\odot$. The black hole is estimated to accrete at an Eddington ratio of $\lambda _{\rm Edd}=L_{\rm bol}/L_{\rm Edd}\approx 0.05-0.24$, depending on whether the bolometric luminosity is measured assuming bolometric corrections from the H$\beta$ luminosity or from the optical continuum (L$_{5100}$).
    
    \item The narrow component of H$\beta$ is extremely narrow,  unresolved (conservatively, $\sigma < 47$ \kms). This, together with the compact size of the galaxy, was used to set an tight,  conservative upper limit on the dynamical mass of $M_{\rm dyn}<4\times 10^8~M_\odot$, which is one order of magnitude lower than the stellar mass of the galaxy inferred when assuming that the Balmer break and optical continuum are of stellar origin ($4\times 10^9~M_\odot$). Our finding indicates that the stellar light cannot contribute more than 10\% (very conservatively) to the Balmer break and to the optical continuum.
    
    \item We have shown that the Black Hole in \target\ is overmassive, by $\sim 1-2$ orders of magnitude relative to both the stellar {\it and} dynamical mass of the host galaxy, when compared to local scaling relations. This is the first high-$z$ black hole found to be so overmassive on both scaling relations, and may have important implication understanding the seeding and early growth of black holes.

    \item The Balmer break-like feature in the PRISM spectrum of \target\ can be described by absorption from dust-free dense gas, likely originating in the BLR.
    Specifically, we have shown that the spectrum can be well fit by a model composed of an optical part dominated by AGN continuum attenuated by dust-free gas in the BLR along the line-of-sight, and then by a dust screen outside (or in the outer parts of the BLR). The UV part is dominated by a less attenuated continuum of an AGN (either scattered or transmitted through partial covering) or from a young stellar population.
    Based on our fiducial model, the dense absorbing gas has a density of $n_{\rm H}=10^{10}~{\rm cm^{-3}}$, a column density of at least $N_{\rm H}=10^{24}~{\rm cm^{-2}}$, an ionization parameter of $U=10^{-1.5}$, and a microturbulence velocity of $v_{\rm turb}=120$ \kms.
    Our modeling approach provides a better fit to the spectrum of \target\ compared to previous works that assume a stellar Balmer break.

    \item
    This non-stellar origin for the Balmer break has the advantages of not invoking a high stellar mass that is inconsistent with the inferred dynamical mass and which would result in an extremely high stellar density; it does not require a peculiar shape of the attenuation curve, and does not imply extremely high intrinsic equivalent widths for broad emission lines.
    
    \item Our fiducial model also predicts an absorption line in \hb\ with an equivalent width of $8.0$ \AA\ with respect to the continuum. This value is consistent, within 1.1$\sigma$ with the tentative \hb\ absorption that we measured from the high-resolution ($R\sim 2700$) IFU spectrum of \target, which has an EW of $5.5^{+2.2}_{-1.7}$ \AA. Further tuning of the model parameters can result in even better consistency.
    
    \item The shape of the Balmer break in our model depends on various model parameters. Specifically, a microturbulence velocity of $v_{\rm turb}\sim 100$ \kms\ is needed to recover the smooth Balmer break observed in \target, which is caused by the superpositions of deep Balmer absorption lines near the Balmer limit. The variation in the microturbulence velocity could lead to different shapes of the Balmer break and might explain the diversity of the break seen in other \jwst-discovered LRDs.

    \item Potentially, the Balmer break of other LRDs may be explained with the same scenario of AGN continuum and dense gas absorption, especially given that at least 20\% of the LRDs present evidence of H$\alpha$ absorption. A non-stellar origin of the Balmer break in LRDs would greatly alleviate the extreme stellar mass densities in these systems, as well as the extremely high stellar mass density per cosmic volume, inferred when assuming a stellar origin of the Balmer breaks. Studies such as the one performed here on other LRDs can clarify this aspect.

    \item The rest-frame UV continuum of \target\ likely has a different origin compared to the optical continuum.
    %If we do not include the spectrum from image B in our analysis, which has a background subtraction issue, 
    The UV continuum can be described by an AGN/SF continuum with a small amount of dust attenuation ($A_{\rm V}\sim 0.1$) and a damped Ly$\alpha$ absorber with $N_{\rm HI}\sim 3\times 10^{21}~{\rm cm^{-2}}$, although we cannot fully exclude the scenario where the UV continuum is instead nebular dominated.
    The nebular dominated scenario would require a temperature of $T_e\sim 30,000$ K and the normalization of the continuum model might not be entirely physical.
    %If we include image B, the UV continuum is better fitted with a nebular continuum, although the best-fit temperature ($T_e\sim 30,000$ K) and the normalization of the continuum might not be entirely physical.
    Overall, the UV continuum likely comes from a more extended region compared to the optical part due to the significantly lower gas column density and dust attenuation. Alternatively, the UV could potentially also be AGN light transmitted through a partial covering absorber.
    
    \item The lack of hard X-ray emission in \target\ might be related to the presence of the dense gas along the line-of-sight in our model, whose lower limit on the column density is close to the Compton-thick regime.
    If the X-ray weakness is indeed due to Compton-thick absorption within the BLR, our best-fit covering fraction of the BLR clouds implies 24\% of the \jwst-discovered AGN are X-ray weak, which is significantly lower than the observed fraction of $\sim90$ \%.
    Meanwhile, if we interpret the dense gas obscuration as the necessary condition for LRDs, this result implies 24\% of the \jwst-discovered AGN are LRDs, which is consistent with the value of $\sim 30$ \% currently found in observations.
    It is possible that for the remaining part of non-LRD AGN (about 70\% of them), the X-ray weakness is intrinsic due to, for example, high accretion rates.
    %, as for instance in the high accretion rate scenarios.
    
    \item Super-Eddington accretion can explain the X-ray weakness and potentially the weakness of \heii$\lambda 4686$ in the PRISM spectrum of \target. However, this scenario requires an overestimation of the black hole mass by a factor of $\gtrsim 3-20$ and the extreme UV part of the super-Eddington SED remains to be examined carefully.

    \item We have found evidence for a higher equivalent width of the Balmer lines in one of the three lensed images of \target, specifically image C, relative to the other two images. We note that a variation in EW cannot be ascribed to any calibration issues. This finding can be explained in terms of a drop of the continuum flux in image C and the fact that the BLR has not yet responded to the inonizing flux variation because of its size. This interpretation is supported by the finding that the flux observed in image C a few months before the spectroscopic observation was about a factor of two higher. 

    \item We used this information to attempt a first reverberation mapping at such an early epoch, which would make this source consistent with the relation between $R_{\rm BLR}$ and $L_{5100}$ found in local reverberation studies.

    \item
    \target has the lowest EW(\oiii$\lambda 5008$) among all \jwst-discovered AGN, and also the weakest H$\beta$ narrow relative to the broad component. These features may indicate that the AGN has exerted extremely strong feedback, removing most of the ISM in the host galaxy and possibly quenching star formation.

    \item
    \target has the lowest flux ratio between \oiii$\lambda 5008$ and the narrow H$\beta$ among all high-z AGN, with \oiii$\lambda 5008$/H$\beta _{\rm narrow} = 0.60\pm 0.18$. This indicates that the host galaxy of this black hole is very metal poor (possibly $Z<10^{-2}~Z_\odot$). Alternatively, very high electron densities ($n_{\rm e}>10^6~{\rm cm}^{-3}$) in the ISM may suppress \oiii$\lambda 5008$ relative to H$\beta$.
\end{itemize}

As a concluding remark, the case study of \target\ provides clues for deciphering the physical conditions of LRDs and AGN discovered by \jwst, and demonstrates the need for deep and high spectral resolution observations that help differentiate AGN and stellar scenarios.
It remains to be explored in future work whether the SED modeling approach proposed in this work can be generalized to other LRD AGN, and whether the X-ray weakness in these sources is related to the inferred properties.
Furthermore, it would be important to perform a statistical analysis to understand how the inferred physical conditions are changing with redshift and whether this explains the significantly reduced number of LRDs at low redshift.

%\section{TODO}

%\begin{itemize}
%    \item dd
%\end{itemize}

\section*{Acknowledgements}
\realredtxt{We thank the anonymous referee, whose thoughtful suggestions improved the clarity of this work.}
We thank Gary Ferland and T.~Taro Shimizu for helpful discussions.
We thank L. Furtak for providing the combined spectrum of the three images presented in their work.
XJ and RM acknowledge ERC Advanced Grant 695671 “QUENCH” and support by the Science and Technology Facilities Council (STFC) and by the UKRI Frontier Research grant RISEandFALL.
RM acknowledges funding from a research professorship from the Royal Society.
%H\"U acknowledges support through the ERC Starting Grant 101164796 ``APEX''.
\redtxt{H\"U acknowledges funding by the European Union (ERC APEX, 101164796). Views and opinions expressed are, however, those of the authors only and do not necessarily reflect those of the European Union or the European Research Council Executive Agency. Neither the European Union nor the granting authority can be held responsible for them.}
MP acknowledges grant PID2021-127718NB-I00 funded by the Spanish Ministry of Science and Innovation/State Agency of Research (MICIN/AEI/10.13039/501100011033). 
MP also acknowledges support through the grant RYC2023-044853-I, funded by  MICIU/AEI/10.13039/501100011033 and FSE+.
KI acknowledges support from the National Natural Science Foundation of China (12073003, 12003003, 11721303, 11991052, 11950410493), 
and the China Manned Space Project (CMS-CSST-2021-A04 and CMS-CSST-2021-A06).
AJB acknowledges funding from the “FirstGalaxies” Advanced Grant from the European Research Council (ERC) under the European Union’s Horizon 2020 research and innovation program (Grant agreement No. 789056).
SC, EP, and GV acknowledge support by European Union’s HE ERC Starting Grant No. 101040227 - WINGS.

This work is based on observations made with the NASA/ESA/CSA James Webb Space Telescope. The data are available at the Mikulski Archive for Space Telescopes (MAST) at the Space Telescope Science Institute, which is operated by the Association of Universities for Research in Astronomy, Inc., under NASA contract NAS 5-03127 for JWST. These observations are associated with the programmes, BlackTHUNDER (PID 5015; PIs: H.~\"Ubler, R.~Maiolino) and UNCOVER (PID 2561; PIs: I.~Labbé, R.~Bezanson).

\section*{Data Availability}

%All \jwst\ data used in this paper are available through the MAST portal.
All analysis results of this paper will be shared on reasonable request to the corresponding author.
The NIRCam cutouts of Figure~\ref{fig:spec_qso1} were made from publicly available mosaics from UNCOVER DR4 \citep{uncover_dr4}.
The NIRSpec observations used in this paper are available through the MAST portal.

\bibliographystyle{mnras} % style aa.bst
\bibliography{ref} % your references ref.bib

\begin{appendix}

\section*{Affiliations}

$^{1}$Kavli Institute for Cosmology, University of Cambridge, Madingley Road, Cambridge, CB3 0HA, UK\\
$^{2}$Cavendish Laboratory, University of Cambridge, 19 JJ Thomson Avenue, Cambridge, CB3 0HE, UK\\
$^{3}$Department of Physics and Astronomy, University College London, Gower Street, London WC1E 6BT, UK\\
$^{4}$Max-Planck-Institut für extraterrestrische Physik, Gie{\ss}enbachstra{\ss}e 1, 85748 Garching\\
$^{5}$Center for Astrophysics $|$ Harvard \& Smithsonian, 60 Garden St., Cambridge, MA 02138, USA\\
$^{6}$Centro de Astrobiolog\'{\i}a (CAB), CSIC-INTA, Ctra. de Ajalvir km 4, Torrej\'on de Ardoz, E-28850, Madrid, Spain\\
\redtxt{$^{7}$Scuola Normale Superiore, Piazza dei Cavalieri 7, I-56126 Pisa, Italy\\
$^{8}$University of Oxford, Department of Physics, Denys Wilkinson Building, Keble Road, Oxford OX13RH, UK\\}
$^{9}$Sorbonne Universit\'e, CNRS, UMR 7095, Institut d'Astrophysique de Paris, 98 bis bd Arago, 75014 Paris, France\\
$^{10}$INAF - Osservatorio Astrofisico di Arcetri, largo E. Fermi 5, 50127 Firenze, Italy\\
$^{11}$European Southern Observatory, Karl-Schwarzschild-Strasse 2, 85748 Garching, Germany\\
$^{12}$Steward Observatory, University of Arizona, 933 N Cherry Avenue, Tucson, AZ 85721, USA\\
$^{13}$Institute of Astronomy, University of Cambridge, Madingley Road, Cambridge CB3 0HA, UK\\
$^{14}$Kavli Institute for Astronomy and Astrophysics, Peking University, Beijing 100871, China\\
$^{15}$Waseda Research Institute for Science and Engineering, Faculty of Science and Engineering, Waseda University, 3-4-1, Okubo, Shinjuku, Tokyo 169-8555, Japan\\
$^{16}$AURA for European Space Agency, Space Telescope Science Institute, 3700 San Matin Drive, Baltimore, MD 21218, USA\\
$^{17}$Dipartimento di Fisica e Astronomia, Università di Bologna, Via Gobetti 93/2, I-40129 Bologna, Italy\\
$^{18}$INAF – Osservatorio di Astrofisica e Scienza dello Spazio di Bologna, Via Gobetti 93/3, I-40129 Bologna, Italy\\
$^{19}$Department of Astronomy and Astrophysics University of California, Santa Cruz, 1156 High Street, Santa Cruz CA 96054, USA\\
$^{20}$Dipartimento di Fisica, Sapienza Universita di Roma, Piazzale Aldo Moro 5, 00185 Rome, Italy\\
$^{21}$INAF/Osservatorio Astronomico di Roma, Via Frascati 33, 00040 Monte Porzio Catone, Italy\\
$^{22}$INFN, Sezione Roma1, Dipartimento di Fisica, ``Sapienza'' Universita di Roma, Piazzale Aldo Moro 2, 00185, Roma, Italy\\
$^{23}$Sapienza School for Advanced Studies, Viale Regina Elena 291, 00161 Roma, Italy\\
$^{24}$Como Lake Center for Astrophysics, DiSAT, Universit$\grave{a}$ degli Studi dell'Insubria, via Valleggio 11, 22100, Como, Italy\\
$^{25}$NRC Herzberg, 5071 West Saanich Rd, Victoria, BC V9E 2E7, Canada\\
$^{26}$ Department of Astronomy, University of Geneva, Chemin Pegasi 51, 1290 Versoix, Switzerland\\
$^{27}$Cosmic Dawn Center (DAWN), Copenhagen, Denmark\\
$^{28}$Niels Bohr Institute, University of Copenhagen, Jagtvej 128, DK-2200, Copenhagen, Denmark

\section{Lens modelling}
\label{sec:lens_model}

The lens magnification factors and time delays are not the focus of this paper, and our results are largely unaffected by the details of the lens model. However, in this section we very briefly summarize the results of previous lens models for the three images of \target, and we also present an independent analysis.

\cite{furtak_abell2744_2023} and \cite{furtak2024} use the wealth of data available for this cluster to model the lensing of the three images of \target. They obtain that the magnification factors for the three images are $\mu _A = 6.15^{+0.77}_{-0.39}$,
$\mu _B = 7.29^{+0.36}_{-2.18}$, and 
$\mu _C = 3.55^{+0.25}_{-0.24}$. They also estimate the time delays between the three images: image A follows image C by about $\sim 19$ years, while image B follows image A by about $\sim 3$ years \citep[although these values were estimated by][ based on the photometric redshift of the source]{furtak_abell2744_2023}.

We have independently validated these quantities, for each of the three images, by
using the high-precision strong lens model derived from \jwst\ observations by \citet{bergamini23}.
We obtain the following lens magnifications: $\mu_A\,=\,5.8\substack{+ 0.4\\ -0.2}$, $\mu_B\,=\,9.1\substack{+ 0.9\\ -0.8}$ and $\mu_C\,=\,3.2\substack{+ 0.1\\ -0.1}$ respectively. 
Additionally,
we find a delay between the arrival of images C and A of $\tau\,=\,17.9\substack{+ 0.3\\ -0.5}\,\textrm{yrs}$ and $\tau\,=\,2.2\substack{+ 0.1\\ -0.1}\,\textrm{yrs}$ between images A and B.
% The results of the two lens models are broadly consistent, but there are also some differences at the $\sim 2 \sigma$ level, indicating that there are uncertainties in the  models beyond the formal errors.
While these results are broadly consistent with those of the \citet{furtak_abell2744_2023} and \cite{furtak2024} analysis, they underscore the inherent uncertainties and systematic biases that can arise from the choice of the lens model.

\section{Fits based on models with different parameters}
\label{appendix:other_params}

\begin{figure*}
    \centering
    \includegraphics[width=0.9\textwidth]{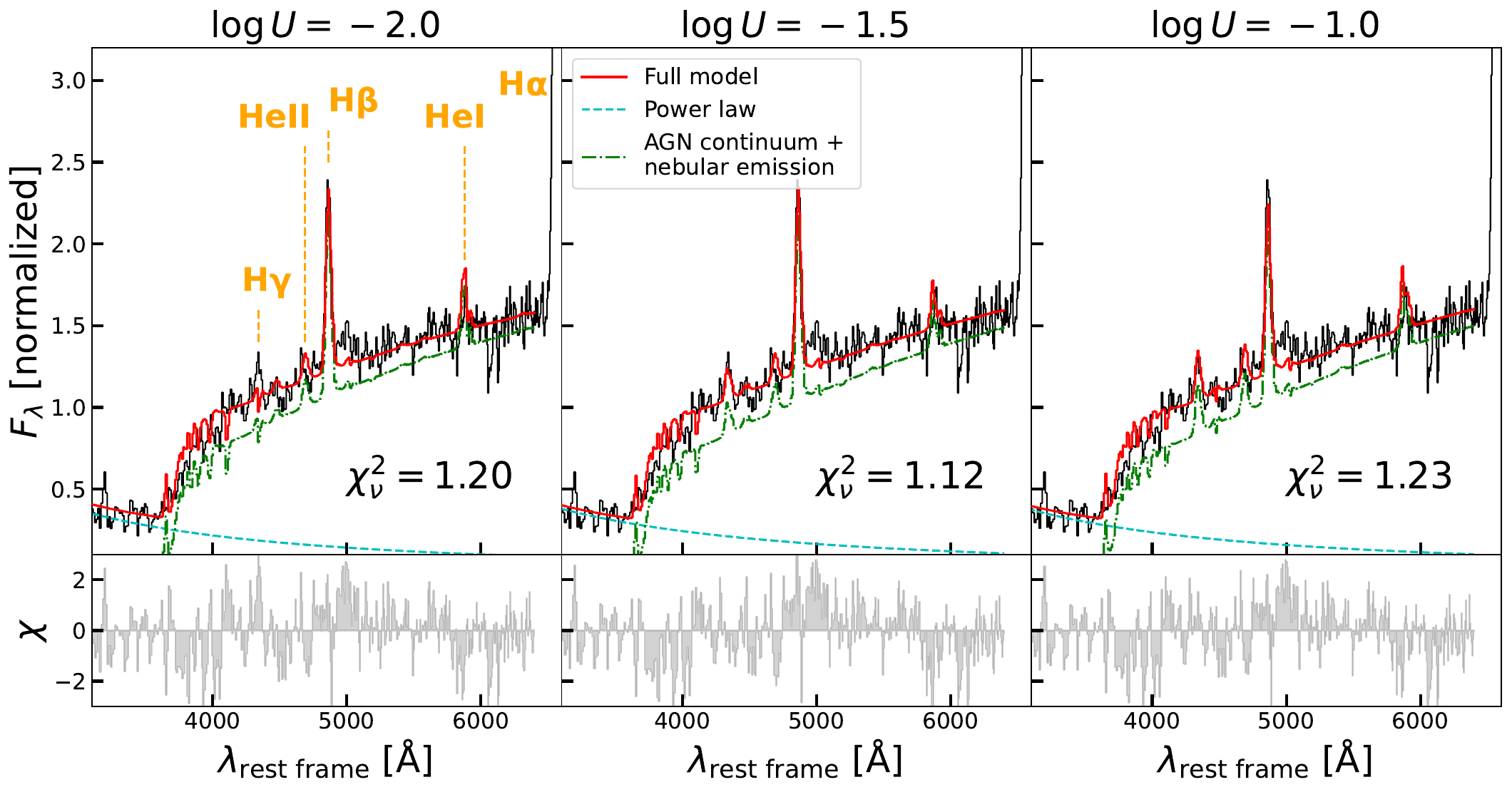}

    \caption{Best-fit models for the combined spectrum of images A+C of \target\ with the GTO reduction in the rest-frame wavelength range of $1600-6400$ \AA.
    We compare three different ionization parameters for the dense gas that absorbs the AGN continuum, which are $U=10^{-2}$, $10^{-1.5}$, and $10^{-1}$.
    The observed spectrum is plotted as the solid black line.
    In each panel, we show the resulting $\chi$ of the fit.
    }
    \label{fig:lgu_demo}
\end{figure*}

\begin{figure*}
    \centering
    \includegraphics[width=\textwidth]{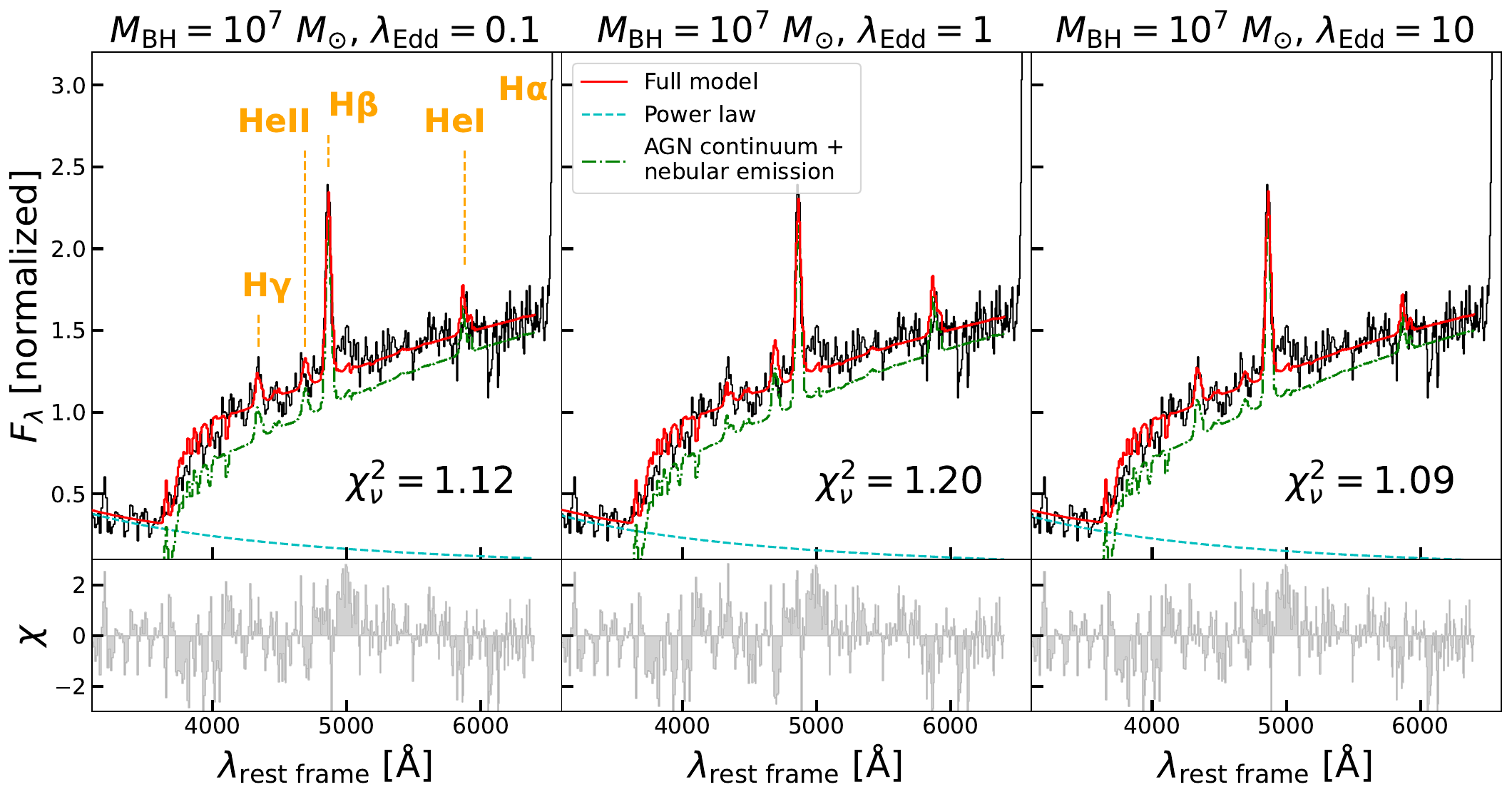}
    
    \caption{Comparison between three fits to the PRISM spectrum of \target\ using different AGN SEDs with $\lambda _{\rm Edd}=0.1,1,\rm and~ 10$ zoomed in the rest-frame optical. The observed spectrum is from the GTO reduced A+C images and is plotted in solid black. The bottom panels show $\chi$ of the fits calculated for each pixel.
    In each panel, we also show the reduced $\chi ^2$ of the fit.
    The main differences between the fits are the nebular emission in H$\gamma$, \heii$\lambda 4686$, and \hei$\lambda 5876$.
    Overall the super-Eddington case with $\lambda _{\rm Edd}=10$ provides the best fit, but only marginally relative to the sub-Eddington fit with $\lambda _{\rm Edd}=0.1$.
    }
    \label{fig:fit_ledd}
\end{figure*}

In this section, we show additional fitting results based on different model assumptions.
In Section~\ref{sec:balmer_break}, we introduced a range of ionization parameters to describe the properties of the warm absorber that produces the Balmer break as well as the tentative \hb\ asborption in \target.
In Figure~\ref{fig:lgu_demo}, we compare best-fit models with different ionization parameters in a range of $U=10^{-2}-10^{-1}$ for the warm absorber.
All the other parameters are fixed to the fiducial values listed in Table~\ref{tab:models}.
One can see the main effect of the ionization parameter is to change the relative strength of the nebular emission with respect to the continuum. At $U=10^{-2}$, the ionization parameter is relatively low and the strength of H$\gamma$ is clearly underpredicted.
In contrast, at $U=10^{-1}$, H$\beta$ is slightly weaker compared to the observation.
The choice of the ionization parameter also impacts the best-fit covering fraction, $C_f$.
For example, at the highest ionization parameter of $U=10^{-1}$, the best-fit covering fraction increases to $C_f=0.34$ compared to the fiducial case of $C_f=0.24$ at $U=10^{-1.5}$.
In comparison, at $U=10^{-2}$, the best-fit covering fraction slightly decreases to $C_f=0.23$.
Overall, the model with out fiducial ionization parameter, $U=10^{-1.5}$, produces the best fit, although the differences between the $\chi^2_{\nu}$ of the fit is not large.

Figure~\ref{fig:fit_ledd} compares fitting results with three different Eddington ratios, $\lambda _{\rm Edd}=0.1$, 1, and 10, for the input SEDs.
The most notable difference is in the nebular emission.
When $\lambda _{\rm Edd}=0.1$, the strengths of \heii$\lambda 4686$ and \hei$\lambda 5876$ from the BLR is slightly overpredicted.
When $\lambda _{\rm Edd}=1$, \heii$\lambda 4686$ and \hei$\lambda 5876$ are further overpredicted but H$\gamma$ is underpredicted.
The smallest reduced $\chi^2$ is achieved at $\lambda _{\rm Edd}=10$.
The above difference is mainly driven by the hardness of the ionizing radiation, which is generally too hard for \target\ at the sub-Eddington regime.
This is consistent with the recent finding of \citet{lambrides2024}, who claimed that \jwst-selected Type 1 AGN generally show softer SEDs compared to those expected at the sub-Eddington regime.
Still, the difference in the reduced $\chi^2$ between the super-Eddington case and the sub-Eddington case is only 0.05. Also, based on our derivation in Section~\ref{sec:mass_acc}, \target\ is more likely to host a sub-Eddington AGN.
In addition, according to \citet{inayoshi_spedd_2024}, the UV spectrum of a super-Eddington AGN should actually become harder due to the emission from the hotter part of the accretion disk within the photon-trapping radius.
This is in contrast to the SED model we took from \citet{pezzulli2017}, which does not include the emission within the photon-trapping radius.
Finally, we note that different from the fiducial case ($\lambda _{\rm Edd}=0.1$) where the best-fit covering fraction is 0.24, the near-Eddington case ($\lambda _{\rm Edd}=1$) yields a best-fit covering fraction of 0.22, and the super-Eddington case a best-fit covering fraction of 0.46.

\section{Fitting of spectra with different reductions}
\label{appendix:other_reductions}

\begin{figure*}
    \centering
    \includegraphics[width=0.9\textwidth]{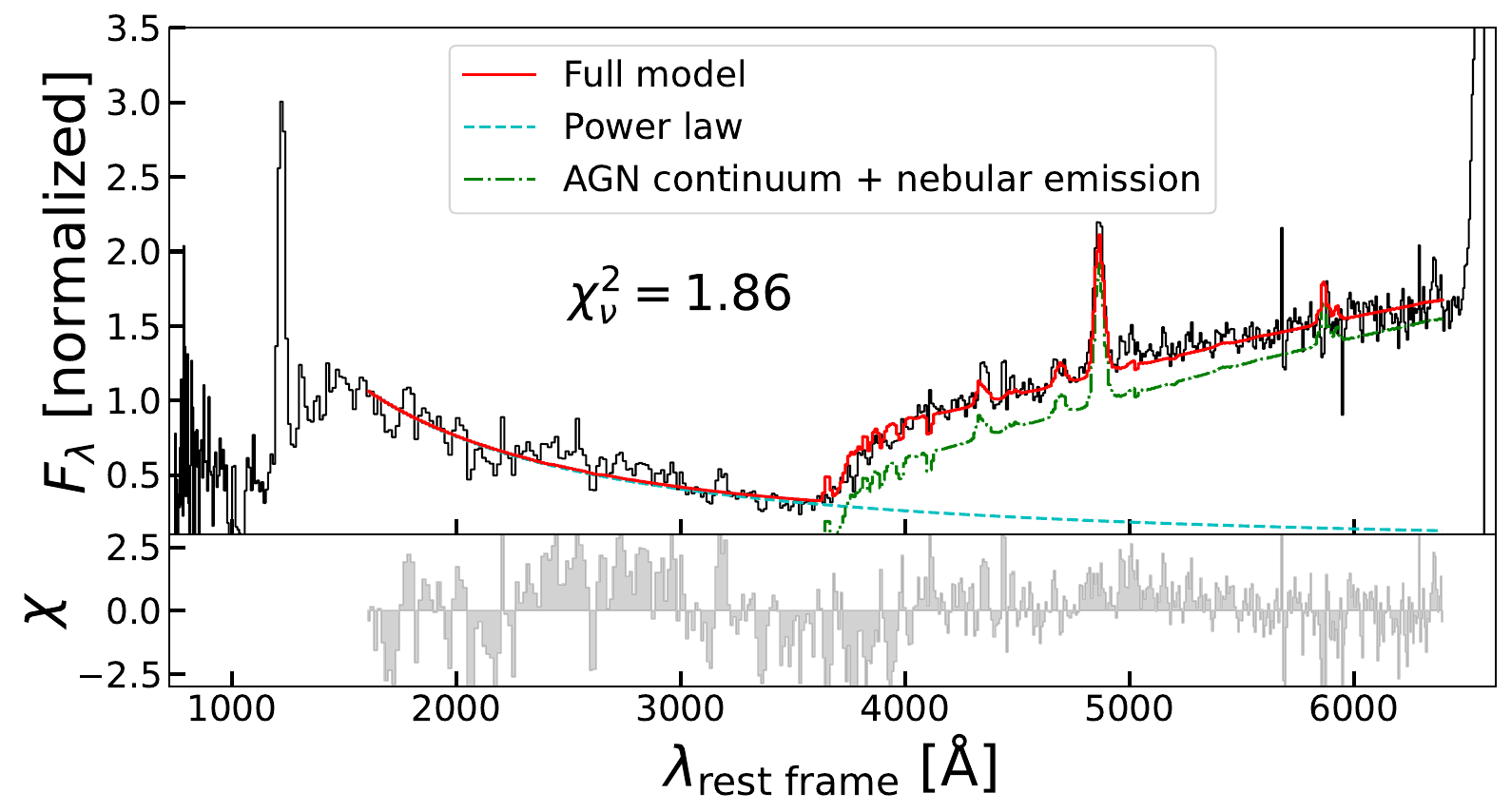}

    \caption{Fiducial fit for the combined images of A+B+C reduced by \citet{furtak2024} in the rest-frame wavelength range of $1600-6400$ \AA.
    The observed PRISM spectrum is normalized to the flux density at $\lambda = 4260$ \AA\ in the rest frame of \target.
    Given the apparently unphysical bad pixels in the optical, we added $5\sigma$-clipping to the fit.
    %\redtxt{[increasing U fits Hg better; bc the spectrum has a redder optical slope]}
    }
    \label{fig:furtak_demo}
\end{figure*}

\begin{figure*}
    \centering\includegraphics[width=\columnwidth]{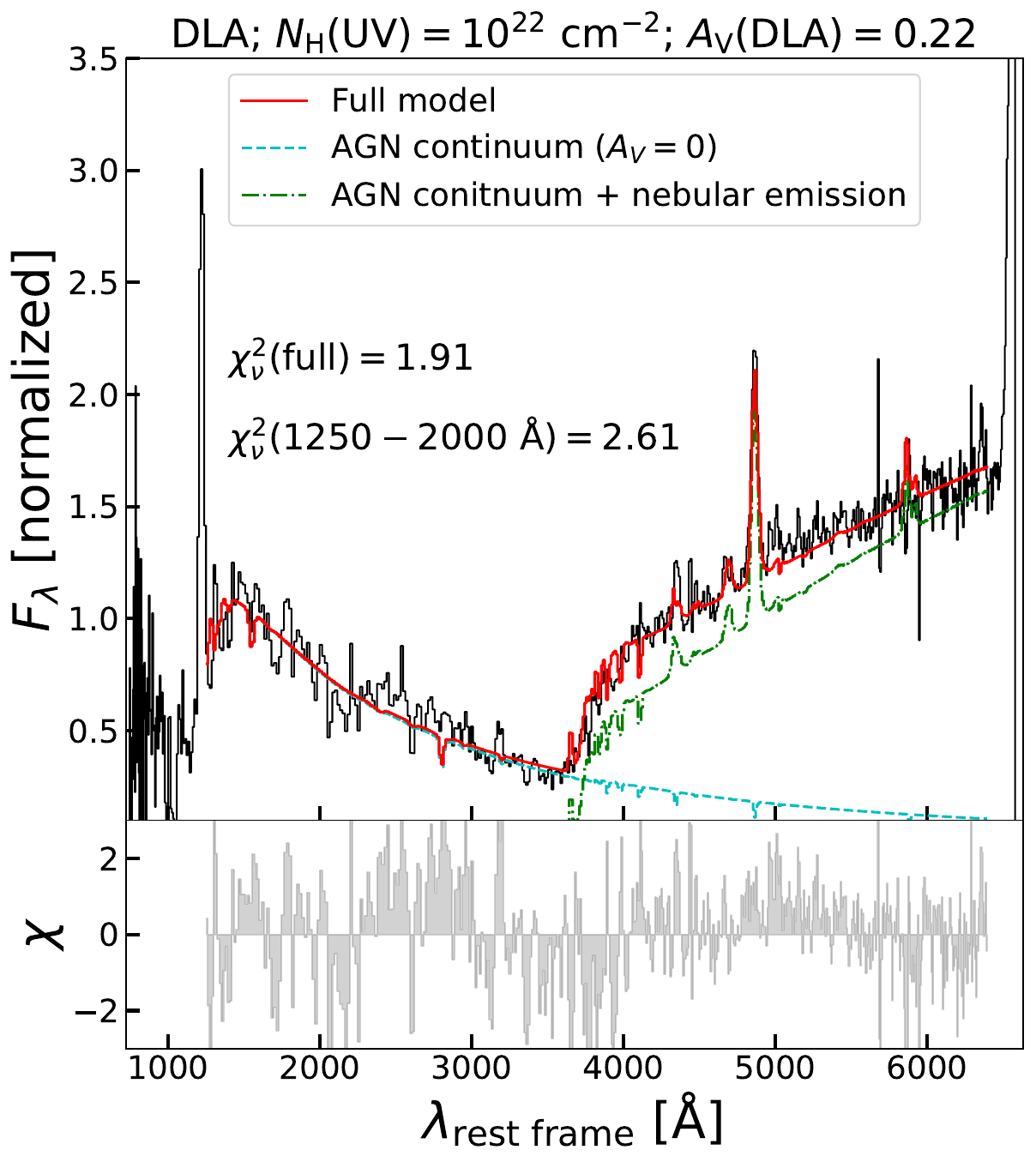}
    \includegraphics[width=\columnwidth]{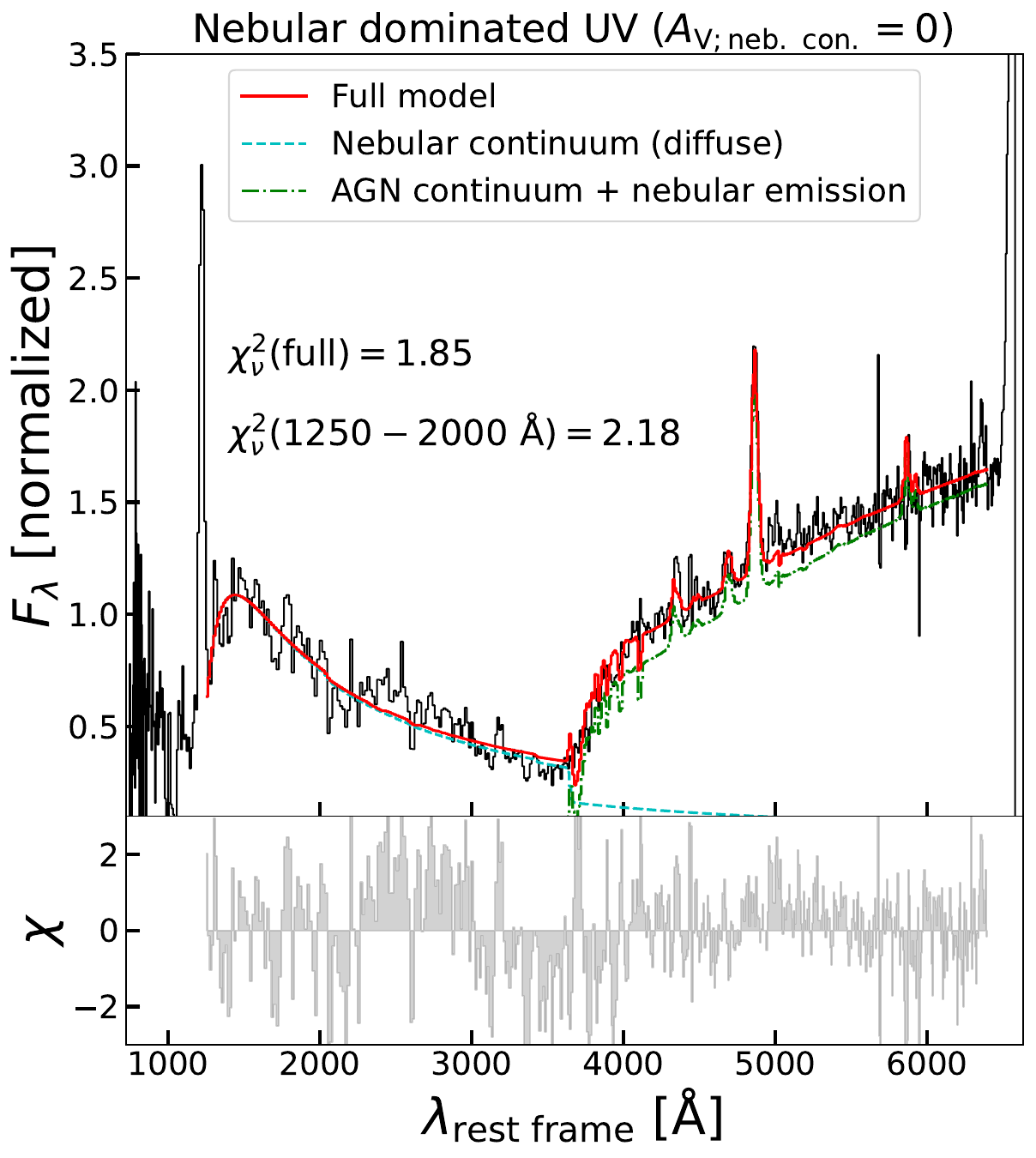}

    \caption{Best-fit models for the PRISM combined images of A+B+C reduced by \citet{furtak2024} in the rest-frame wavelength range of $1250-6400$ \AA\ with different assumptions in the UV.
    \textit{Left:} the UV continuum is assumed to be the AGN continuum emission attenuated by a slab of gas with $N_{\rm H}=10^{22}~{\rm cm^{-2}}$.
    \textit{Right:} the UV continuum is assumed to be a nebular continuum composed of a two-photon continuum, a hydrogen free-bound continuum, and a hydrogen free-free continuum.
    Both models assume no dust attenuation in the UV.
    The nebular continuum model provides a better fit compared to the DLA model.
    }
    \label{fig:furtak_uv}
\end{figure*}

\begin{figure}
    \centering
    \includegraphics[width=\columnwidth]{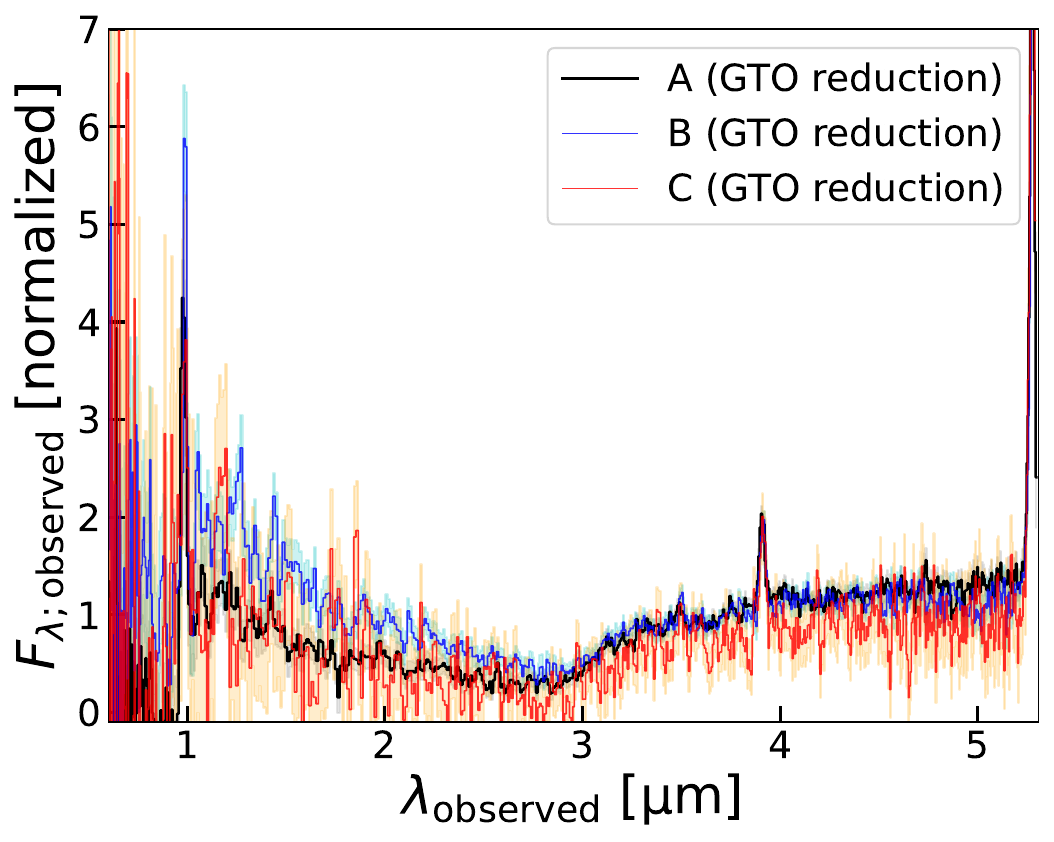}
    \caption{Comparison between three MSA PRISM spectra from the three images of \target. The spectra are reduced with the GTO pipeline and are normalized to the flux density at $\lambda = 3.6~{\rm \mu m}$. The rest-frame UV continuum of the spectrum from image B is bluer compared to the other two spectra, possibly due to a background subtraction issue.
}
    \label{fig:3img_com}
\end{figure}

\begin{figure*}
    \centering
    \includegraphics[width=0.9\textwidth]{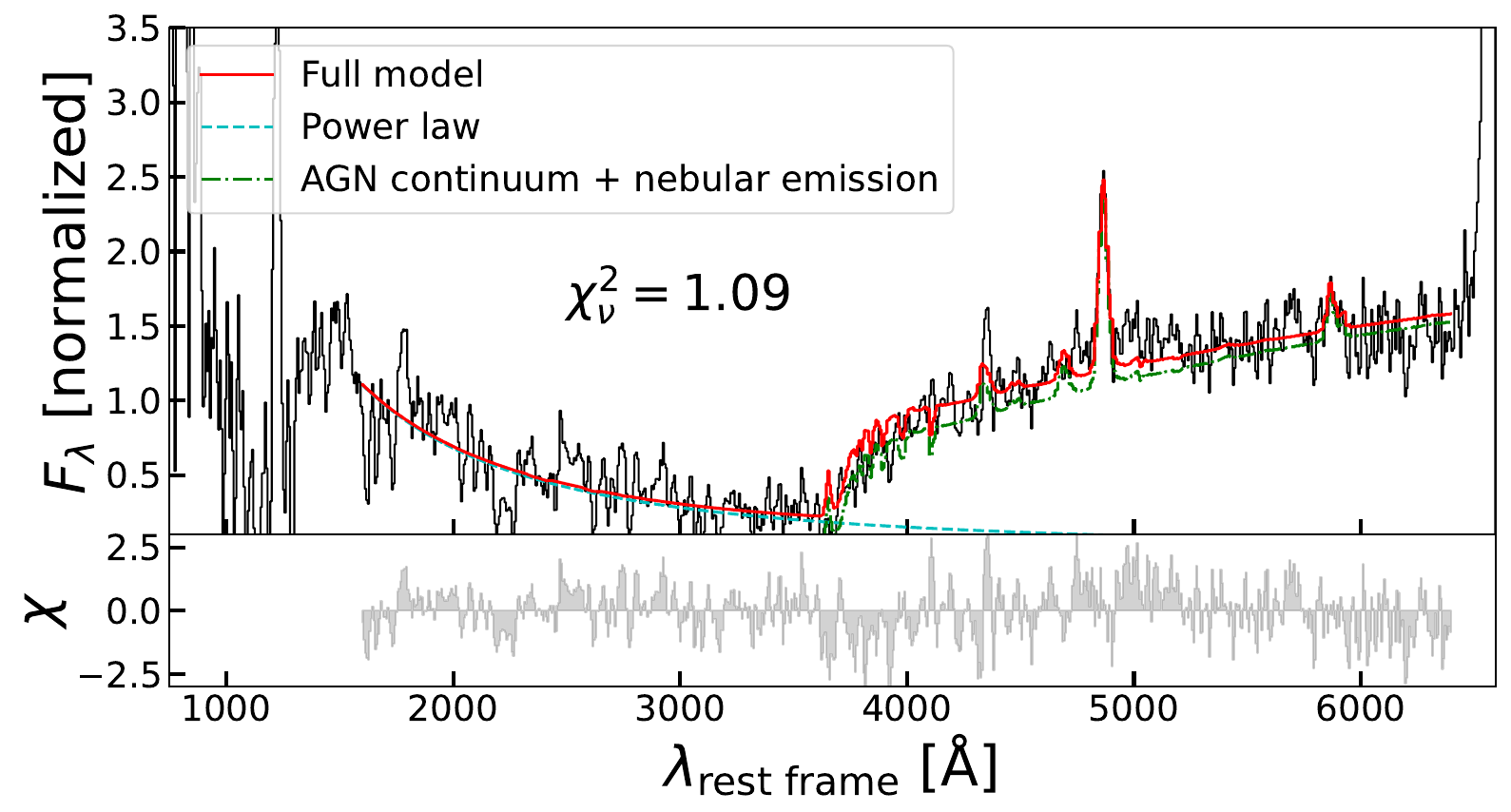}

    \caption{Fiducial fit for the integrated PRISM IFU spectrum of image A from \blackthunder\ in the rest-frame wavelength range of $1600-6400$ \AA.
    The observed PRISM spectrum is normalized to the flux density at $\lambda = 4260$ \AA\ in the rest frame of \target.
    }
    \label{fig:blackthunder_demo}
\end{figure*}

\begin{figure*}
    \centering\includegraphics[width=\columnwidth]{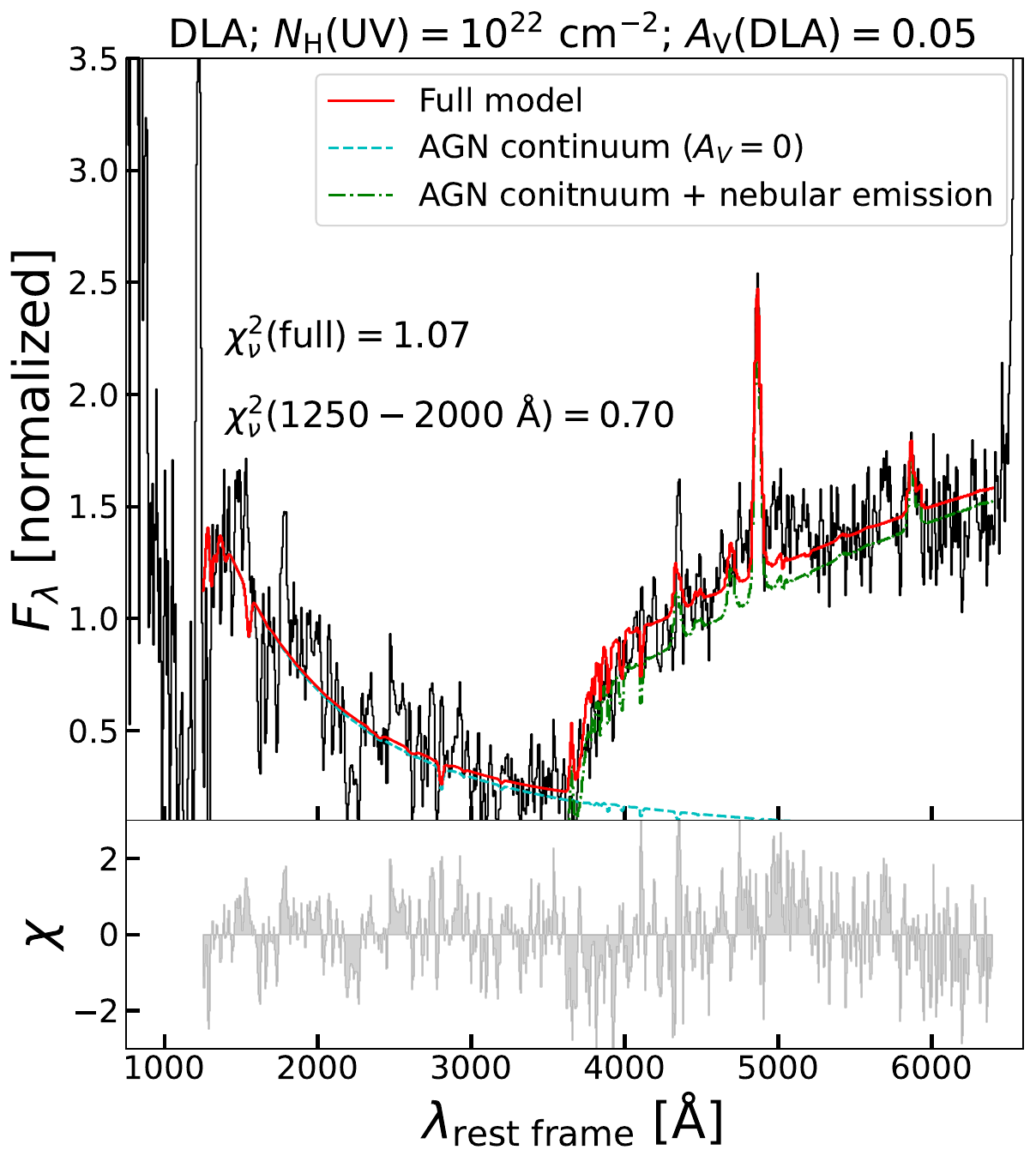}
    \includegraphics[width=\columnwidth]{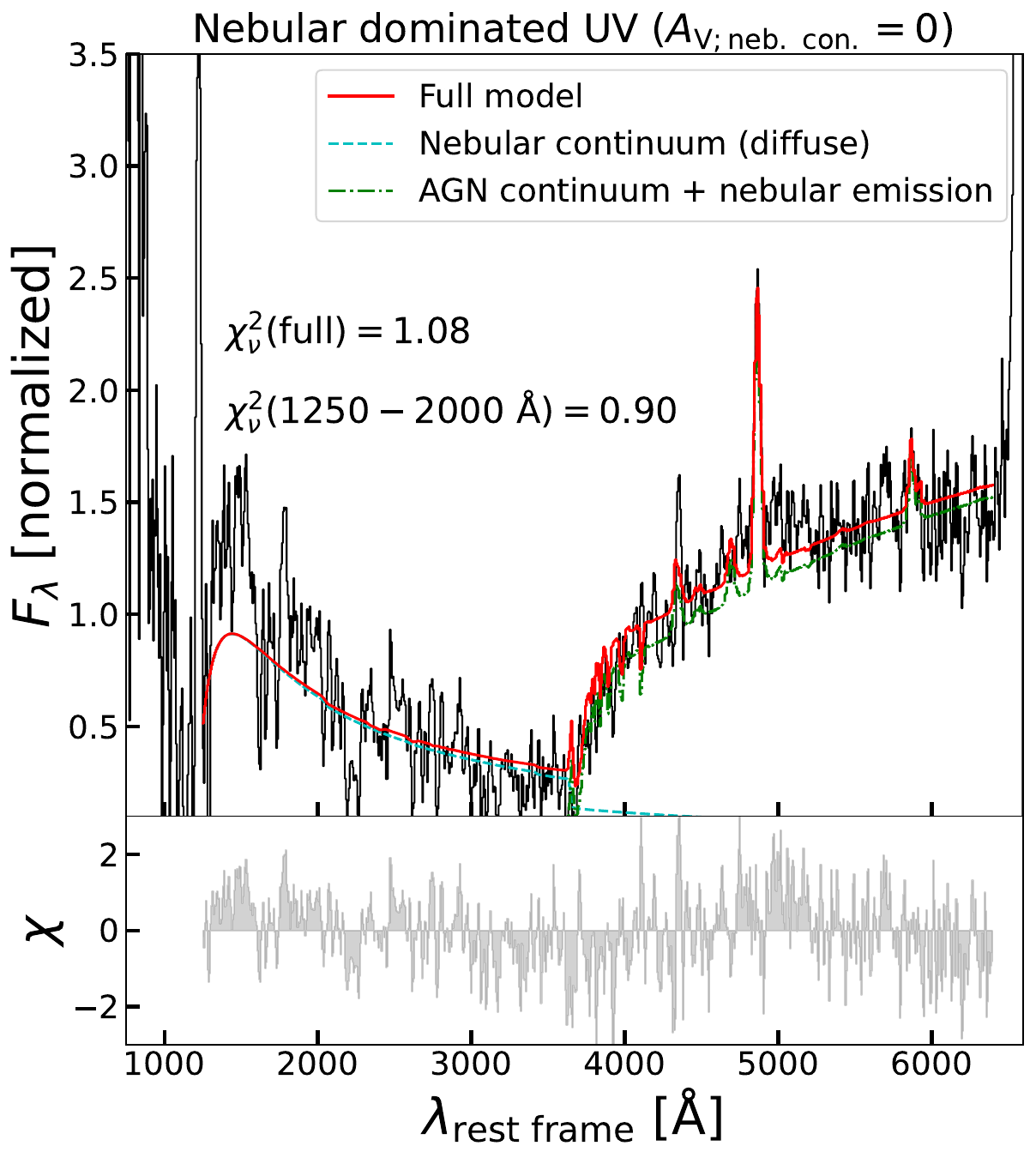}

    \caption{Fiducial fit for the integrated PRISM IFU spectrum of image A from \blackthunder\ in the rest-frame wavelength range of $1250-6400$ \AA\ with different assumptions in the UV.
    \textit{Left:} the UV continuum is assumed to be the AGN continuum emission attenuated by a slab of gas with $N_{\rm H}=10^{22}~{\rm cm^{-2}}$.
    \textit{Right:} the UV continuum is assumed to be a nebular continuum composed of a two-photon continuum, a hydrogen free-bound continuum, and a hydrogen free-free continuum.
    Both models assume no dust attenuation in the UV.
    The nebular continuum model provides a worse fit compared to the DLA model.
    }
    \label{fig:blackthunder_uv}
\end{figure*}

\begin{figure*}
    \centering
    \includegraphics[width=0.9\textwidth]{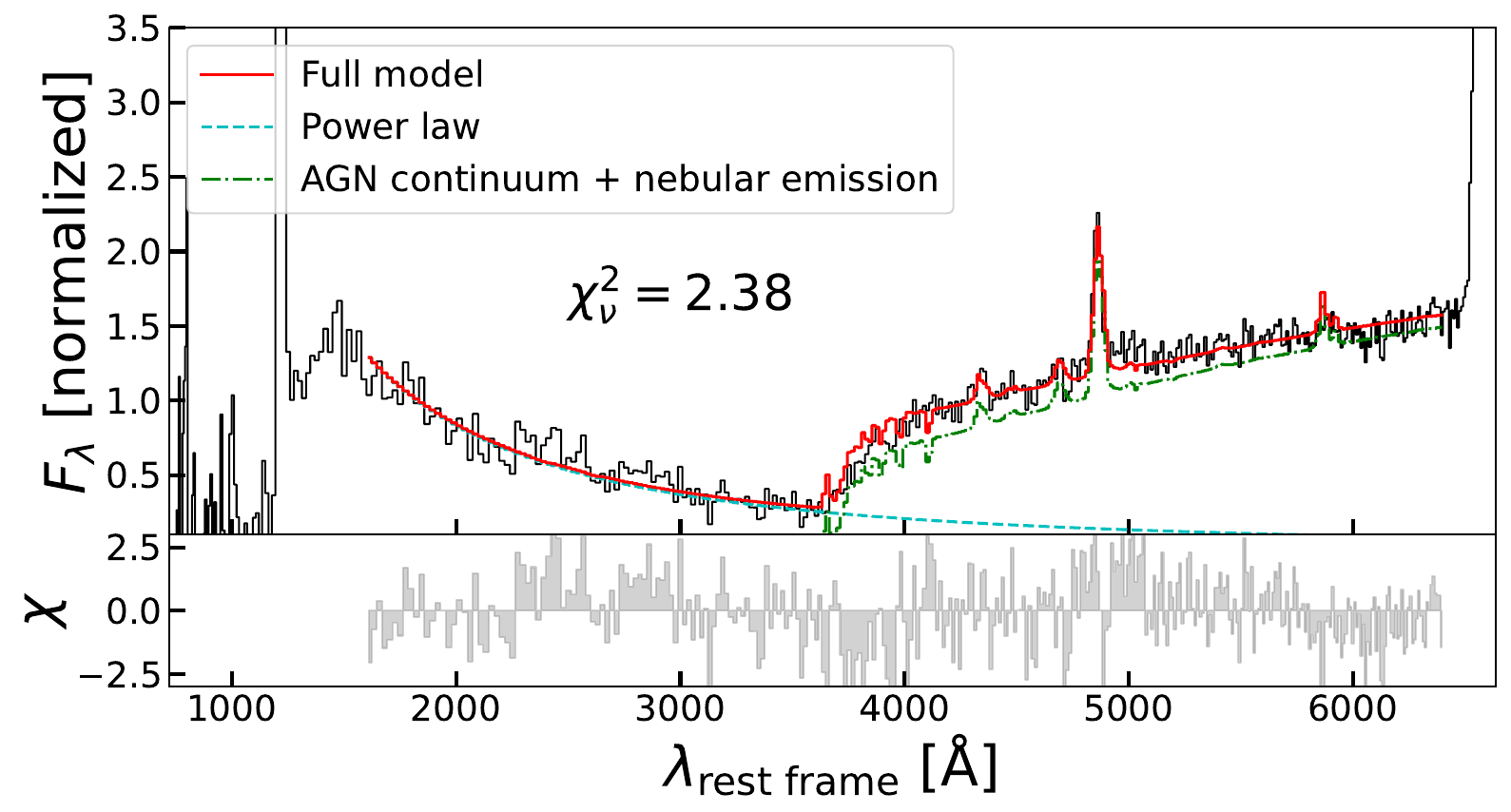}

    \caption{Fiducial fit for the image A of \target\ from the UNCOVER data release in the rest-frame wavelength range of $1600-6400$ \AA.
    This is the same spectrum that was fitted by \citet{mayilun_lrd_2024}, who obtained $\chi^{2}_{\nu}=2.74-4.32$ based on their models.
    The observed PRISM spectrum is normalized to the flux density at $\lambda = 4260$ \AA\ in the rest frame of \target.
    We adopted $5\sigma$-clipping during the fit.
    }
    \label{fig:ma_demo}
\end{figure*}

\begin{figure*}
    \centering\includegraphics[width=\columnwidth]{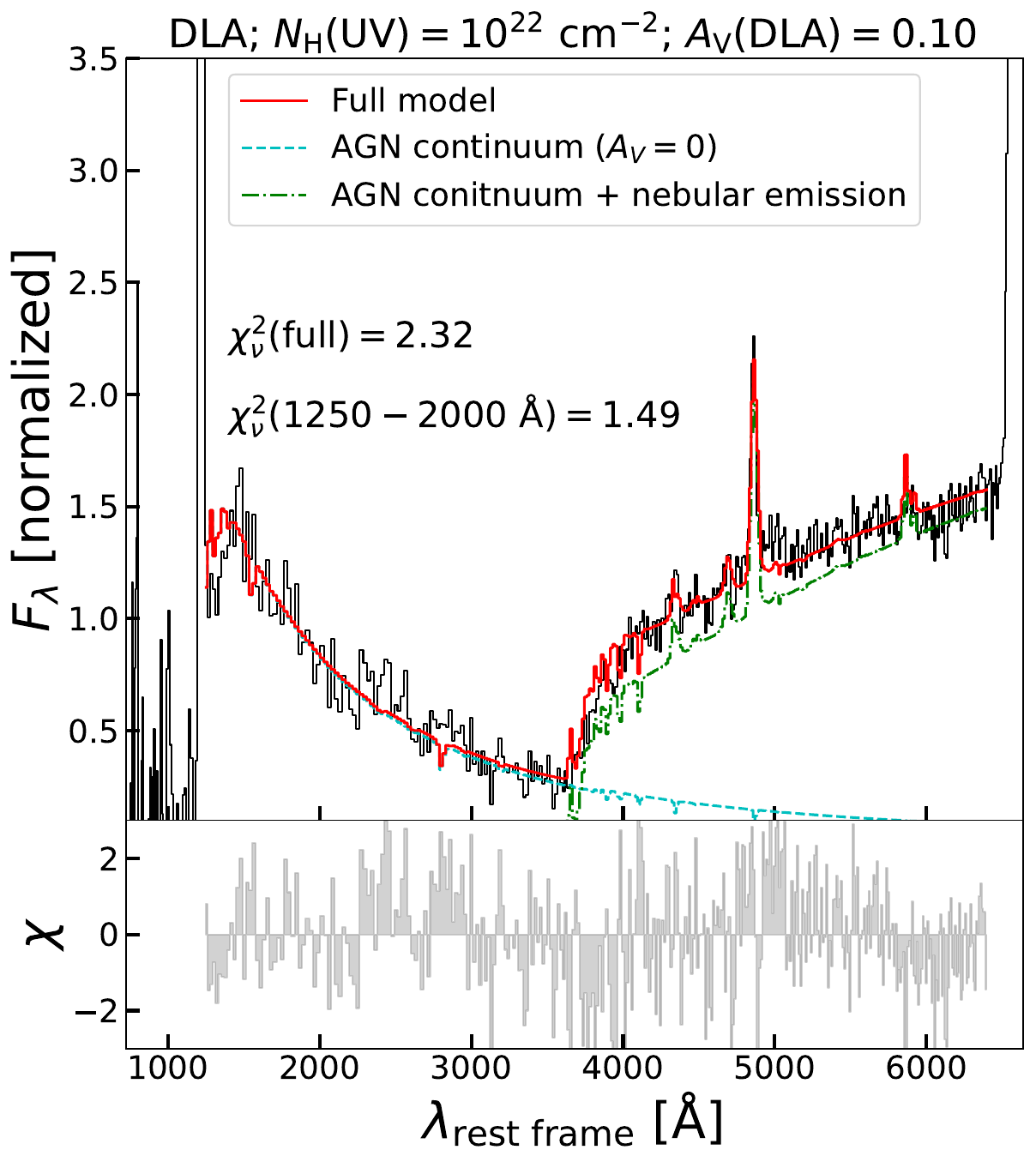}
    \includegraphics[width=\columnwidth]{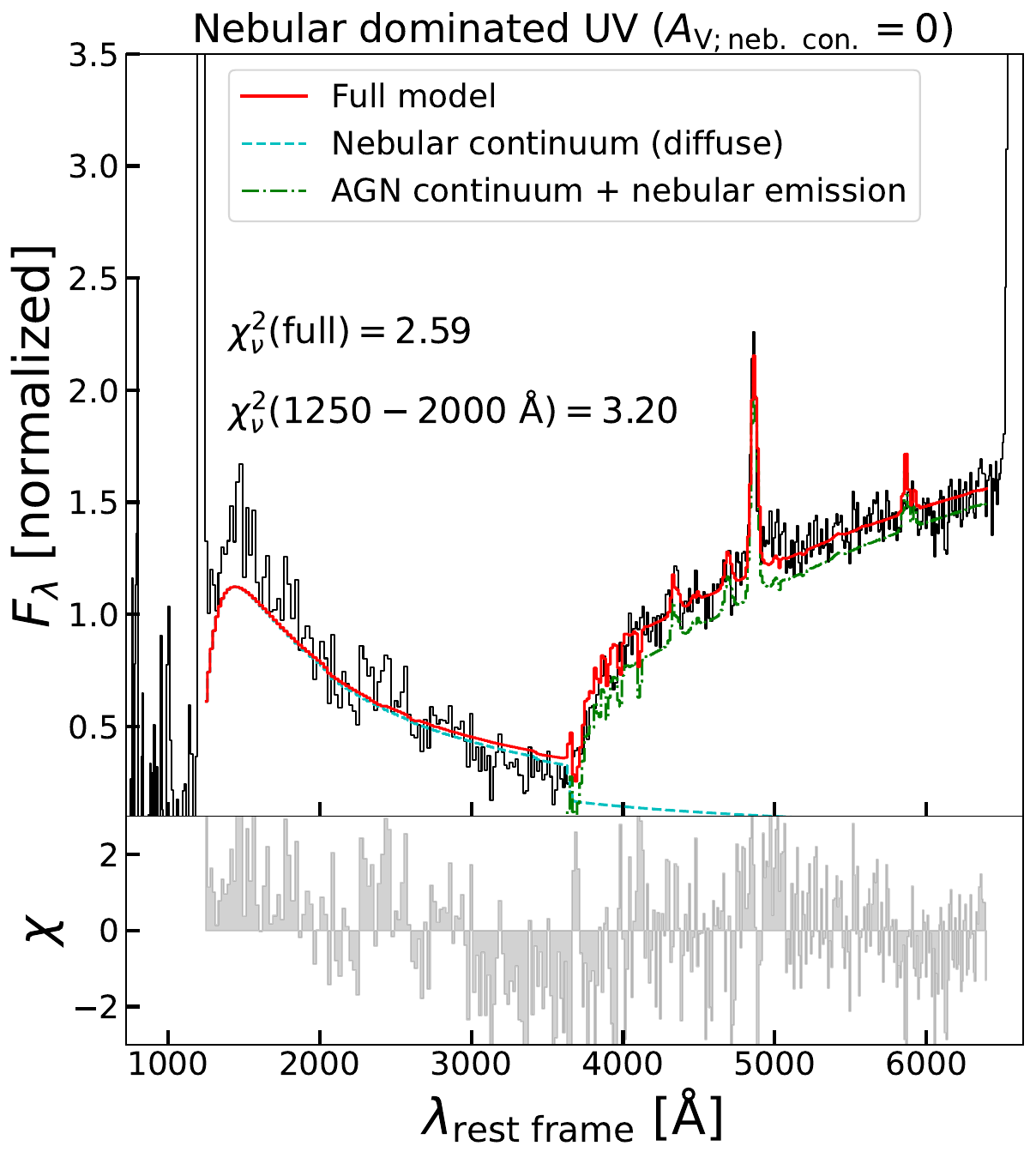}

    \caption{Fiducial fit for the spectrum of image A from UNCOVER DR4 in the rest-frame wavelength range of $1250-6400$ \AA\ with different assumptions in the UV.
    \textit{Left:} the UV continuum is assumed to be the AGN continuum emission attenuated by a slab of gas with $N_{\rm H}=10^{22}~{\rm cm^{-2}}$.
    \textit{Right:} the UV continuum is assumed to be a nebular continuum composed of a two-photon continuum, a hydrogen free-bound continuum, and a hydrogen free-free continuum.
    Both models assume no dust attenuation in the UV.
    The nebular continuum model provides a worse fit compared to the DLA model.
    }
    \label{fig:ma_uv}
\end{figure*}

In this section we show the fitting results for NIRSpec MSA and IFU PRISM spectra with different data reductions.
The first spectrum we used is the same spectrum adopted by \citet{furtak2024}, which combines MSA PRISM spectra from three images of \target\ following the reduction process detailed in \citet{furtak2024}.
The second spectrum we used is the PRISM spectrum integrated from the IFU observation of \blackthunder, of which is data reduction process is given in Section~\ref{sec:data}.
The \blackthunder\ PRISM spectrum is only for the image A of \target.
The third spectrum we used is the same spectrum adopted by \citet{mayilun_lrd_2024}, which is the MSA PRISM spectrum for image A of \target\ from the UNCOVER DR4.
The data reduction is given by \citet{uncover_dr4} and also summarized by \citet{mayilun_lrd_2024}.
We summarize the best-fit model parameters for these spectra in Table~\ref{tab:best_fit}.

Figure~\ref{fig:furtak_demo} shows our fiducial fit to the spectrum reduced by \citet{furtak2024}.
The photoionization model parameters we used to compute the attenuated AGN continuum are the fiducial set given in Table~\ref{tab:models}.
Compared to the fit to the A+C images, the best-fit model for the spectrum reduced by \citet{furtak2024} yields a larger $\chi^2_\nu$ (which is still significantly lower than the $\chi^2_\nu\sim 3$ reported by \citealp{mayilun_lrd_2024} using alternative models).
The UV part appears less well fitted by a featureless power law and shows continuous flux excess around 2000--3000 \AA.
The resulting UV slope is shallower than the one fitted for the A+C images, leading to a longer tail in the optical and thus requiring an $A_{\rm V}$ 0.26 magnitudes higher than that of the A+C images (cf., Table~\ref{tab:best_fit}).
Figure~\ref{fig:furtak_uv} shows fits to the \citet{furtak2024}'s spectrum with different UV models for a wavelength range of $1250-6400$ \AA.
As can be seen in Figure~\ref{fig:spec_qso1}, \citet{furtak2024}'s spectrum shows a smoother DLA-like feature around Ly$\alpha$.
This actually makes the DLA model with $N_{\rm H}=10^{22}~{\rm cm^{-2}}$ with a small amount dust attenuation [$A_{\rm V}{\rm (DLA)}=0.22$] fail to describe the UV part well (left panel), assuming the intrinsic UV spectrum has an AGN origin.
In comparison, a nebular dominated UV continuum with $T_{\rm e}=30,000$ K is able to the fit the UV part better, especially the turn-over feature around Ly$\alpha$.

The different result from the \citet{furtak2024}'s spectrum is likely caused by the inclusion of image B.
As shown in Figure~\ref{fig:3img_com}, when the spectra from three images are normalized to the rest-frame optical, the UV continuum of image B shows a different slope compared to the other two images.
As we have noted in Section~\ref{sec:data}, the MSA PRISM spectrum from image B (included in the combined spectrum of \citealp{furtak2024}) possibly has a background subtraction issue, which might affect the DLA-like feature in the UV and lead to the different fitting result.
In addition, we caution that in our analyses, the normalization of the nebular continuum is set to a free parameter during the fit.

Figures~\ref{fig:blackthunder_demo} and \ref{fig:blackthunder_uv} show the corresponding results for the integrated IFU PRISM spectrum from the observations of \blackthunder.
Here the spectrum is taken from the central $0\farcs2$ of image A only.
In Table~\ref{tab:best_fit}, one can see that the best-fit model has the steepest UV slopes among all the fits.
The best-fit $A_{\rm V}$ is 0.09 magnitudes lower than that of the A+C images.
Similar to the results of the A+C images, a DLA model is favored over the nebular dominated model for the UV part of the spectrum.

Finally, we show our fit to the spectrum from the UNCOVER DR4 adopted by \citet{mayilun_lrd_2024} in Figure~\ref{fig:ma_demo}.
This reduced spectrum produces the largest $\chi^2_{\nu}$, which is still significantly lower than the fitting results from \citet{mayilun_lrd_2024}.
The comparison between the DLA fit and the nebular continuum fit is shown in Figure~\ref{fig:ma_uv}.
Similar to the result based on image A from \blackthunder, the DLA model provides a better fit compared to the nebular continuum model.
This further indicates the different result obtained from the PRISM spectrum of \citet{furtak2024} might be caused by the spectrum from image B.

\section{Additional analyses and information on variability}
\label{appendix:var}

In this appendix, we provide some additional multi-epoch analyses on the variability of \target.
We also discuss the impact of the data reduction on the variability analysis.

\subsection{Comparison of IFU and MSA spectra of image A}

\begin{figure}
    \centering
    \includegraphics[width=\columnwidth]{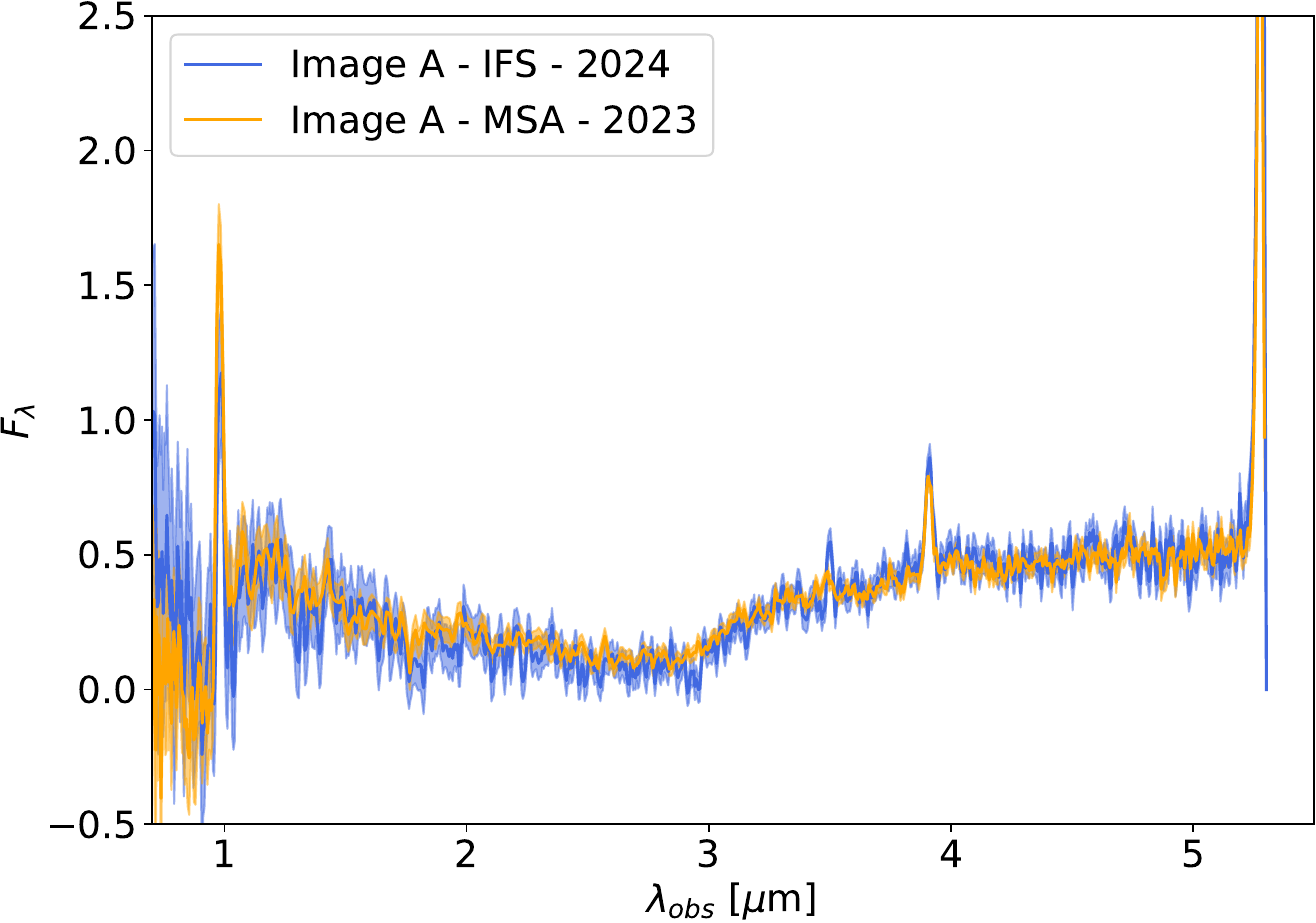}
    \caption{
    Comparison of the two low resolution spectra of Image A of Abell2744-QSO1 obtained with the MSA in November 2023 (orange, reduced with the GTO pipeline) and with the IFU in Nov 2024 through the BlackTHUNDER programme (blue).
    There are not significant variations between the two spectra.
    }
    \label{fig:Comp_epochs}
\end{figure}

As already discussed in the main text, the BlackTHUDER IFU spectrum taken in November 2024, provides an additional spectroscopic epoch of image A, relative to the MSA spectrum taken in July 2023 (i.e., 1.7 months earlier in the source rest-frame).
Figure~\ref{fig:Comp_epochs} compares the PRISM spectrum of image A extracted from the IFU Cube 
with the PRISM MSA spectrum. The two spectra are consistent with each other within the errors, implying that the source has probably been stable within the $\sim 2$ months in the rest frame covered by the two observations, although intra-epoch variability cannot be excluded. 

It is, however, intriguing that some changes are seen in the emission lines. The H$\beta$ is slightly higher in the IFU spectrum, although within the uncertainties (both Poissonian and calibration). However, the variation is not seen in terms of equivalent width (Section~\ref{sec:variability}; i.e., when the spectra are normalized to the continua), meaning that the slight variation of H$\beta$ may be simply due to a calibration effect.

%More puzzling is the variation of H$\gamma$, which seems significantly stronger in the IFS spectrum, although the variation is still at the 2$\sigma$ level. 
Also, H$\gamma$ seems stronger in the IFU spectrum. However, this is still only at the $2\sigma$ level, and it is possible that this is due to an artifact in the IFU PRISM cube.
Alternatively, it may trace real variability in the absorbing medium (either in the Balmer absorption or dust absorption, if located in the outer parts of the BLR). Or, the IFU may be tracing gas on more extended scales, and less obscured, that is missed by the MSA. There is no clear evidence for extension of the Balmer lines, but it is also true that image A is located at the edge of the shutter, hence some structural variation close to the resolution limit may result in flux differences between the MSA and the IFU.

Some flux variation, again only at the $\sim 2\sigma$ level is seen in Ly$\alpha$.
Once again, variability of the Ly$\alpha$ emission lines between the two epochs cannot be excluded.
% but the significance of the variation indicates that it may also be a noise fluctuation. 
However, Ly$\alpha$ is spectrally unresolved or marginally resolved, hence the apparent drop in the peak flux of Ly$\alpha$ is mostly due to the MSA having higher resolution (hence the line more peaky) than in the IFU spectrum.

\subsection{Issues associated with custom-extraction from the 2D spectra}

\begin{figure}
    \centering
    \includegraphics[width=\columnwidth]{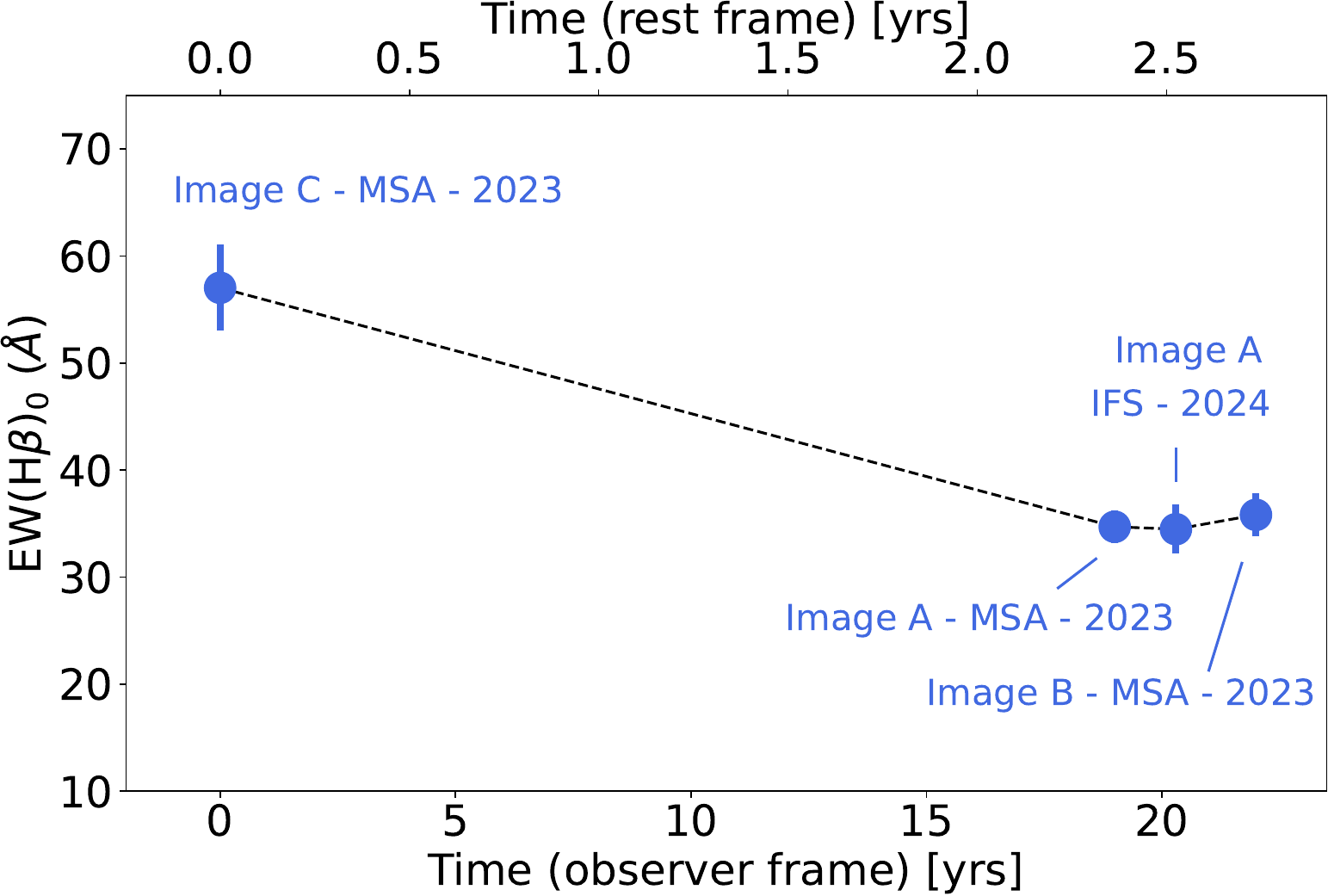}
    \caption{
    Same as Figure~\ref{fig:EWs}, but where now the spectra are not obtained from the full-shutter pipeline extraction, which takes properly into account the path- and diffraction-losses, but by extracting the spectrum from 3-pixel apertures from the 2D spectra (which does not take into account for path- and diffraction-losses). In this case the EW variation of H$\beta$ is still highly significant, but reduced. The effect is likely arising from the inappropriate spectral extraction in this case.
    }
    \label{fig:EWs_3pix}
\end{figure}

As already pointed out in Section~\ref{sec:archival_data},
the correction of path- and diffraction-losses is very accurate for unresolved sources like this one. The pipeline extraction of the spectra from the full-shutter takes into account of these aspects.
On the contrary, extracting the 1D spectra with bespoke apertures on the 2D spectra is deprecated if one wants to carefully account for the flux calibration, as this procedure does not take into account the path- and diffraction-losses, and this can introduce issues when looking for variability. Especially for sources located close to the edge of the shutter, using extraction from the 2D spectra can easily lead to inappropriate flux calibration, and the effect can also be more pronounced at the location of strong emission lines, whose PSF and diffraction losses may not be captured by a fixed or bespoke aperture on the 2D spectrum. As stated, the rigorous approach is to take the full-shutter spectra extracted by the pipeline. In order to illustrate the effect of using the bespoke extraction from the 2D, Figure~\ref{fig:EWs_3pix} shows the quantities same as Figure~\ref{fig:EWs}, i.e. the variation of EW among different observations, but where the EWs are measured on spectra extracted from 3-pixel aperture on the 2D spectrum (i.e. similar to the optimized extraction used in other works), instead of the pipeline full-shutter extraction. The variation in EW in image C is still highly significant, but it is now reduced to about 50\%-60\%. This is not due to an extended component of H$\beta$ (which might result in picking more H$\beta$ in the full shutter relative to the 3 pixel extraction). Indeed, we have checked in the IFU cube that the EW(H$\beta$) changes less than 2\% using spectra extracted from $0\farcs2$ and $0\farcs5$ apertures. The difference is instead likely originating from the fact that the bespoke extraction from the 2D spectrum does not incorporates path- and diffraction-losses, which may be particularly problematic around strong emission lines and especially for sources close to the edge of the spectrum.

\end{appendix}

%   \bibliographystyle{mnras} % style aa.bst
%   \bibliography{ref} % your references ref.bib

% Don't change these lines
\bsp	% typesetting comment
\label{lastpage}
\end{document}